\documentclass[letterpaper]{sig-alternate-10pt}

\usepackage{epsfig, amsmath, amssymb, subcaption, color, cite, url, multirow, setspace, epstopdf, algorithm, mathptm, relsize, listings, enumitem}

\usepackage{balance}

\usepackage{microtype} 
\usepackage[normalem]{ulem}
\newcommand\redsout{\bgroup\markoverwith{\textcolor{red}{\rule[0.5ex]{2pt}{1pt}}}\ULon}

\usepackage{array}
\newcolumntype{L}{>{\centering\arraybackslash}m{1cm}}

\usepackage[noend]{algorithmic}

\usepackage[T1]{fontenc}
\usepackage[font=small,labelfont=bf,tableposition=top]{caption}

\global\def\@maketitle{%
  \newpage
  \begin{center}%
  \let \footnote \thanks
  \null
    \vskip -.3em%
    {\ttlfnt \@title \par}%
    \vskip 1em%
    {\large
      \lineskip .5em%
      \begin{tabular}[t]{c}%
        \@author
      \end{tabular}\par}%
    \vskip 1em%
    {\large \@date}%
  \end{center}%
  \par
  \vskip 2em}

\DeclareCaptionLabelFormat{andtable}{#1~#2  \&  \tablename~\thetable}

\newcommand{\name}[0]{{\tt V-Health}}
\newcommand{\nameS}[0]{{\tt V-Health} }

\newcommand{\numCycles}[0]{{13,377} }
\newcommand{\numTests}[0]{{50} }
\newcommand{\numBatt}[0]{{15} }
\newcommand{\cumMonth}[0]{{72} }

\begin{document}

\title{
How Long Will My Phone Battery Last? 
}
\numberofauthors{2} 
\author{
\alignauthor
Liang He\\
       \affaddr{University of Colorado Denver}\\
       \affaddr{1380 Lawrence Street}\\
       \affaddr{Denver, CO, 80217}\\
       \email{liang.he@ucdenver.edu}
\alignauthor Kang G. Shin\\
       \affaddr{University of Michigan at Ann Arbor}\\
       \affaddr{2260 Hayward St.}\\
       \affaddr{Ann Arbor, MI, 48109}\\
       \email{kgshin@umich.edu}
}

\maketitle

\begin{sloppypar}
\begin{spacing}{0.99}

\begin{abstract}
Mobile devices are only as useful as their battery lasts.
Unfortunately, the operation and life of a mobile device's battery degrade over time and usage. 
The state-of-health (SoH) of batteries quantifies their degradation, but mobile devices are
unable to support its accurate estimation --- despite its importance --- due mainly to their limited 
hardware and dynamic usage patterns, causing various problems such as unexpected device 
shutoffs or even fire/explosion. 
To remedy this lack of support, we design, implement and evaluate \name, a low-cost 
user-level SoH estimation service for mobile devices based only on their battery voltage, 
which is commonly available on all commodity mobile devices. 
\nameS also enables four novel use-cases that improve mobile users' experience 
from different perspectives.
The design of \nameS is inspired by our empirical finding that the relaxing voltages of a device 
battery {\em fingerprint} its SoH, and is steered by extensive measurements with \numBatt batteries used 
for various commodity mobile devices, such as Nexus 6P, Galaxy S3, iPhone 6 Plus, etc. 
These measurements consist of \numCycles battery discharging/charging/resting cycles 
and have been conducted over \cumMonth months cumulatively.
\nameS has been evaluated via both laboratory experiments and field tests 
with multiple Android devices over $4$--$6$ months, showing $<$$5\%$ error in SoH estimation. 
\end{abstract}

\maketitle

\section{Introduction}
\label{sec:introduction}

Apple announced a free-replacement program of iPhone 6S batteries in Nov.~2016
\cite{iphone6sbatteryreplacement}, due to frequent users' complaints on the phone shutoffs even 
when showing $10$--$30\%$ remaining power, and concluded faster-than-normal battery degradation 
to have caused the problem~\cite{iphone6sshutdown}. 
Similar unexpected phone shutoffs also occurred on devices such as Nexus 6P~\cite{nexus6p}, 
Galaxy S4~\cite{galaxys4}, iPhone 5~\cite{Iphone51}, to name a few.

\begin{figure}[t]
\begin{minipage}{1\columnwidth}
\centering
{\includegraphics[width=1\columnwidth]{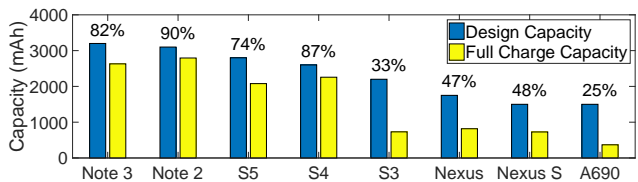}}
\caption{{\bf Deficient SoH information on mobile devices:} Android's BatteryManager
specifies all of these batteries to be in {\tt good} health even though up to $75\%$ capacity 
degradation is observed.}
\label{fig:ReducedCapacity2}
\end{minipage}
\end{figure}

These incidents imply the inability to accurately answer a simple question ``{\em how long will 
my phone battery last?}", which means (i) the remaining battery life (e.g., relative to battery degradation 
and thus its warranty period) or (ii) remaining device operation time until the battery runs out 
(i.e., the operation time with a single charge).
The answer relies on the quantification of battery's capacity degradation, which is traditionally 
captured by its {\em state-of-health} (SoH), defined as the ratio of the battery's full charge 
capacity to the designed capacity~\cite{bolunthsis,B,Plett20112319}.
Unfortunately,  mobile devices are not equipped with the capability necessary for 
accurately quantifying its battery's SoH.
For example, Android only specifies battery health as {\tt good} or {\tt dead}, without any quantified 
information~\cite{batterymanager}. Fig.~\ref{fig:ReducedCapacity2} plots our measurements on 
the battery SoH of $8$ Android phones with a battery tester: all of these batteries are tagged 
as {\tt good} although their capacities are observed to have degraded by as much as $75\%$.

The non-existence of quantified battery SoH introduces errors in estimating the devices' 
remaining power (i.e., state-of-charge (SoC)) 
and thus shutting them off prematurely or unexpectedly~\cite{Hoque,Pathak}, as happened to 
iPhone 6S and other devices, because batteries' SoC, by definition, is grounded on 
their SoH~\cite{A,D,impedancetrack}.
It also prevents the comparison of a device's battery life against its warranty period, as users will 
not know whether the shortened device-operation is due to system updates\footnote{For example, 
Android 6.0 Marshmallow is reported to reduce device operation when first launched~\cite{Marie}.} 
and app installations~\cite{Min:2015:PPS:2809695.2809728,7835690}, or because of battery degradation.
Last but not the least, this inaccurate SoC easily leads to battery over-charge/deep-discharge, 
accelerating SoH degradation and thus increasing the SoC estimation error~\cite{A,Hoque,K}, 
thus forming a positive feedback loop between the two~\cite{B}.

The deficiency of health information on mobile devices' batteries stems from the non-existence of
compatible methods to estimate their SoH. 
Most existing SoH estimation methods require either battery parameters, determination 
of which is beyond mobile devices' 
capability due to hardware limitation (e.g., impedance~\cite{L,Eddahech2012487,Widodo201111763} 
and ultrasonic echo~\cite{K}), or specific applicable conditions that do not always hold due to devices' 
dynamic usage patterns (e.g., small current to fully charge and discharge~\cite{G, impedancetrack, max17040}). 
Moreover, even Coulomb counting --- the most widely-deployed SoH estimation method 
via current integration~\cite{impedancetrack,Ng20091506} --- is not supported well on mobile devices 
because (i) not all power management ICs~(PMICs) of mobile devices support electric current 
sensing~\cite{vedge}, making Coulomb counting infeasible; (ii) the PMIC-provided current information 
is too coarse and lacks real-time capability, even when available~\cite{max17047,Giovino}. 
Such unreliable current information on mobile devices is also reported by Ampere,
a current sensing app with millions of downloads~\cite{ampere}.

\begin{figure}[t]
\begin{minipage}{1\columnwidth}
\centering
{\includegraphics[width=1\columnwidth]{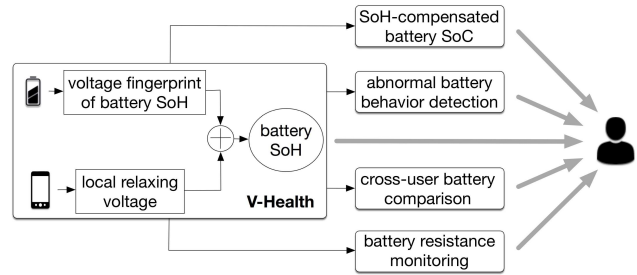}}
\caption{{\bf \nameS Overview:} estimating mobile devices' battery SoH based only on voltage information, and enabling 4 novel use-cases.}
\label{fig:overview_function}
\end{minipage}
\end{figure}

To remedy the above problems, we propose \name, a {\em user-level} SoH estimation service for 
mobile devices based solely on their battery {\em voltage}, and is thus compatible to all commodity mobile 
devices with voltage sensing and processing capabilities, such as smartphones, tablets, smartwatches, 
and even electric vehicles.
With the thus-estimated SoH, \nameS also enables 4 novel use-cases that improve 
user experience: (i) SoH-compensated SoC estimation that alleviates unexpected device shutoffs, 
(ii) detection of abnormal battery behaviors that reduces safety risks such as thermal runaway, 
(iii) cross-user battery comparison that identifies battery-friendly/harmful usage patterns, and 
(iv) battery resistance monitoring. Fig.~\ref{fig:overview_function} presents an overview of \name.

The design of \nameS is inspired by our empirical finding: the relaxing battery voltages --- a time 
series of battery voltages when resting it after its charge/discharge --- {\em fingerprinting} its SoH, 
and this voltage--SoH relationship holds reliably for all same-model batteries.  
We uncover and validate this property via machine learning and based on 
extensive measurements with \numBatt batteries used 
for various mobile devices, such as Nexus 6P, Nexus 5X, Xperia Z5, Galaxy S3, iPhone 6 Plus, etc., 
consisting of a total of \numCycles discharging/charging/resting cycles and have been collected 
over \cumMonth months cumulatively. 

Resting battery to collect its relaxing voltages is not always feasible for mobile devices because they draw 
dynamically changing amounts of current from batteries {\em continuously}, 
even when idle~\cite{backgroundactivities,7530866}. 
\nameS exploits over-night device charging to collect the relaxing voltages, which (i) rests device battery 
after fully charging it~\cite{Yevgen, iCharge}, (ii) offers stable battery conditions in both device operation and 
thermal environment, (iii) masks the disturbances caused by device usage behaviors, and 
(iv) is frequently done by users~\cite{Ferreira,deviceanalyzer,Nilanjan} --- our dataset 
of $976$ device charging cases collected from $7$ users shows that $34\%$ of them are over-night charge 
lasting $6$+ hours and are long enough to rest the battery once fully charged. 
This way, \nameS does not degrade user experience, as the external charger supplies 
the power needed for information reading/logging.  

This paper makes the following main contributions:
\begin{itemize}[noitemsep]
\item Discovery of the correlation between relaxing battery voltages and their SoH, 
uncovering the feasibility of voltage-based SoH estimation (Sec.~\ref{sec:overview});
\item Design and implementation of \name, an SoH estimation service for mobile devices via voltage 
fingerprinting, neither requiring additional hardware support nor incurring energy overhead that degrades 
user experience (Secs.~\ref{sec:fingerprintmp} and \ref{sec:voltagecollection}); 
\item Evaluation of \nameS using both laboratory experiments and field-tests on multiple devices 
over $4$--$6$ months, showing $<$$5\%$ SoH estimation error (Sec.~\ref{sec:evaluations}); 
\item Demonstration of 4 novel use-cases enabled by \nameS (Sec.~\ref{sec:usecases}).
\end{itemize}

\begin{figure*}
\centering
\begin{minipage}{0.66\columnwidth}
\centering
{\includegraphics[width=1\columnwidth]{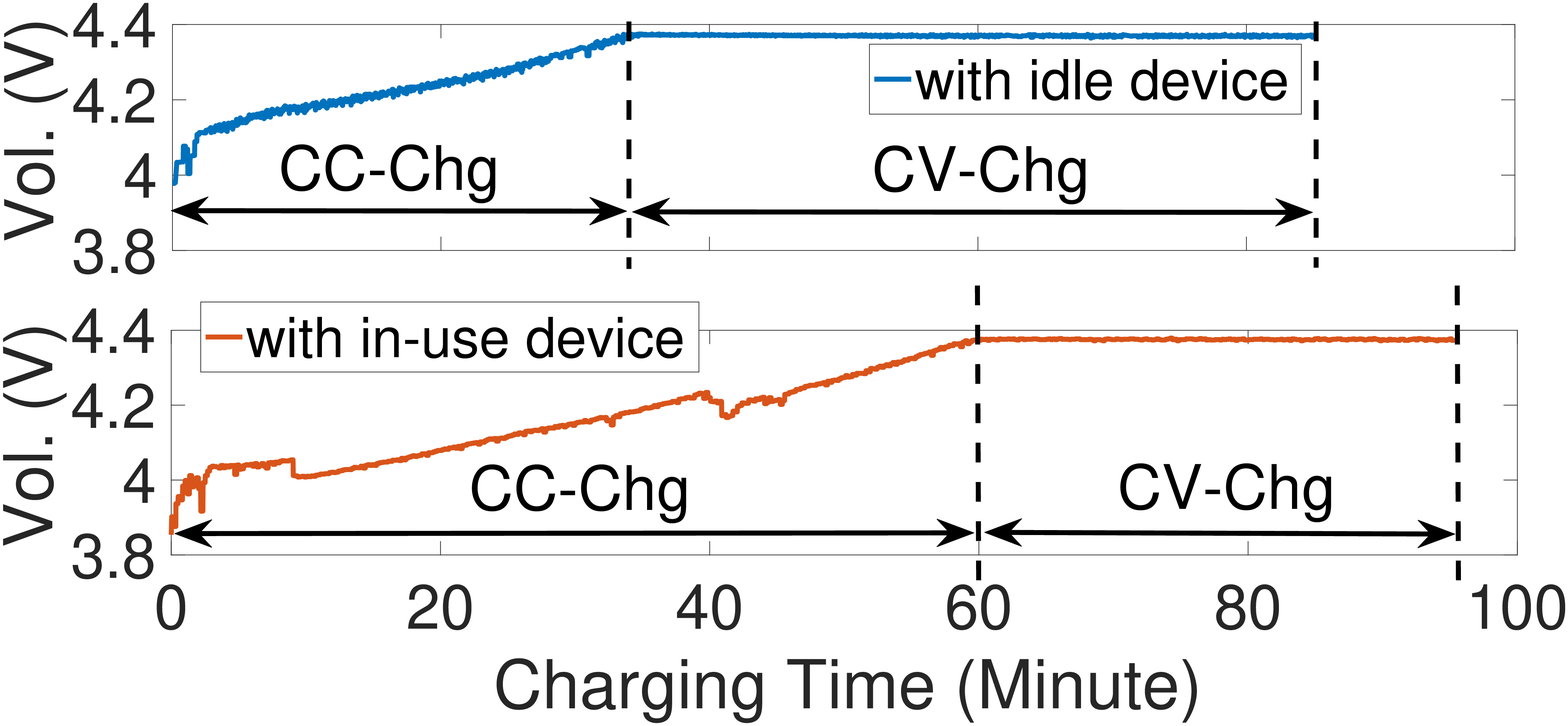}}
\caption{{\bf Device usage behavior during charging matters to \cite{C}:} battery voltage rises 
much slower when the phone is in active use, degrading the SoH estimation accuracy of~\cite{C}.}
\label{fig:voltageduringCCChg}
\end{minipage}
\hfill
\begin{minipage}{1.34\columnwidth}
\centering
\begin{subfigure}{0.49\columnwidth}
{\includegraphics[width=1\columnwidth]{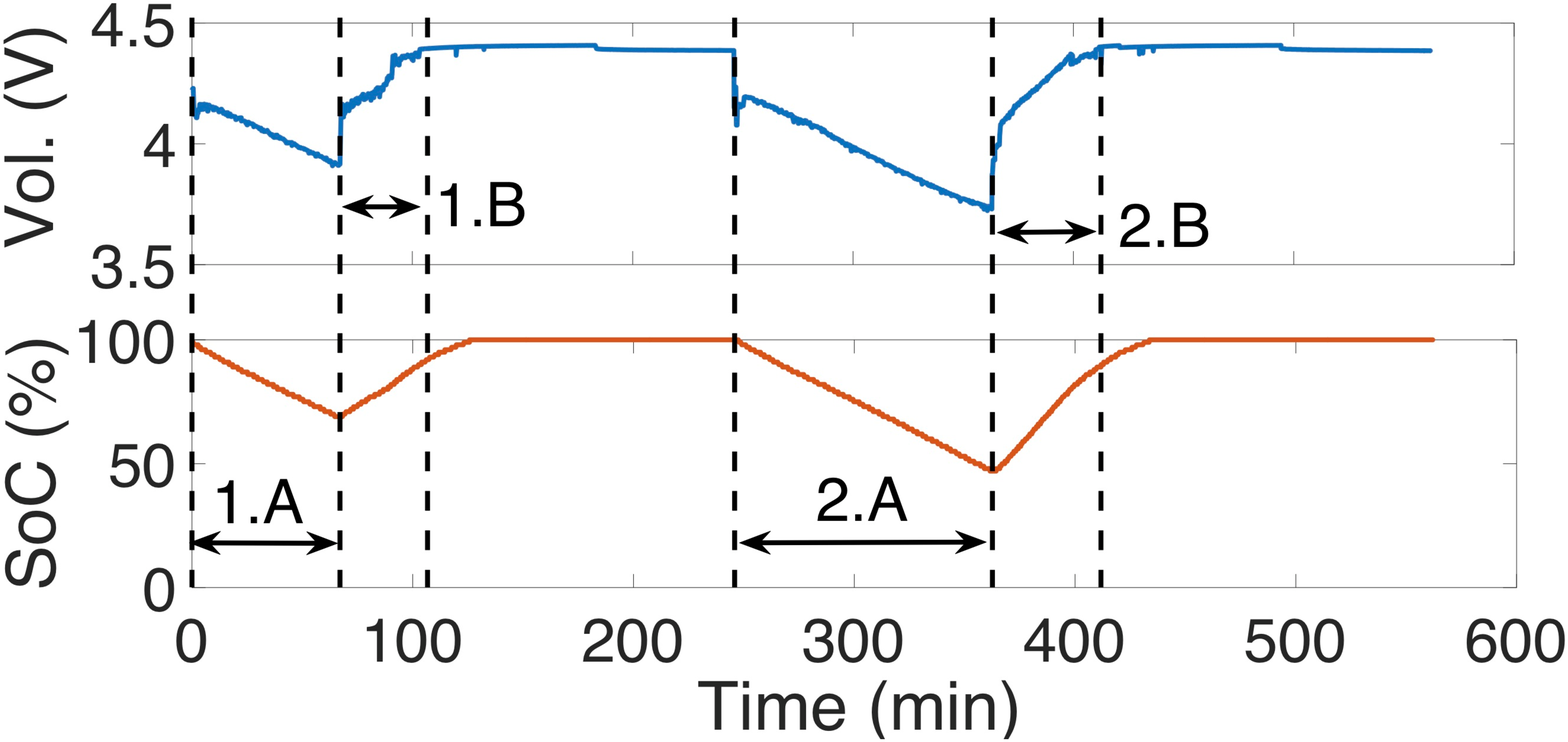}}
\caption{Two consecutive charges of an idle phone}
\end{subfigure}
\hfill
\begin{subfigure}{0.49\columnwidth}
{\includegraphics[width=1\columnwidth]{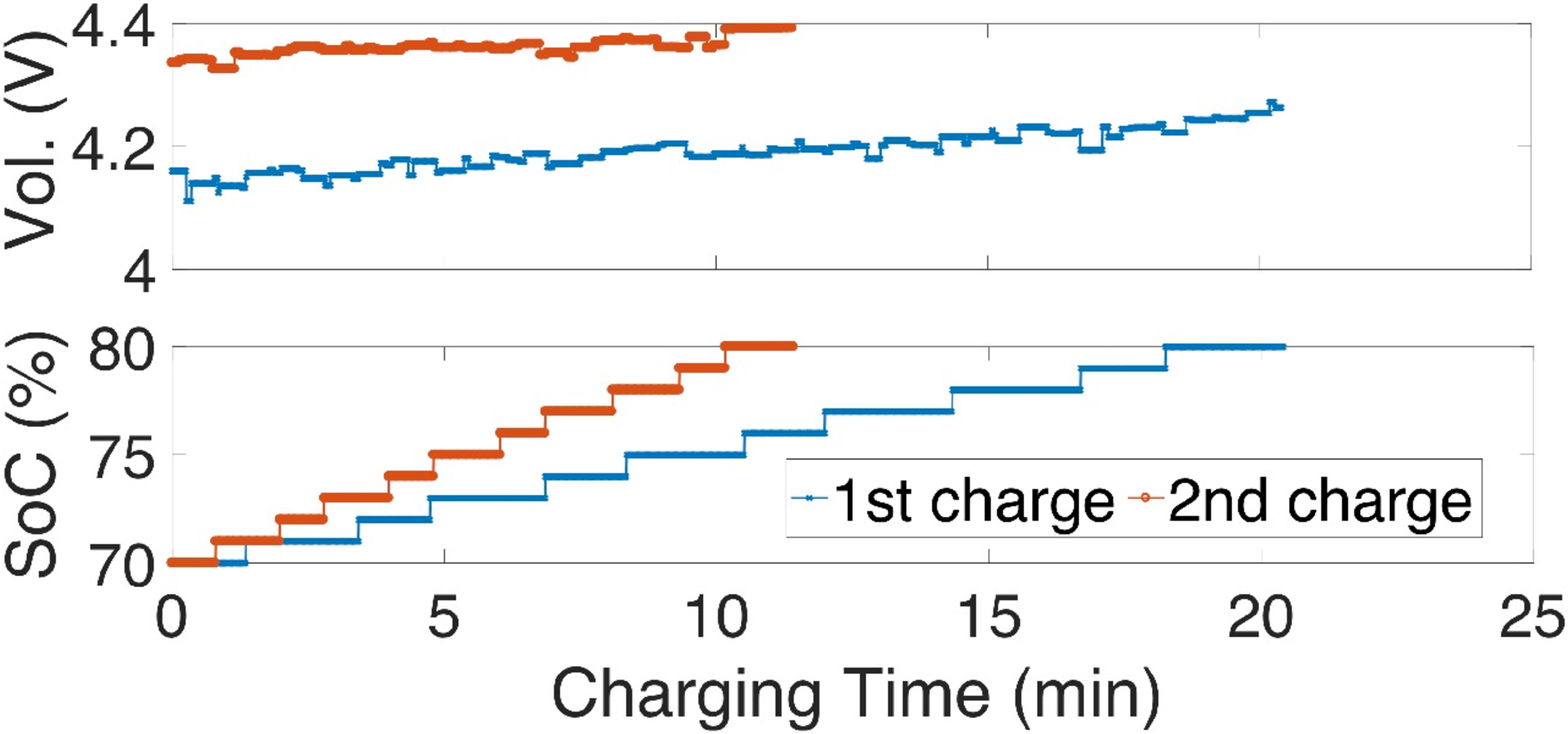}}
\caption{Both voltages and charging durations vary}
\end{subfigure}
\caption{{\bf  Device usage behavior before charging matters to \cite{C}}: (a) two consecutive charges 
of an idle Nexus 6P phone after discharged to different SoCs (1.A and 2.A); (b) the voltage--time relationship 
varies (part of 1.B and 2.B), 
degrading the SoH estimation accuracy of~\cite{C}.}
\label{fig:beforechargeoperation}
\end{minipage}
\end{figure*}

\section{Related Work}
\label{sec:relatedwork}

Accurate SoH estimation is crucial for battery management~\cite{Waag2014321,Zhang20116007}, 
which has been studied extensively based on various battery parameters such as voltage~\cite{A,B,E}, current~\cite{N,O,P,Hoque,J}, 
open-circuit-voltage~(OCV)~\cite{D,F,G}, SoC~\cite{H,M}, resistance~\cite{Q}, impedance~\cite{L,Eddahech2012487,Widodo201111763}, and even ultrasonic echo~\cite{K}.
These SoH estimation methods, albeit reported to be accurate, cannot be deployed on mobile devices 
due to their limited hardware support and dynamic operating conditions.

Mobile devices offer limited hardware support for sensing, rendering some of the needed battery information unavailable.
For example, the battery impedance needed in~\cite{L,Eddahech2012487,Widodo201111763} requires 
a specialized equipment to collect, costing as much as  \$5,000 apiece.
Actually, even the relatively easy-to-measure electric current --- the foundation of the most widely-deployed 
SoH estimation method, {\em Coulomb counting} --- is not always available on mobile devices~\cite{vedge}, 
and suffers from poor accuracy and lacks timeliness even when available. We will elaborate more on 
the insufficient current sensing on mobile devices in Sec.~\ref{sec:motivation}.
Also, battery information such as OCV and SoC requires specific conditions to be met for their accurate estimation. 
For example, OCV and SoC can be accurately estimated only when the battery has been charged/discharged 
with small current (e.g., less than $0.05$C) for a long period of time~\cite{G, impedancetrack, resistancemeasurements},
which does not always hold due to devices' dynamic usage patterns, thus yielding poor estimation accuracy, e.g., an SoC error of $\pm 25\%$ is specified for Qualcomm's PM8916~\cite{PM8916}. 
We will make two existing solutions requiring SoC and OCV in \cite{H,M} adopt the over-night charge 
to improve reliability, and use them as the baselines for comparison in Sec.~\ref{sec:evaluations}.

In contrast, voltage is the most pervasively/easily available battery information on mobile devices, 
and hence we choose its use for SoH estimation, i.e., \name.
To the best of our knowledge, the closest to \nameS are~\cite{C} and \cite{iccps17}.

Guo et al. \cite{C} estimates battery SoH based on its voltage--time relationship during charging. 
Such a voltage--time relationship, however, depends strongly on device usage behavior, 
making it unreliable on mobile devices. 
First, usage behavior during charging affects the voltage--time relationship.
Fig.~\ref{fig:voltageduringCCChg} plots the voltage curves during two consecutive charges of a Galaxy 
S6 Edge phone --- the phone is left idle during the first charge and operates actively during 
the second, showing clear dependency of the voltage curve on device operation.
Second, the usage behavior before device charge affects the voltage--time relationship, 
making \cite{C} unreliable even when only applying it during over-night charge, as \nameS does.
Fig.~\ref{fig:beforechargeoperation}(a) plots two consecutive charges of an idle Nexus 6P phone 
after discharging it to 69\% (1.A) and 31\% SoC (2.A), respectively. Their charging phases during the [70\%, 80\%] SoC 
range (part of 1.B and 2.B in Fig.~\ref{fig:beforechargeoperation}(a)) are compared 
in Fig.~\ref{fig:beforechargeoperation}(b), showing significant differences in both durations and voltage levels
and thus dependency on before-charging device usage.

He et al.~\cite{iccps17} explores the voltage-based SoH estimation based on (i) a power model of battery voltage (i.e., $v(t) = a\cdot t^b + c$) and (ii) a linear model between the power factor $b$ and battery SoH. Clearly, the accuracy of~\cite{iccps17} depends on the model accuracy and the empirically-identified model parameters, which we observed to vary over battery aging.  
\nameS reduces such model dependency with a machine learning approach, which is further assisted by a set of data pre-processing techniques including filtering, smoothing, and dimension reduction. 
We will use \cite{iccps17} as another baseline method for comparison in~Sec.~\ref{sec:evaluations}.  

In summary, existing SoH estimation methods are not applicable to, or inaccurate for, mobile devices because of 
the non-existence of required battery information or the inability of meeting the required 
conditions.\footnote{As an alternative, some commodity phones use a counter of battery's complete 
charge/discharge cycles to indicate its health. This, however, is not reliable as battery 
degradation depends heavily on how it is cycled, such as charging/discharge rates, discharge depth, 
and temperature~\cite{Yancheng,Haishen,Gang,Long}. For example, Choi {\em et al.}~\cite{Choi2002130} 
showed that battery health, even with the same cycle count, could differ as much as 3x due to different discharge rates.} 
To remedy this problem, we propose \nameS which estimates SoH based only on voltage information 
and is enabled on mobile devices with the common usage pattern of over-night charge.

\begin{figure}
\begin{minipage}{1\columnwidth}
\centering
{\includegraphics[width=1\columnwidth]{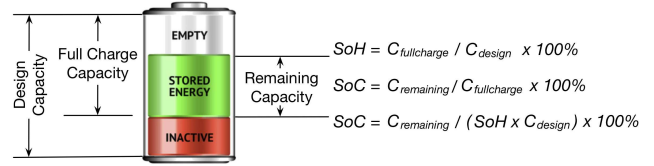}}
\caption{{\bf Battery SoH:} quantifies its capacity degradation and is required for SoC estimation.}
\label{fig:SoCandSoH}
\end{minipage}
\end{figure}

\section{Motivation}
\label{sec:motivation}

This section provides the motivation behind \name.

\subsection{What is Battery SoH?}

SoH is one of the most critical battery parameters (Fig.~\ref{fig:SoCandSoH}), quantifies battery's
capacity degradation, and is defined as the ratio of battery's full charge capacity $C_{\rm fullcharge}$ to 
its designed levels $C_{\rm design}$~\cite{bolunthsis,B,Plett20112319}, i.e.,
\begin{equation}
SoH = C_{\rm fullcharge}/C_{\rm design} \times 100\%.
\label{equ:soh}
\end{equation}
SoH is also the key in estimating a battery's real-time SoC:
\begin{equation}
SoC =  C_{\rm remaining}/(SoH \times C_{\rm design}) \times 100\%.
\label{equ:socandsoh}
\end{equation}
where $C_{\rm remaining}$ is the real-time remaining capacity.

$C_{\rm fullcharge}$ is the foundation of SoH estimation, which is usually estimated via Coulomb
counting~\cite{impedancetrack,impedancetrackforphones}, i.e., integrating the current when 
discharging/charging the battery between two SoC levels to calculate the discharged/charged capacity as
\begin{equation}
\Delta C = \int_{t(SoC_1)}^{t(SoC_2)} i(t) dt,
\nonumber
\end{equation}
where $i(t)$ is the current at time $t$. This way we know
\begin{equation}
C_{\rm fullcharge} =\Delta C/|SoC_1 - SoC_2|.
\nonumber
\end{equation}

\begin{figure}[t]
\begin{minipage}{1\columnwidth}
\centering
{\includegraphics[width=1\columnwidth]{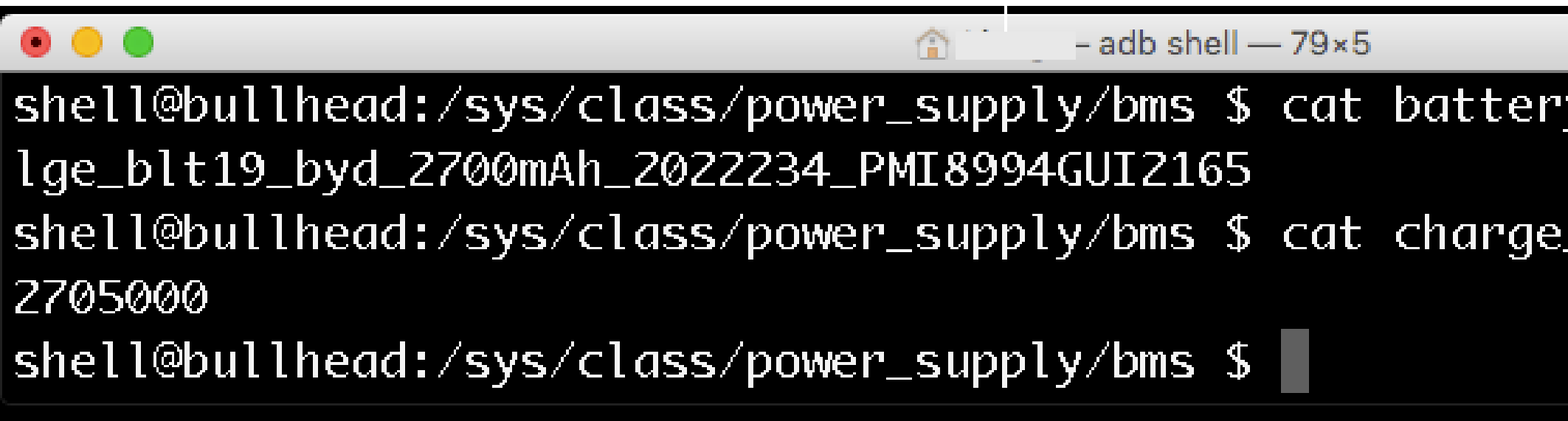}}
\caption{{\bf Inaccurate SoH information on Nexus 5X:} showing $2,705$mAh full-charge capacity and thus about 100\% 
SoH even though the phone has been used extensively for 14 months and observed to have a clearly shortened operation time.}
\label{fig:Nexus5XfullchargeC}
\end{minipage}
\end{figure}

\begin{figure}
\begin{minipage}{0.37\columnwidth}
\centering
{\includegraphics[width=1\columnwidth]{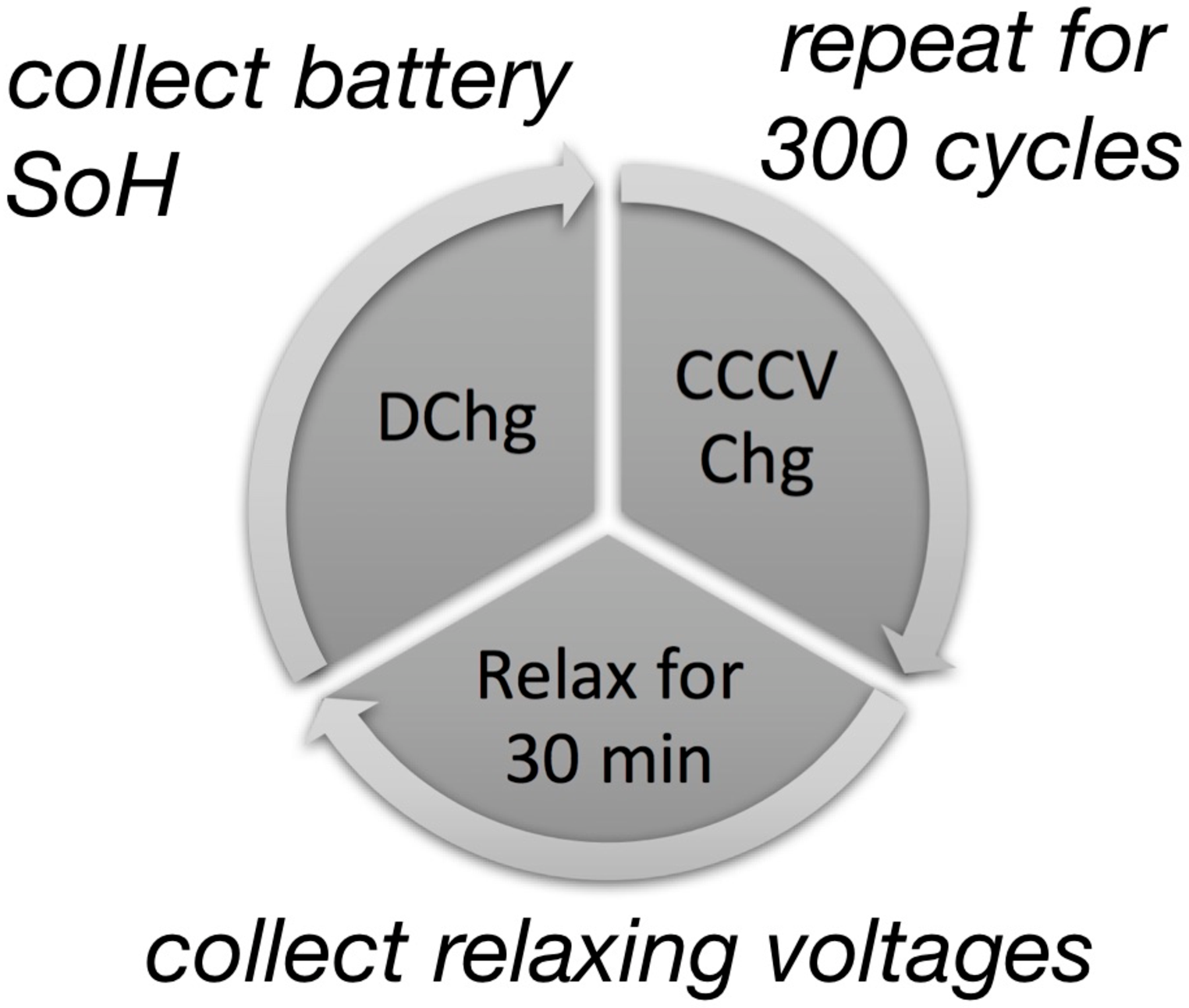}}
\caption{{\bf Cycling measurement:} charge/rest/discharge for $300$ cycles.}
\label{fig:cycletest}
\end{minipage}
\hfill
\begin{minipage}{0.59\columnwidth}
\centering
{\includegraphics[width=1\columnwidth]{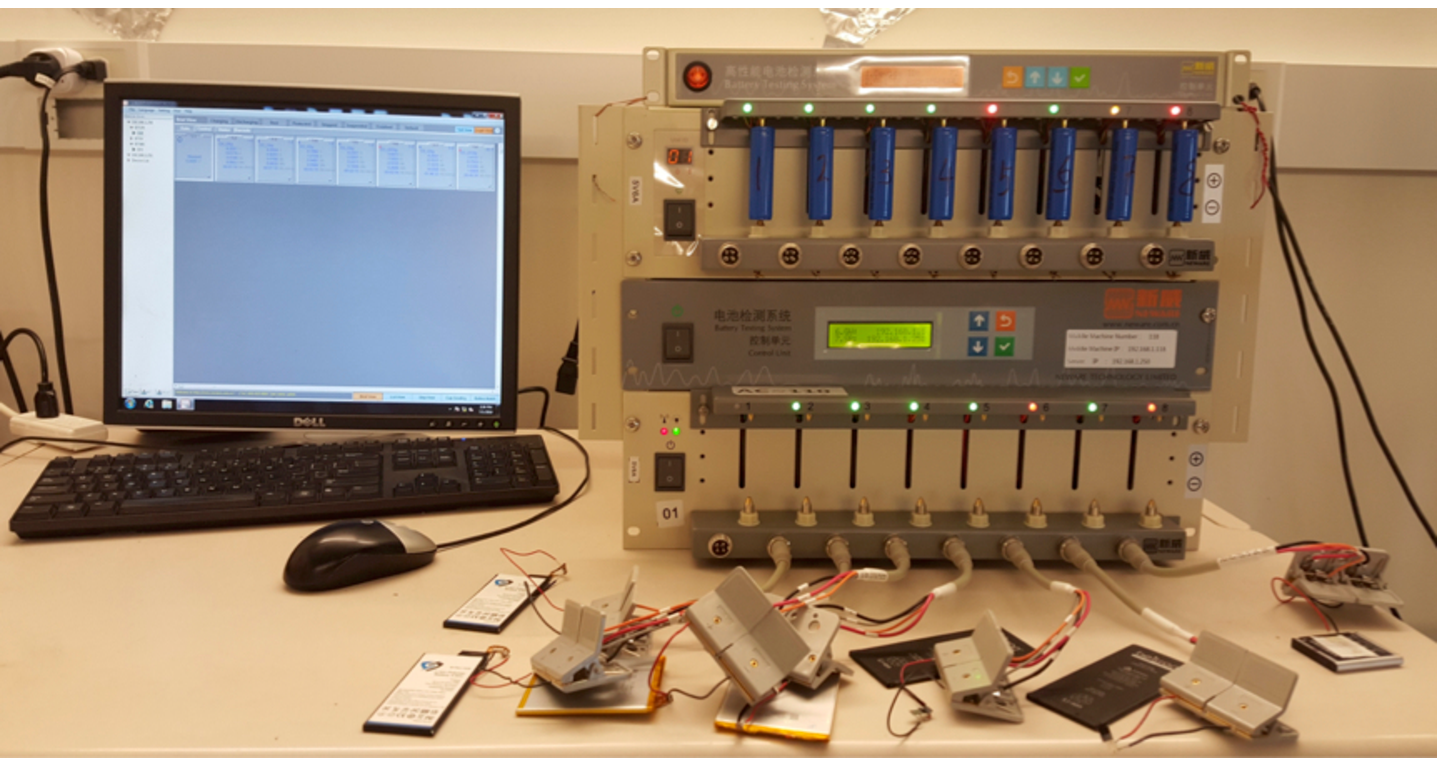}}
\caption{{\bf BTS4000 battery tester:} controls battery charge/discharge 
with less than $0.5\%$ error and logs at up to $10$Hz~\cite{neware}.}
\label{fig:BTS}
\end{minipage}
\end{figure}

\begin{figure*}[t]
\centering
\begin{minipage}{2\columnwidth}
\begin{subfigure}{0.33\columnwidth}
{\includegraphics[width=1\columnwidth]{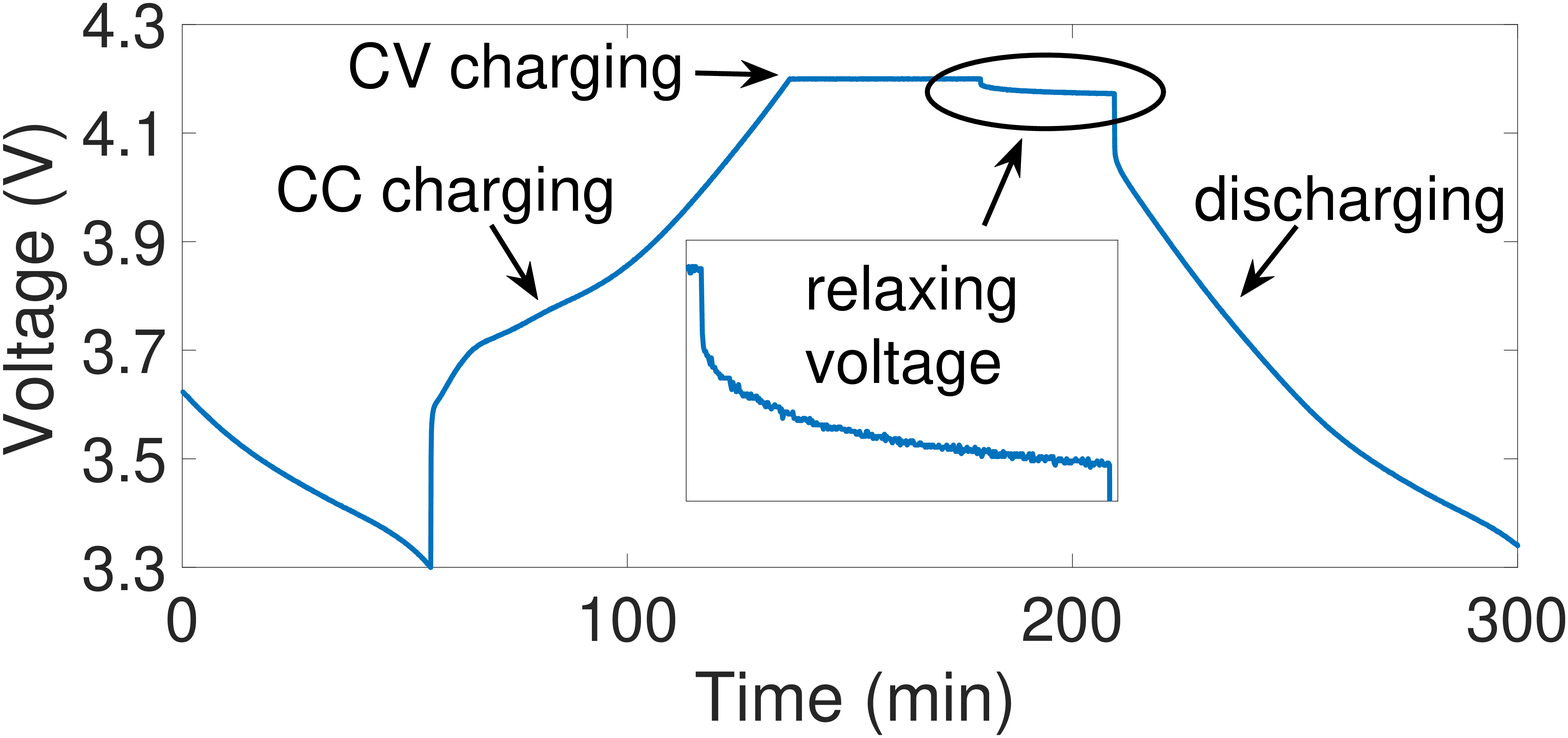}}
\caption{Voltage during one cycle}
\label{fig:relaxingvoltageexample}
\end{subfigure}
\hfill
\begin{subfigure}{0.33\columnwidth}
{\includegraphics[width=1\columnwidth]{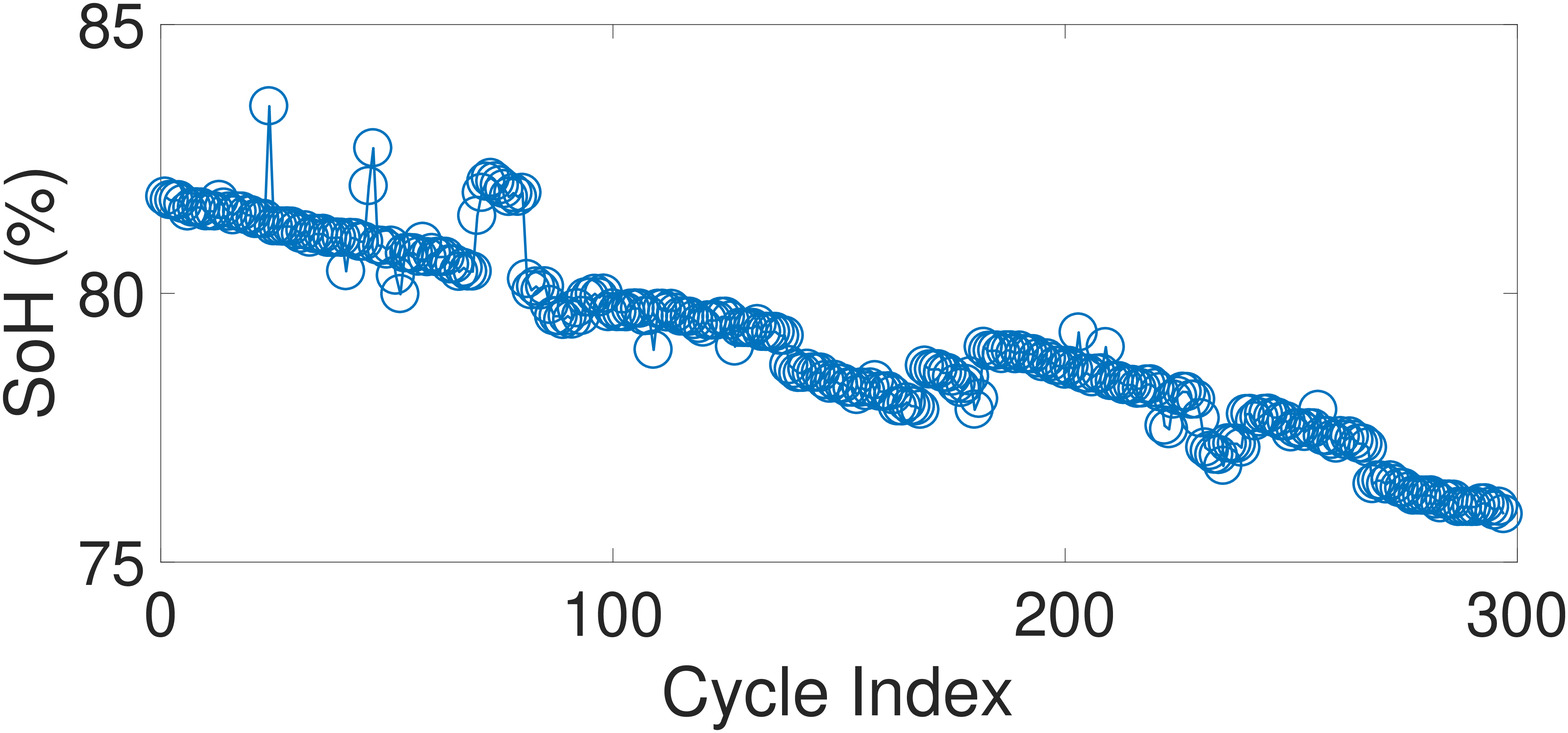}}
\caption{SoH degrades}
\end{subfigure}
\hfill
\begin{subfigure}{0.33\columnwidth}
{\includegraphics[width=1\columnwidth]{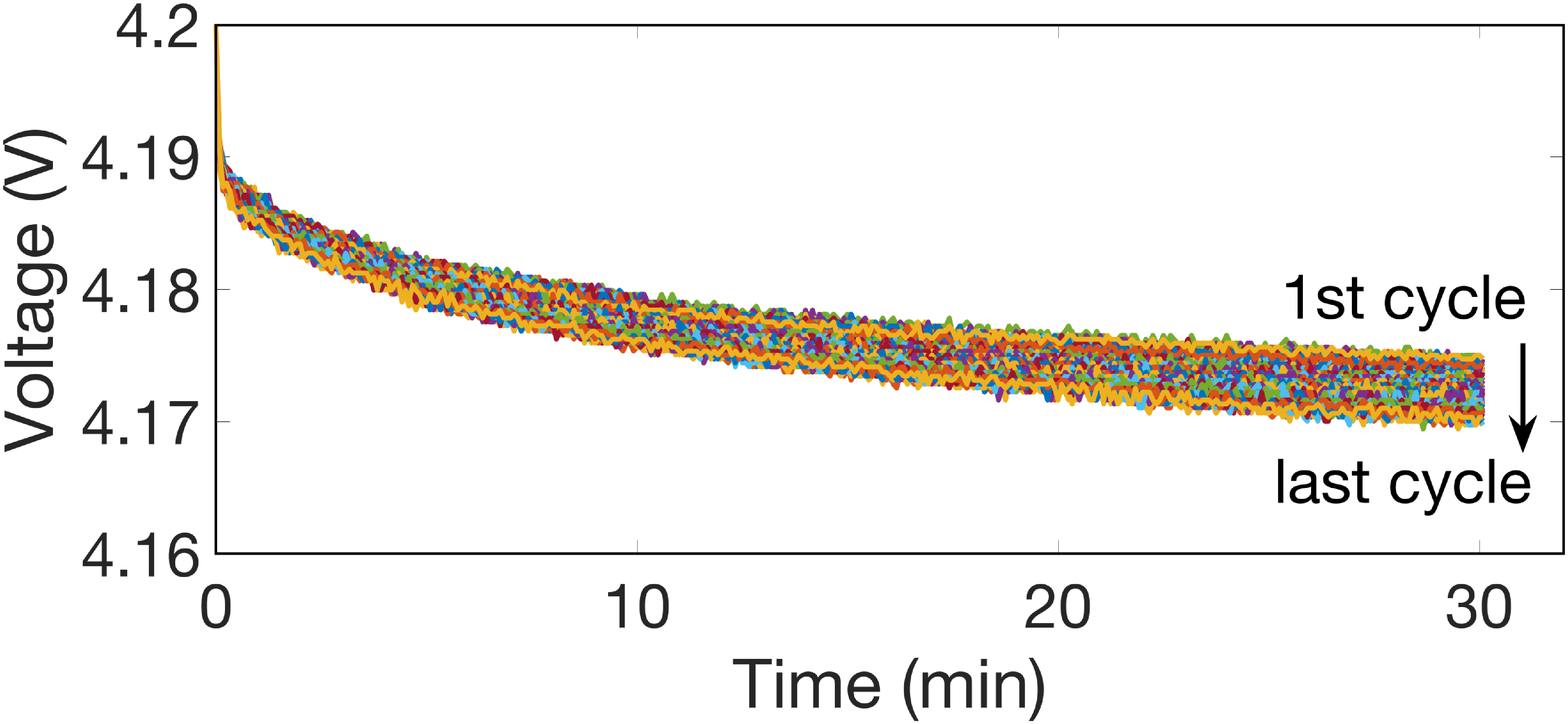}}
\caption{Relaxing voltage lowers}
\end{subfigure}
\caption{{\bf Relaxing voltage fingerprints battery SoH:} (a) voltage curve during one charging/resting/discharging 
cycle and the relaxing voltage during resting; (b) battery SoH degrades during the measurements; 
(c) the relaxing voltage decreases during the measurements.}
\label{fig:VoltageandSoH}
\end{minipage}
\end{figure*}

\begin{figure*}[ht!]
\centering
\begin{minipage}{2\columnwidth}
\centering
{\includegraphics[width = 0.85\columnwidth]{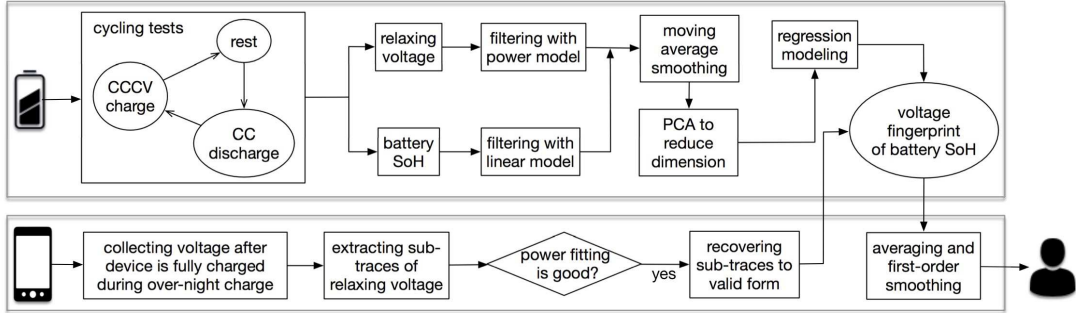}}
\caption{{\bf \nameS summary:} collecting relaxing battery voltages on mobile devices and checking 
with the fingerprint map for SoH estimation.}
\label{fig:overview_detailed}
\end{minipage}
\end{figure*}

\subsection{Why is SoH Absent from Mobile Devices?}

Commodity mobile devices do not support Coulomb counting well in terms of availability, 
accuracy, and timeliness, thus making it difficult to estimate their battery SoH.
First, not all the PMICs, or more specifically their fuel gauge components~\cite{SOSP15}, of mobile devices 
support current sensing~\cite{vedge, iccps17}.
Moreover, the PMIC-provided current information, even when available, is very coarse~\cite{Giovino}.
Our measurement with a Nexus 5X phone shows that its PMIC's current reading deviates from the true 
value --- collected with the Monsoon power meter --- by an average of 4\% even at room temperature.
Last but not the least, the current information may lack timeliness, which is crucial for Coulomb 
counting because of devices' dynamic currents, i.e., varying from tens 
to thousands of milliamps in a few milliseconds~\cite{7530866}.
A $47\%$ counting error due to insufficient sampling rates is reported in~\cite{iccps17}.

As a real-life evidence of mobile devices' deficiency in supporting Coulomb counting and their limited SoH information, 
Fig.~\ref{fig:Nexus5XfullchargeC} shows the full-charge capacity of a Nexus 5X phone provided by 
its fuel-gauge chip, saying its battery, with a design capacity of $2,700$mAh, can still deliver $2,705$mAh 
capacity upon being fully charged and thus an SoH of about $100\%$, even though 
the phone has been used extensively for $14$ months and observed to have a shorter operation time. 
This motivates us to explore {\em current-free} SoH estimation, i.e., \name.

\section{Overview}
\label{sec:overview}

\nameS is built on our key finding that batteries' relaxing voltages {\em fingerprint} their SoH.\footnote{Relaxing voltages also facilitate battery SoC estimation~\cite{Robert2,Thomas}.}
We demonstrate this finding with a $2,200$mAh Galaxy S3 battery.
Specifically, we test the battery by (i) fully charging it with a constant-current constant-voltage (CCCV) 
profile of $<$0.5C, 4.2V, 0.05C$>$$_{\rm cccv}$ as commonly specified in Li-ion battery 
datasheet~\cite{tenergydatasheet,NCR18650B},\footnote{CCCV is widely used 
to charge mobile devices~\cite{Hoque}, described by $<$$I_{\rm cc}, V_{\rm full}, I_{\rm cutoff}$$>$$_{\rm cccv}$: 
charge the battery with a constant current $I_{\rm cc}$ until its voltage reaches $V_{\rm full}$ (i.e., CC-Chg), 
and then charge it further with a constant voltage $V_{\rm full}$ until the current 
reduces to $I_{\rm cutoff}$ (i.e., CV-Chg), as observed in Fig.~\ref{fig:voltageduringCCChg} with the Galaxy S6 Edge phone. 
Also, battery charge/discharge is often expressed in {\em C-rate}: 
at 1C-rate, the current drains the battery completely in $1$ hour, e.g., $2,200$mA for 
the battery used here.} (ii) resting it for $30$ minutes, (iii) fully discharging it at $0.5$C-rate until 
reaching a cutoff voltage of $3.3$V, at which mobile devices normally shut off, 
and (iv) repeating the process for $300$ cycles, as summarized in Fig.~\ref{fig:cycletest}. 
This measurement is made with the NEWARE BTS4000 battery tester~\cite{neware} 
as shown in Fig.~\ref{fig:BTS}, and the cycling 
process (i.e., current, voltage, timestamp) is logged at $1$Hz. 
Fig.~\ref{fig:VoltageandSoH}(a) plots the battery voltage during one such charging/resting/discharging cycle, 
and highlights the relaxing voltages during resting.
The relaxing voltage drops instantly upon resting and then decreases gradually further until it converges. 

We collect the battery's full charge capacity (and thus its SoH according to Eq.~(\ref{equ:soh})) 
via Coulomb counting during each discharge,  thus recording its degradation process 
during the cycling measurement, as shown in Fig.~\ref{fig:VoltageandSoH}(b). 
Also, $300$ time series of relaxing voltages are collected, each during one of the $30$-minute 
resting period (Fig.~\ref{fig:VoltageandSoH}(c)).
Comparison of Figs.\ref{fig:VoltageandSoH}(b) and \ref{fig:VoltageandSoH}(c) shows that the 
battery SoH degrades over the cycling measurement due to its capacity degradation, while during the same measurement, 
its relaxing voltage decreases, exhibiting the possibility to fingerprint battery SoH with the relaxing voltages.

 \nameS exploits this voltage--SoH relationship to estimate the SoH of device batteries
 by checking their relaxing voltages with an offline-constructed fingerprint map.
 Fig.~\ref{fig:overview_detailed} presents an overview of \name, which we will 
 elaborate in the next two sections.

\begin{table*}[t]
\centering
  \caption{\nameS is steered and validated by \numCycles empirically collected relaxing voltage traces via \numTests cycling tests with \numBatt phone batteries.}
  \label{table:testsdetails}
  \scriptsize
  \begin{tabular}{|c|*5c|}
    \hline
    \textbf{\textbf{Battery}} & \textbf{Rated Capacity} &\textbf{\# of Tests} & \textbf{\# of Cycles} & \textbf{Per-Cycle Profile} & \textbf{Covered SoH (\%)}\\
    \hline
    \hline
            \multirow{1}{*}{\textbf{Nexus 6P x 1}}  &3,450mAh & 5   & 1,300 & $<$0.50C, 4.35V, 0.05C$>$$_{\rm cccv}$;~30min rest;~0.5C DChg to 3.3V & [0, 93.6] \\
        \hline
             \multirow{1}{*}{\textbf{Nexus 5X x 2}}  &2,700mAh & 3   & 1,104 & $<$0.50C, 4.35V, 0.05C$>$$_{\rm cccv}$;~30min rest;~0.5C DChg to 3.3V & [59.2, 94.0] \\
        \hline
       	    \multirow{1}{*}{\textbf{Nexus S x 1}}  &1,500mAh & 3   & 150 & $<$0.50C, 4.20V, 0.05C$>$$_{\rm cccv}$;~30min rest;~0.5C DChg to 3.2V & [49.9, 54.3] \\
	    \hline
	    \multirow{1}{*}{\textbf{Xperia Z5 x 1}}  &2,900mAh & 5   & 655 & $<$0.50C, 4.20V, 0.05C$>$$_{\rm cccv}$;~30min rest;~0.5C DChg to 3.2V & [12.4, 87.1] \\
	    \hline
	    \multirow{1}{*}{\textbf{iPhone 6 Plus x 1}}  &2,900mAh & 2   & 100 & $<$0.50C, 4.35V, 0.05C$>$$_{\rm cccv}$;~30min rest;~0.5C DChg to 3.3V & [67.6, 79.1] \\	
	   \hline
	    \multirow{1}{*}{\textbf{Galaxy Note 2 x 1}}  &3,100mAh & 5   & 1,350 & $<$0.50C, 4.20V, 0.05C$>$$_{\rm cccv}$;~30min rest;~0.5C DChg to 3.2V & [21, 96.6] \\  
	    \hline
       	    \multirow{1}{*}{\textbf{Galaxy S5 x 1}}  &2,800mAh & 3   & 964 & $<$0.50C, 4.35V, 0.05C$>$$_{\rm cccv}$;~30min rest;~0.5C DChg to 3.3V & [73.1, 91.8] \\
	    \hline
	    \multirow{1}{*}{\textbf{Galaxy S4 x 3}}  &2,600mAh & 8   & 2,374 & $<$0.50C, 4.20V, 0.05C$>$$_{\rm cccv}$;~30min rest;~0.5C DChg to 3.0V & [2.8, 93.2] \\
	    \hline
	    \multirow{1}{*}{\textbf{Galaxy S3 x 4}}  &2,200mAh & 12   & 4,800 & $<$0.50C, 4.20V, 0.05C$>$$_{\rm cccv}$;~30min rest;~0.5C DChg to 3.3V & [69.5, 97.0] \\
	    \multirow{1}{*}{\textbf{}}  & --- & 4   & 580 & $<$0.25C, 4.20V, 0.05C$>$$_{\rm cccv}$;~30min rest;~0.5C DChg to 3.3V &[87.8, 92.3] \\
	\hline
  \end{tabular}
\end{table*}

\section{Voltage Fingerprinting of SoH}
\label{sec:fingerprintmp}

We now empirically characterize the voltage fingerprint map of battery SoH.

\subsection{Data Collection}

Knowledge of batteries' SoH degradation and relaxing voltages is necessary to characterize their 
relationship with extensive battery cycling tests. 
Such tests are readily available for smartphone OEMs, such as Samsung and Apple when testing 
their products,\footnote{This also makes \nameS ideally suitable as an OEM service.} but are not 
available for non-OEM researchers. 
Therefore, we have conducted extensive battery cycling measurements with \numBatt batteries used 
for various mobile devices as summarized in Table~\ref{table:testsdetails} (including the one shown 
in Fig.~\ref{fig:VoltageandSoH}): collecting the relaxing voltages during each resting period and 
logging batteries' SoH degradation based on their capacity delivery during each discharge. 
These measurements consist of \numCycles cycles in total and last over \cumMonth months cumulatively.
In these measurements, the settings of $<$0.5C, 4.2V, 0.05C$>$$_{\rm cccv}$ and $V_{\rm cutoff} = 3.0$V 
are commonly used to specify battery properties in industry during battery 
testing~\cite{C,tenergydatasheet,NCR18650B}, and $V_{\rm max} = 4.35$V 
and  $V_{\rm cutoff}$ of $3.2$--$3.3$V specify more device characteristics: 
mobile devices are normally charged to a maximum voltage of $4.3$--$4.4$V and shut off when 
their battery voltage reduces to $3.2$--$3.3$V~\cite{7530866}.

These \cumMonth-month measurements are long enough to identify the voltage--SoH 
relationship within the SoH range users experience most (e.g., users rarely switch to new 
batteries/devices until the old ones degrade to $0\%$ SoH).
Moreover, the thus-identified voltage--SoH relationship can be extended to the SoH ranges not covered by these measurements, as we will explain later.

\subsection{Construction of Fingerprint}

Next we use $12$ of such measurements based on $4$ Galaxy S3 batteries to elaborate on 
the construction of a voltage-based SoH fingerprint map. 
Each of these $12$ measurements consists of $\approx$$300$ charging/resting/discharging cycles, logged at $1$Hz. 
This way, we collected $12$ SoH-degradation traces, each from one measurement, and also recorded $3,612$ time series of relaxing voltages, each from the resting period within a cycle.
The same approach of fingerprint map construction is applied to all the batteries in 
Table~\ref{table:testsdetails} and evaluated, as we will explain in Sec.~\ref{sec:evaluations}. 

{\bf Data Filtering and Smoothing.}
Variance/noise exists in the measurements of SoH degradation and relaxing voltages 
(as observed in Fig.~\ref{fig:VoltageandSoH}), which are likely due to battery dynamics, especially 
when considering the stable laboratory environment (i.e., with an UPS connected 
and room temperature control) and the battery tester's high accuracy (i.e., less than $0.5\%$ error 
in controlling the cycling processes). Such a variance in battery measurements has also 
been reported in~\cite{C}, necessitating pre-processing (i.e., filtering and smoothing) of data
before constructing the fingerprint map.
The collected data were filtered and smoothed using two empirically established models for 
the SoH degradation and relaxing voltages.

\begin{figure}
\begin{minipage}{0.48\columnwidth}
\centering
{\includegraphics[width=1\columnwidth]{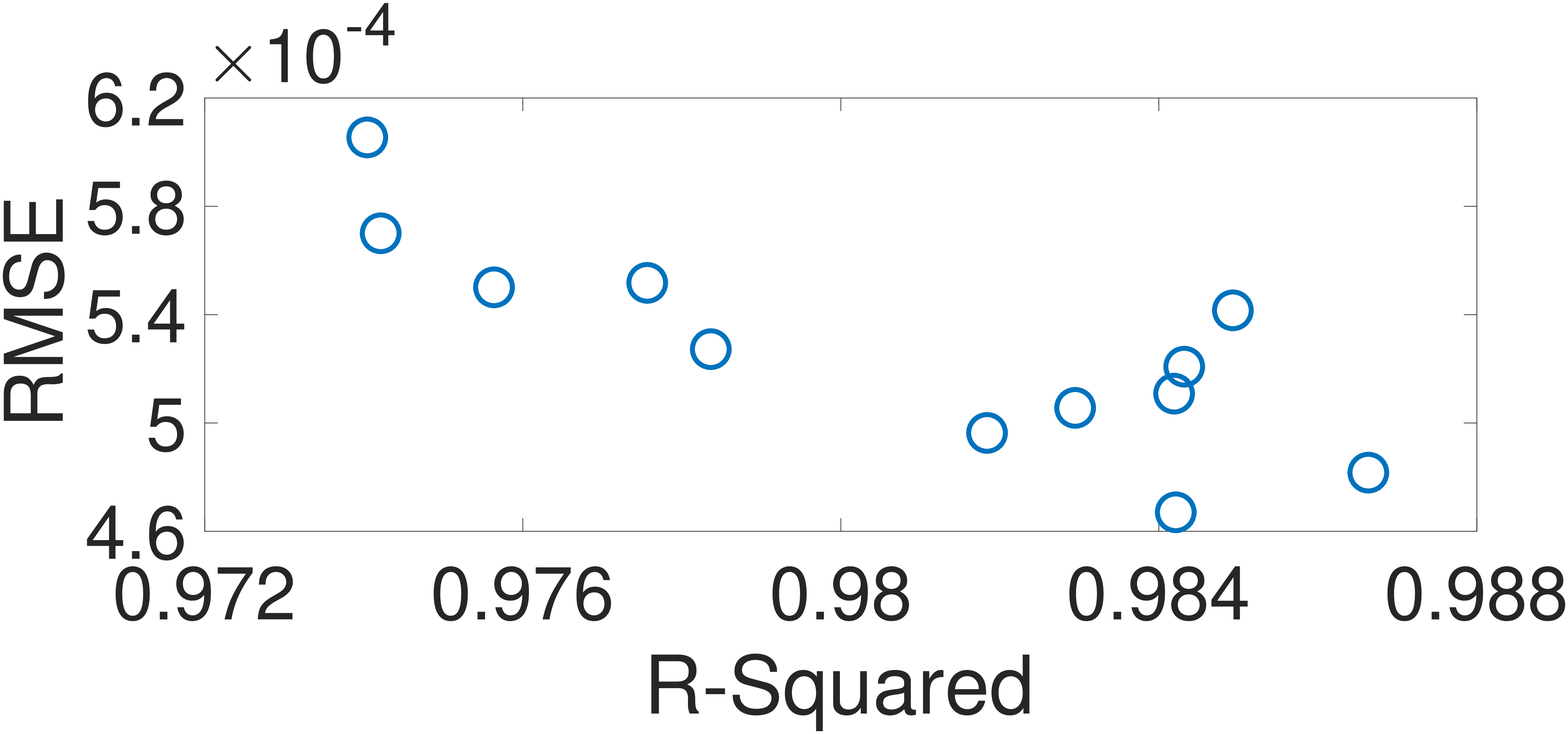}}
\caption{{\bf Linear fitting of SoH degradation:} all the $12$ degradation processes fit linearly 
with RMSE $<$$0.00062$ and R-Squared $>$$0.972$.}
\label{fig:LinearFitofSoH}
\end{minipage}
\hfill
\begin{minipage}{0.48\columnwidth}
\centering
{\includegraphics[width=1\columnwidth]{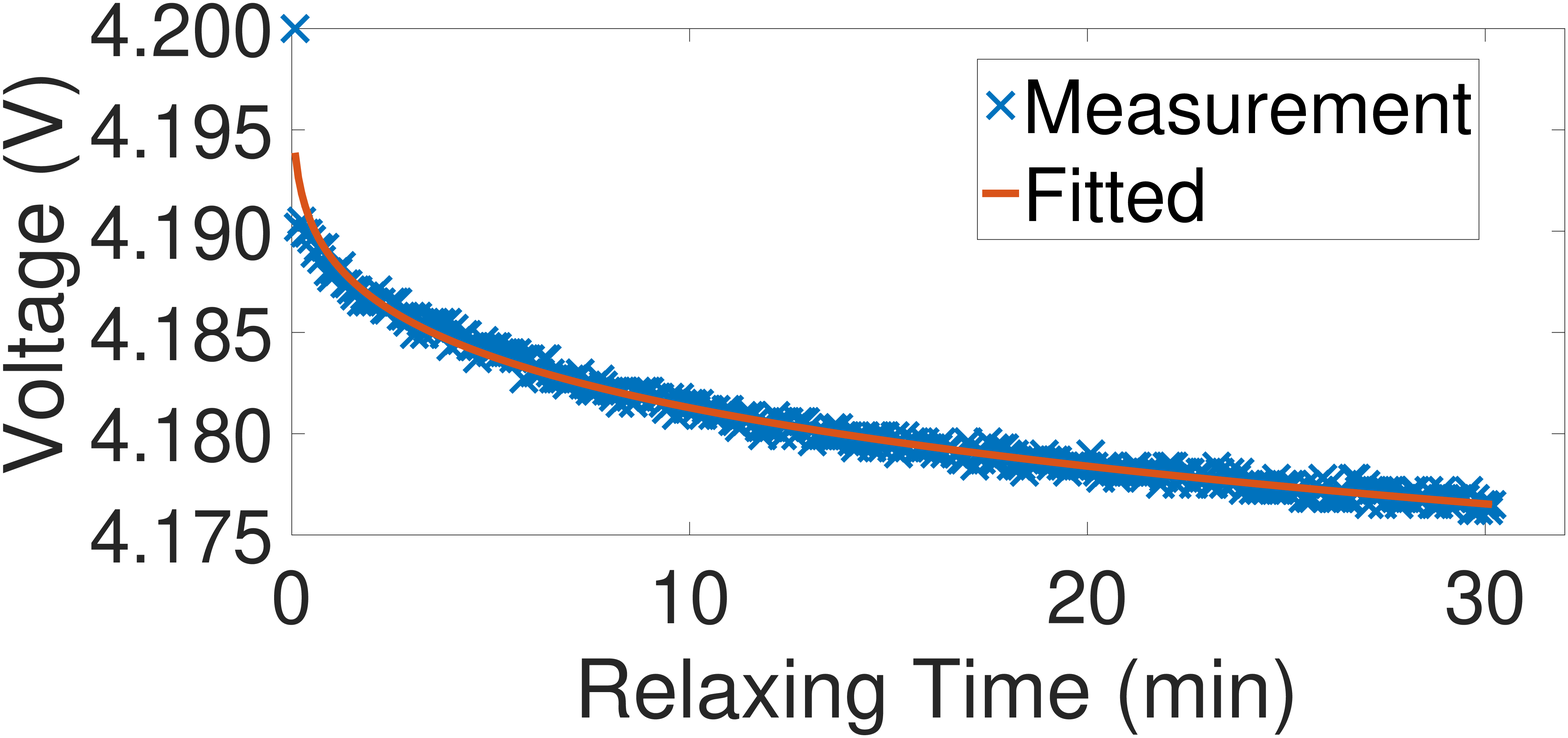}}
\caption{{\bf Relaxing voltages are of power-shape:} fitting the collected relaxing voltages to 
a power function $v(t) = a \cdot t^b +c~~(t \geq 0)$.}
\label{fig:examplepowerfit}
\end{minipage}
\end{figure}

The battery health is shown to degrade approximately linearly (as observed 
in Fig.~\ref{fig:VoltageandSoH}(b)) until it really becomes bad~\cite{Q,Yancheng}. 
To further validate this linear degradation, we tried a linear fit of the $12$ collected SoH degradation processes, 
and all of them have an excellent goodness-of-fit in terms of root-mean-square error (RMSE) 
and R-Squared, as shown in Fig.~\ref{fig:LinearFitofSoH} where each point represents the 
goodness-of-fit for a particular SoH degradation process.
\nameS removes outlier SoH samples based on this linear model --- those SoH samples deviating too 
much from the linear fitting (e.g., $>$$0.5\%$ SoH) are tagged as outliers and removed, 
and then the remaining samples are smoothed with a moving average.

\begin{figure}[t]
\begin{minipage}{1\columnwidth}
\centering
{\includegraphics[width=0.8\columnwidth]{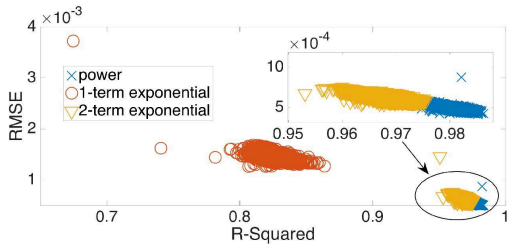}}
\caption{{\bf Goodness of power fitting:} all the $3,612$ relaxing voltage traces have 
RMSE $<$$0.0009$ and R-Squared $>$$0.965$; the power model describes relaxing voltages 
more accurately than the traditional exponential models.}
\label{fig:powerfitofrelaxingvoltage_combined}
\end{minipage}
\end{figure}

\begin{figure}[t]
\centering
\begin{minipage}{1\columnwidth}
\centering
{\includegraphics[width=0.8\columnwidth]{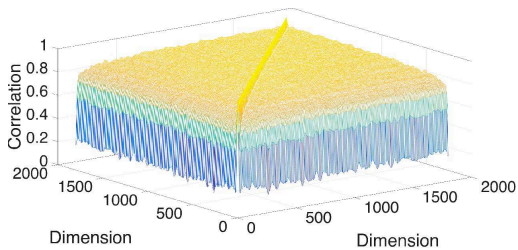}}
\caption{{\bf Different dimensions in relaxing voltage are highly correlated:} $>$$0.8$ correlation 
coefficients are observed for most dimension pairs.}
\label{fig:CorrelationbetweenDimensions}
\end{minipage}
\end{figure}

Similarly, \nameS filters and smooths the relaxing voltages based on another empirical observation 
that the relaxing voltages conform to a power function $v(t) = a \cdot t^b +c~~(t \geq 0)$,
where $t$ is the time since resting, as illustrated in Fig.~\ref{fig:examplepowerfit}.
We apply the power fitting to the $3,612$ collected relaxing voltage traces to statistically verify 
this observation. Fig.~\ref{fig:powerfitofrelaxingvoltage_combined} summarizes the goodness-of-fit --- the fitting 
RMSE is bounded below $0.0009$ and the R-Squared above $0.965$, showing excellent fitting accuracy. 
Note that this power model differs from existing models with exponential-shape relaxing voltages~\cite{HueiPeng}.
Fig.~\ref{fig:powerfitofrelaxingvoltage_combined} also plots the goodness-of-fit when fitting the same set 
of relaxing voltages as $1$-term and $2$-term exponential functions, i.e., $v(t) = a \cdot e^{t \cdot b}~~(t \geq 0)$ 
and $v(t) = a \cdot e^{t \cdot b} + c \cdot e^{t \cdot d}~~(t \geq 0)$, showing reasonably good accuracy, but not as good as the power fitting.
\nameS filters the relaxing voltages with this power model, e.g., tagging the relaxing voltage traces with the bottom 5\% goodness-of-fit as outliers.
The moving average smoother is then used again to smooth the remaining valid relaxing voltage traces.

Note that if an SoH sample is tagged as an outlier, so is the relaxing voltage in the same cycle, and vice versa.
Also, \nameS only filters out the outliers based on these empirical models, instead of using the model fitting 
results to construct the fingerprint map, thus alleviating its dependency on model accuracy --- a clear advantage over~\cite{iccps17}.
As an example, $268$ SoH samples and relaxing voltage traces are selected after the data pre-processing 
from the $300$-cycle measurement shown in Fig.~\ref{fig:VoltageandSoH}. 

\begin{figure}[t]
\centering
\begin{minipage}{1\columnwidth}
\centering
\begin{subfigure}{0.48\columnwidth}
\centering
{\includegraphics[width=1\columnwidth]{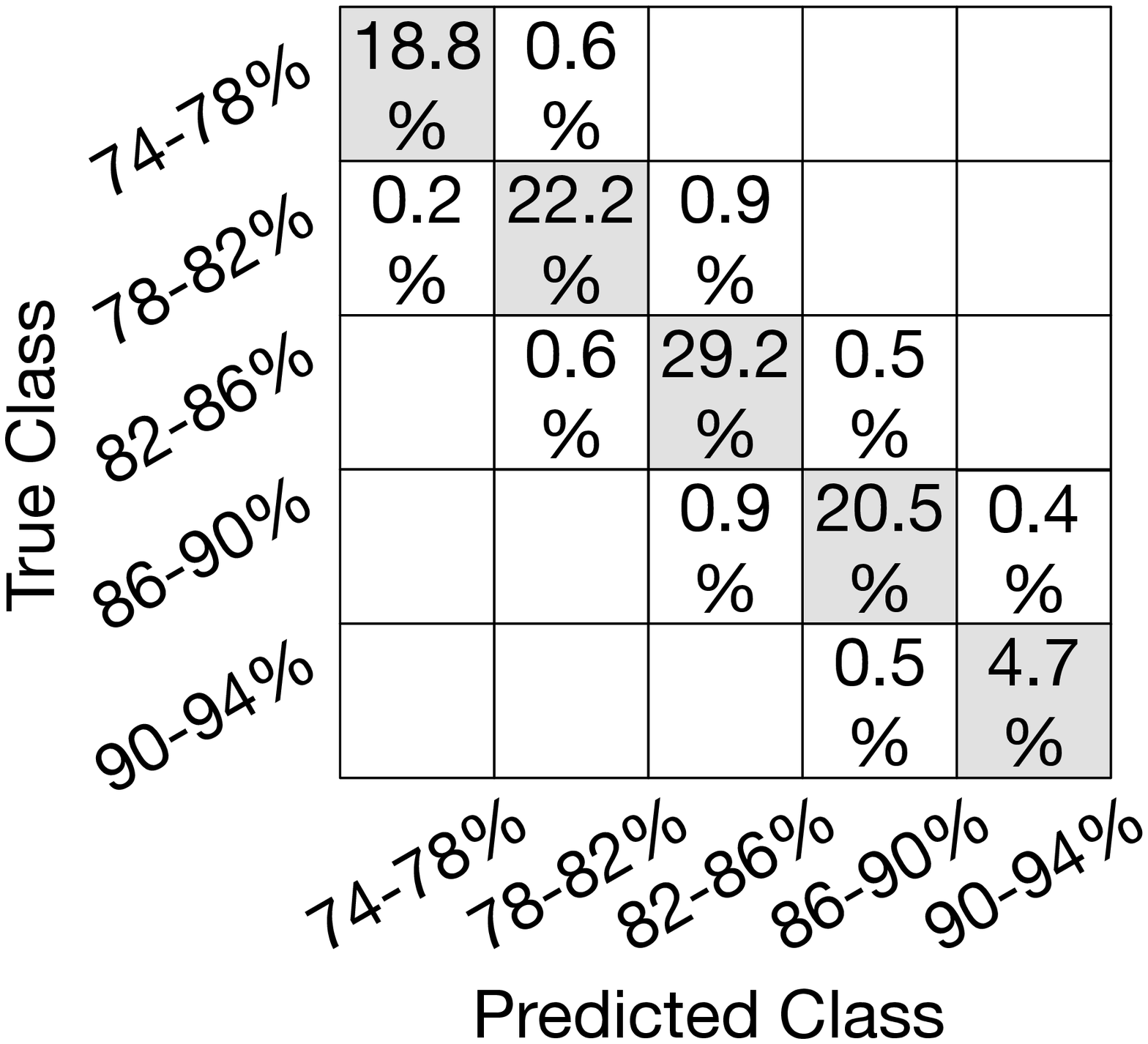}}
\caption{{\bf Battery \#1}}
\end{subfigure}
\hfill
\begin{subfigure}{0.48\columnwidth}
\centering
{\includegraphics[width=1\columnwidth]{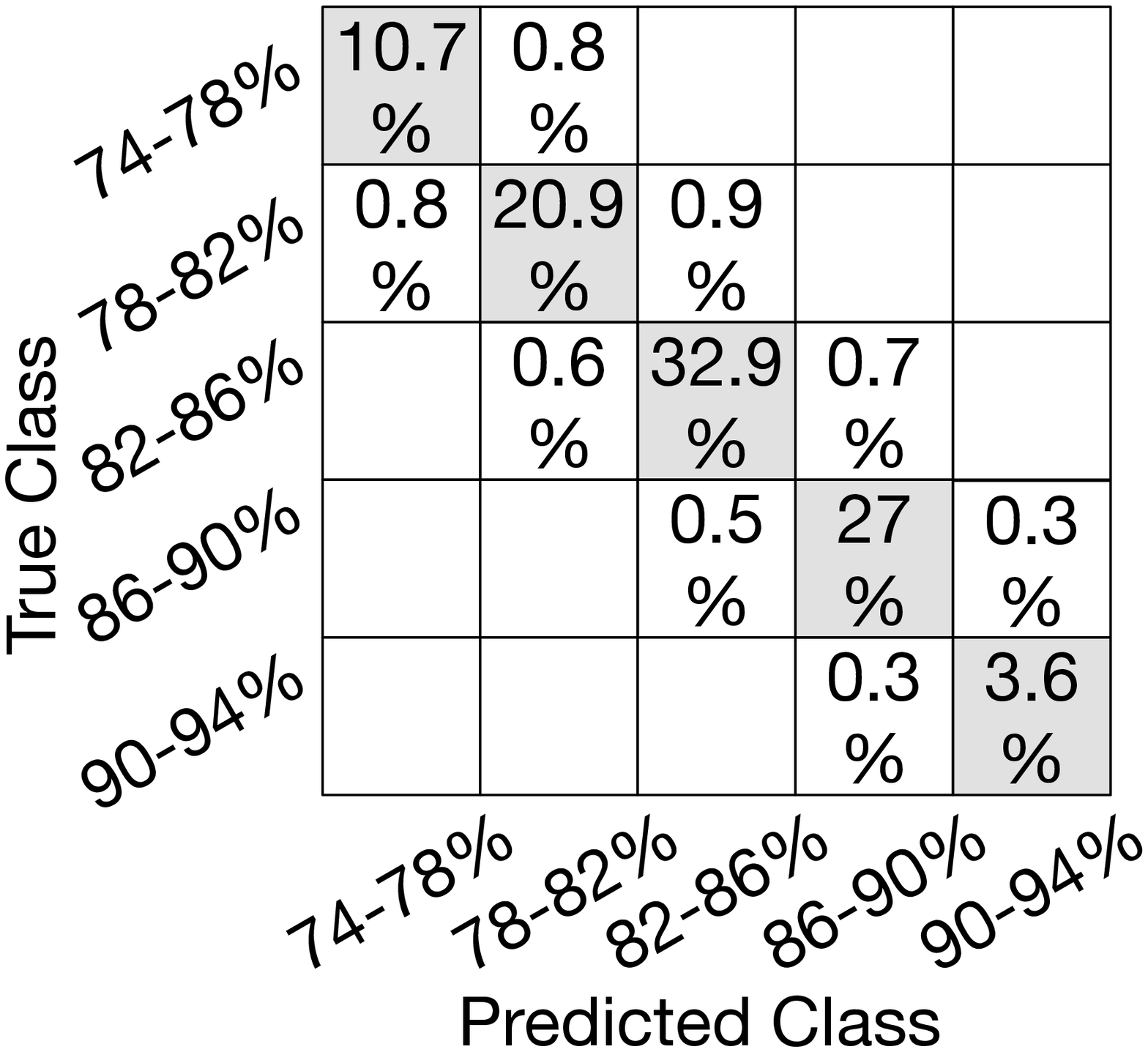}}
\caption{{\bf Battery \#2}}
\end{subfigure}
\vspace{+6pt}

\begin{subfigure}{0.48\columnwidth}
\centering
{\includegraphics[width=1\columnwidth]{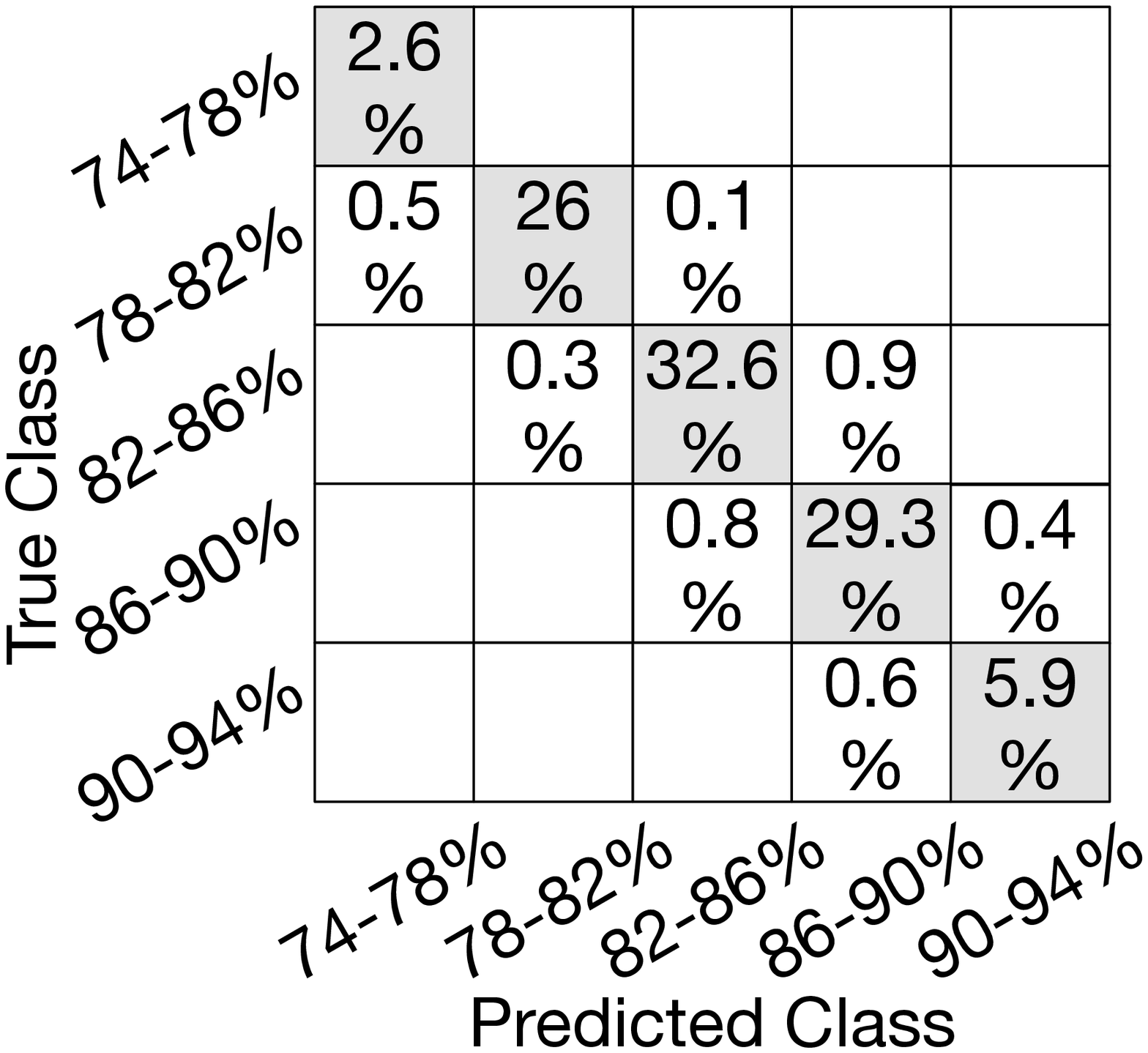}}
\caption{{\bf Battery \#3}}
\end{subfigure}
\hfill
\begin{subfigure}{0.48\columnwidth}
\centering
{\includegraphics[width=1\columnwidth]{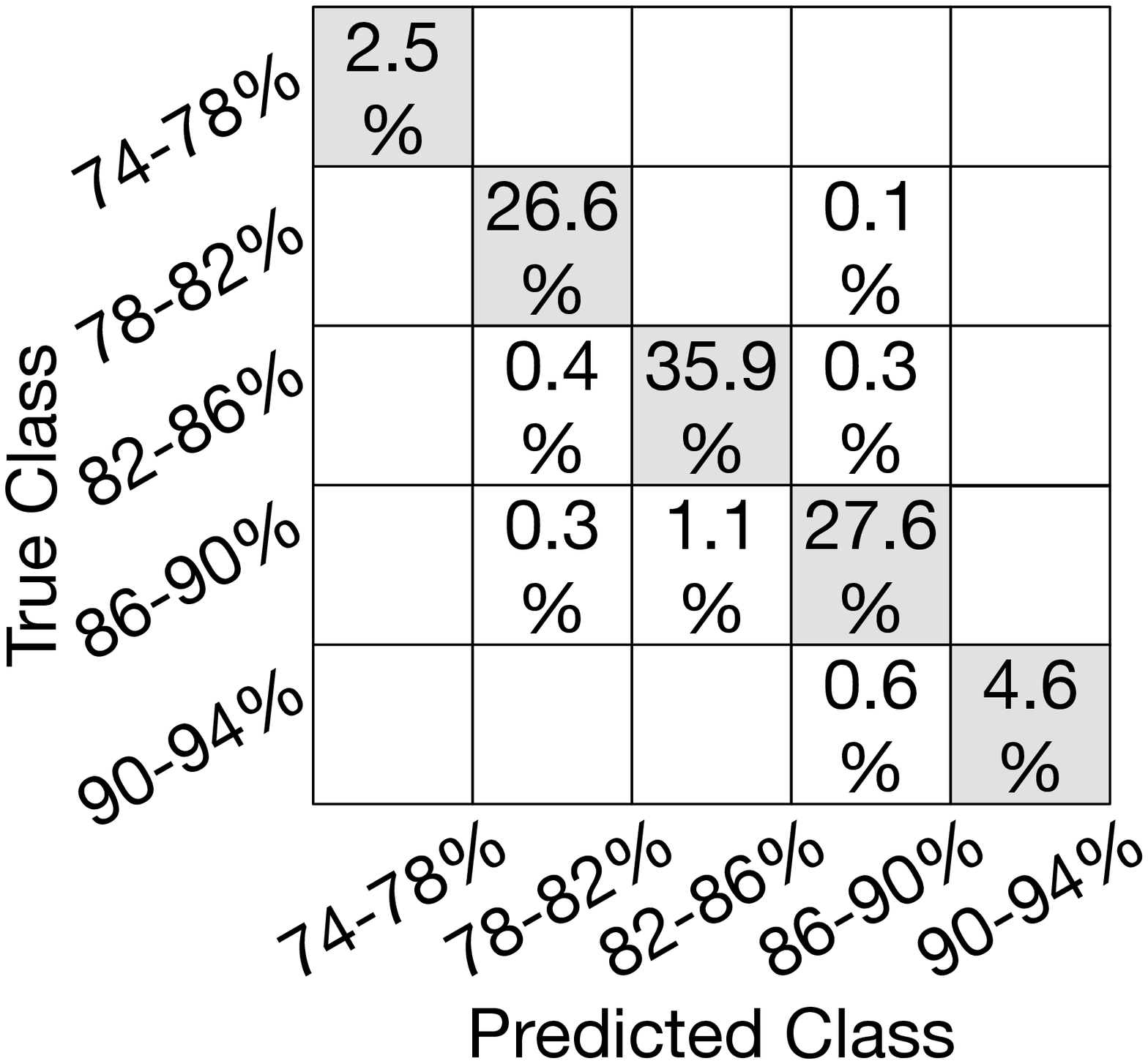}}
\caption{{\bf Battery \#4}}
\end{subfigure}
\caption{{\bf Confusion matrices:} over $95\%$ 
accuracy when forming SoH categories with $4\%$ step-size.}
\label{fig:confusionmatrix}
\end{minipage}
\end{figure}

\begin{table*}
\begin{minipage}{1.4\columnwidth}
  \caption{{Classification accuracy with other regression methods (\%).}}
  \label{table:classificationaccuracy}
  \scriptsize
  \begin{tabular}{|c||*6c||c|}
    \hline
        \textbf{Battery} &\textbf{Linear SVM} & \textbf{Qua. SVM} &\textbf{Cub. SVM} & \textbf{Fine KNN}  & \textbf{Med. KNN} & \textbf{Coarse KNN} & \textbf{Tree}\\
    \hline
    \hline
                    \multirow{1}{*}{\textbf{\#1}}  & 94 & 94  & 90 & 67 & 73 & 71 & 95 \\	    	    
        \hline
                    \multirow{1}{*}{\textbf{\#2}}  & 98 & 94  & 97  &  92 & 95 & 95 & 95\\	    	    
        \hline
                    \multirow{1}{*}{\textbf{\#3}}  & 93 & 92  &  76 & 91 & 94 & 90 & 96\\	    	    
        \hline
                    \multirow{1}{*}{\textbf{\#4}}  & 91 & 84  &  70 & 83 & 92 & 89 & 97\\	       	  
        \hline
  \end{tabular}
  \end{minipage}
  \hspace{+6pt}
  \begin{minipage}{0.6\columnwidth}
\centering
    \caption{{Correlated degradation.}}
  \label{table:correlateddegradation}
  \scriptsize
  \begin{tabular}{|c|*4c|}
    \hline
    \textbf{\textbf{~~~Battery~~~}} & \textbf{~~~\#1~~~} &\textbf{~~~\#2~~~} & \textbf{~~~\#3~~~} & \textbf{~~~\#4~~~}\\
    \hline
    \hline
            \multirow{1}{*}{\textbf{\#1}}  & 1 & 0.99  & 0.98 & 0.98 \\	    	    
        \hline
                    \multirow{1}{*}{\textbf{\#2}}  & 0.99 & 1  & 0.99  &  0.98\\	    	    
        \hline
                    \multirow{1}{*}{\textbf{\#3}}  & 0.98 & 0.99  &  1& 0.98\\	    	    
        \hline
                    \multirow{1}{*}{\textbf{\#4}}  & 0.98 & 0.98  &  0.98 & 1\\	    	    
        \hline
  \end{tabular}
\end{minipage}
\end{table*}

{\bf Dimension Reduction.}
Each of the collected relaxing voltages covers a $30$-minute resting period logged at $1$Hz, 
yielding $30\times 60 = 1,800$ dimensions of data. 
Also, the voltage values in each of these dimensions are correlated. 
Fig.~\ref{fig:CorrelationbetweenDimensions} plots the correlations between each pair of the $1,800$ 
dimensions of the $268$ relaxing voltages selected from Fig.~\ref{fig:VoltageandSoH}, 
where strong correlations (with correlation coefficients $\approx$$0.8$ or higher) are observed in most cases.  
Such highly-correlated, high-dimension relaxing voltages justify  \name's use of the principal component 
analysis (PCA) for reduction of dimensions, lowering the computational effort in constructing 
the fingerprint map.
Again, taking the measurements in Fig.~\ref{fig:VoltageandSoH} as an example, applying PCA reduces
the relaxing voltage dimensions from $1,800$ to $35$ with a variance of $99\%$.

{\bf Regression Modeling.}
Finally, \nameS uses a regression tree to construct the fingerprint map, with the above-obtained principal 
components as predictors and the corresponding SoH as response. 
Fig.~\ref{fig:confusionmatrix} plots the confusion matrices when validating the constructed regression model for each battery, showing over $95\%$ classification accuracy when forming $5$ SoH categories with $4\%$ step-size. 
Note that this $4\%$ step-size is only for visual clarity, and a more fine-grained step-size of $0.1\%$ 
SoH is used for the evaluation of \nameS in Sec.~\ref{sec:evaluations}.
We have also tried other regression methods such as SVM, KNN, and their variations, but
have not observed any clear advantages over the regression tree in accuracy, {as summarized in Table~\ref{table:classificationaccuracy}.}
Thus, the regression tree is used for its simplicity and high interpretability.

\subsection{Generality Analysis}

\begin{figure*}
\centering
\begin{minipage}{2\columnwidth}
\centering
\begin{subfigure}{0.162\columnwidth}
{\includegraphics[width=1\columnwidth]{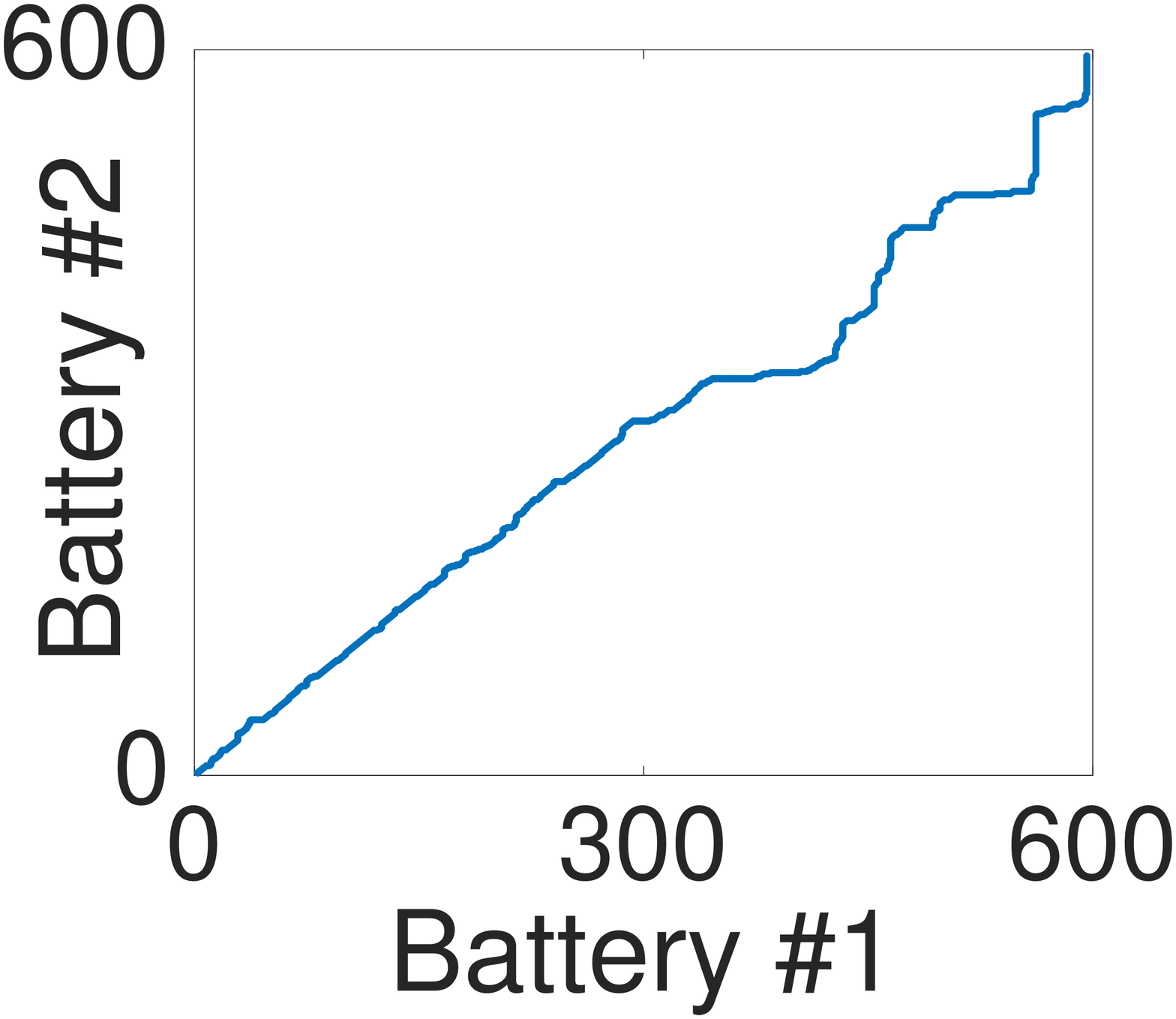}}
\caption{{\bf Betw. \#1 and \#2}}
\end{subfigure}
\hfill
\begin{subfigure}{0.162\columnwidth}
{\includegraphics[width=1\columnwidth]{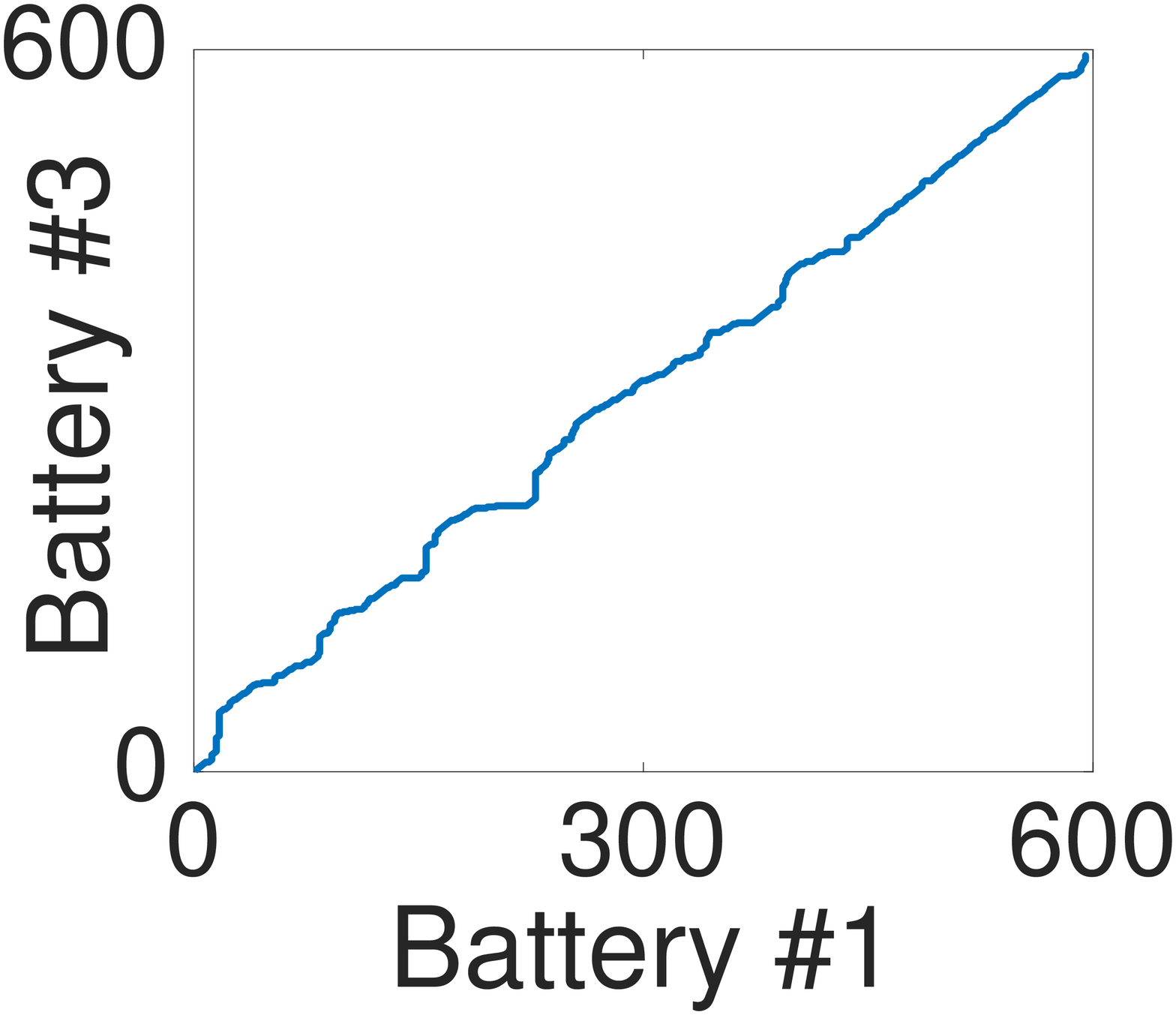}}
\caption{{\bf Betw. \#1 and \#3}}
\end{subfigure}
\hfill
\begin{subfigure}{0.162\columnwidth}
{\includegraphics[width=1\columnwidth]{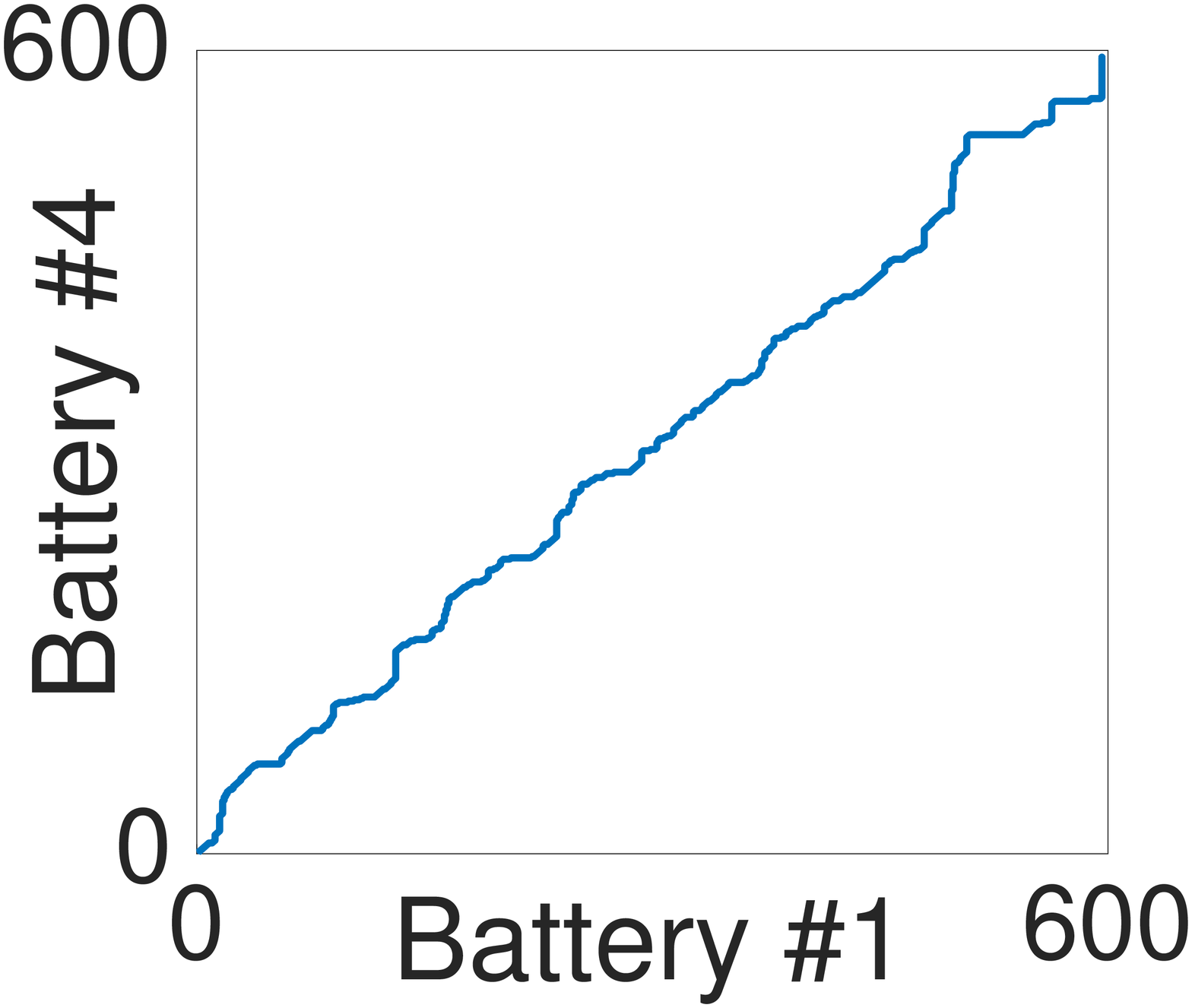}}
\caption{{\bf Betw. \#1 and \#4}}
\end{subfigure}
\hfill
\begin{subfigure}{0.162\columnwidth}
{\includegraphics[width=1\columnwidth]{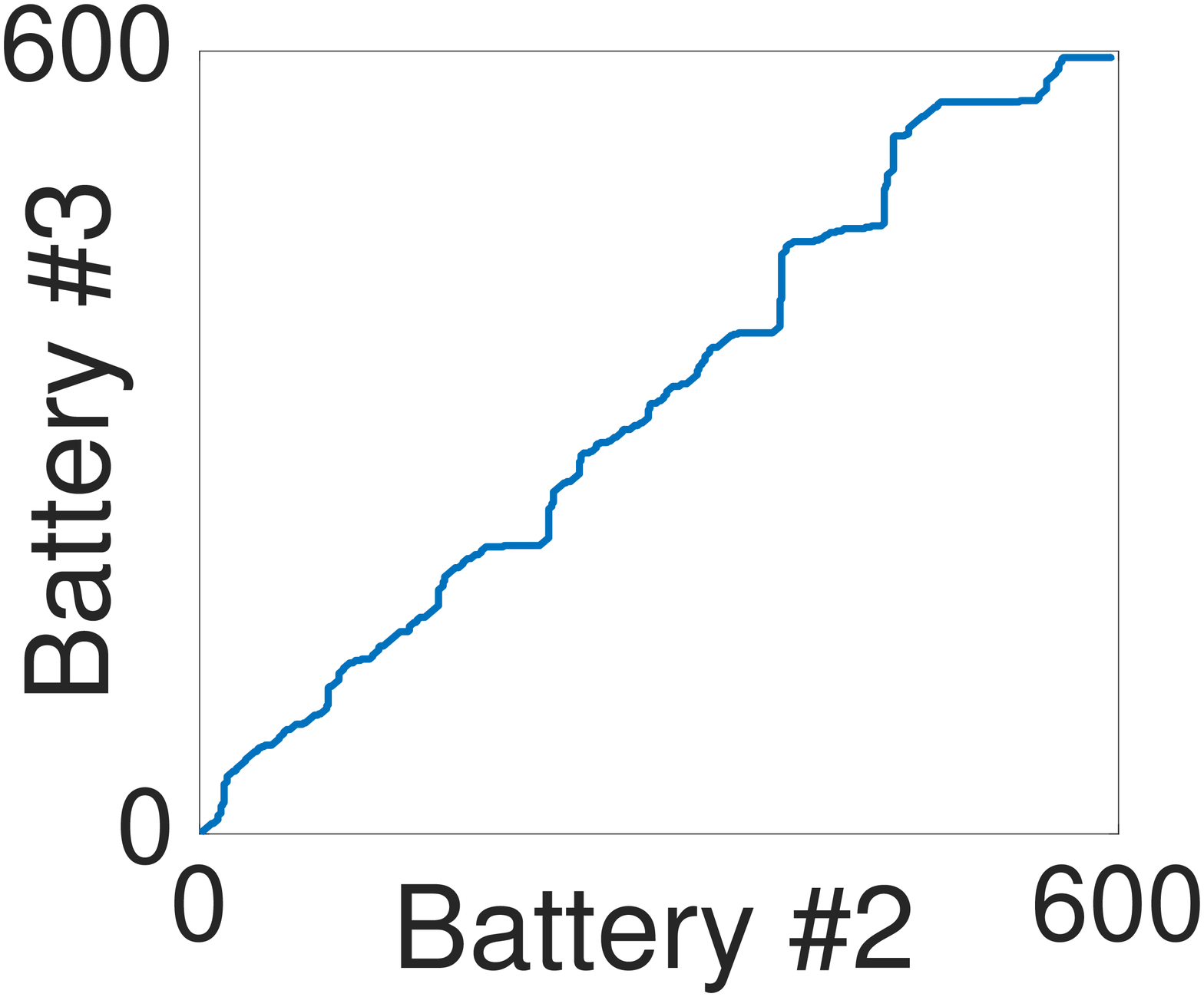}}
\caption{{\bf Betw. \#2 and \#3}}
\end{subfigure}
\hfill
\begin{subfigure}{0.162\columnwidth}
{\includegraphics[width=1\columnwidth]{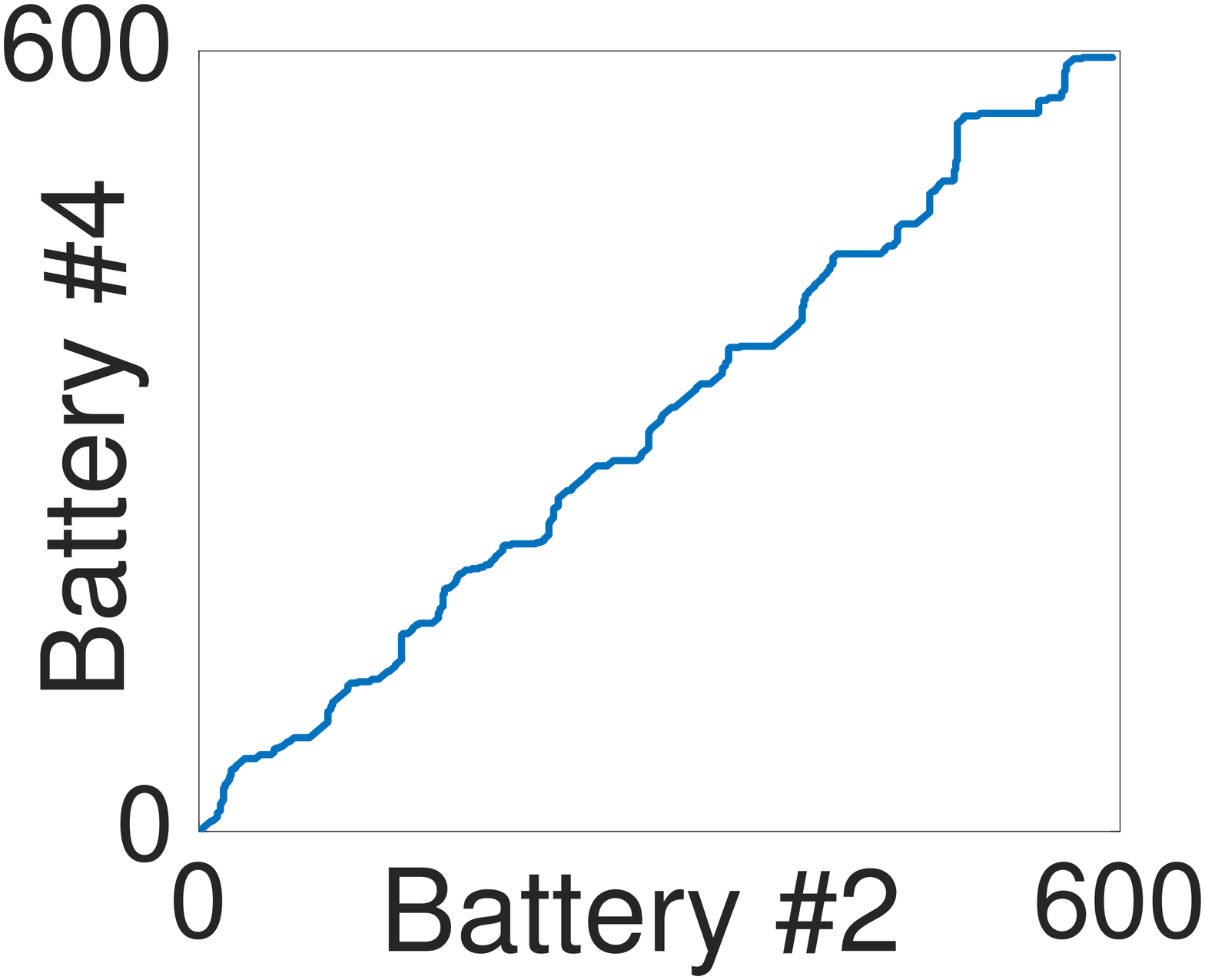}}
\caption{{\bf Betw. \#2 and \#4}}
\end{subfigure}
\hfill
\begin{subfigure}{0.162\columnwidth}
{\includegraphics[width=1\columnwidth]{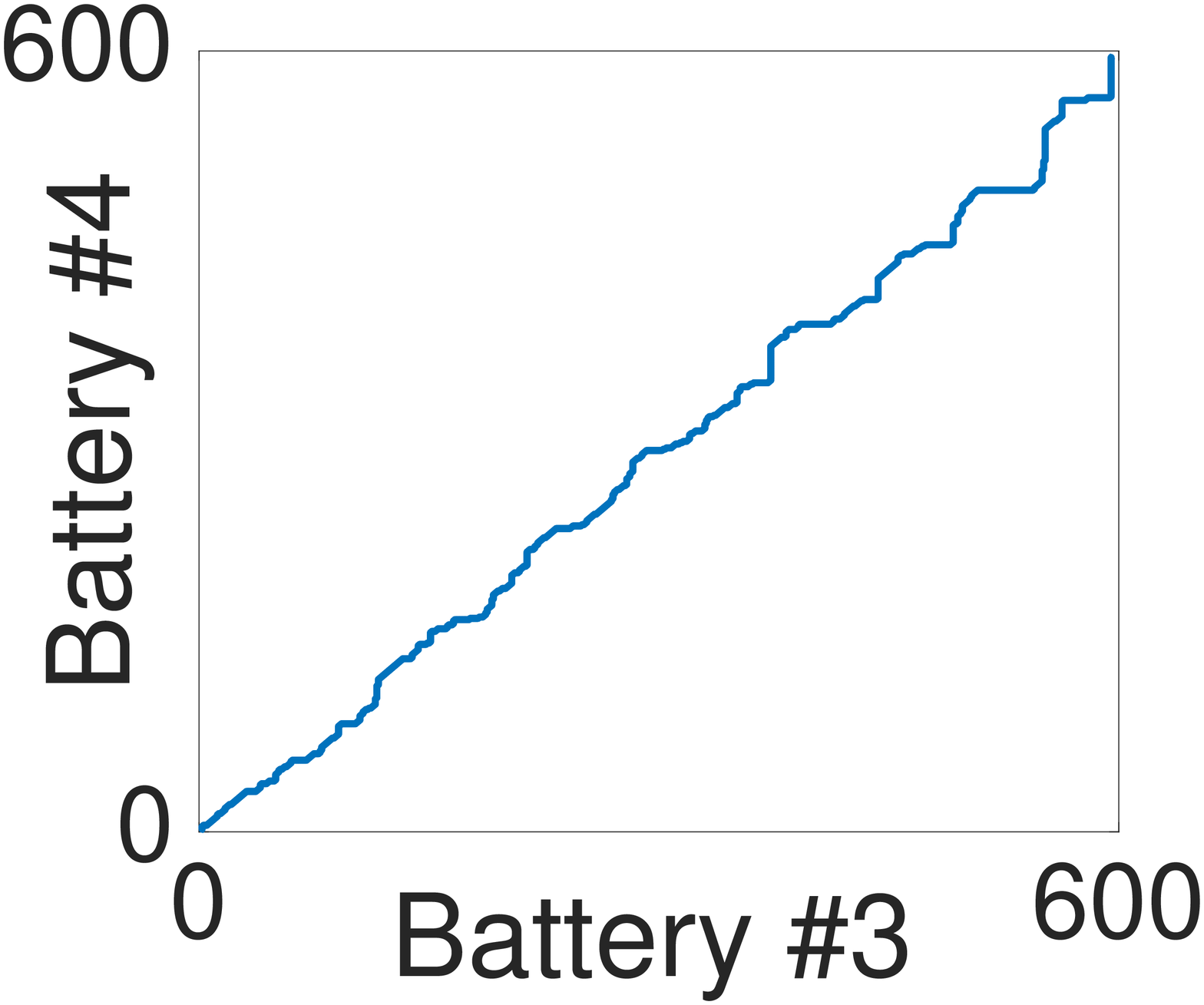}}
\caption{{\bf Betw. \#3 and \#4}}
\end{subfigure}
\caption{{\bf Similarity between degradation processes via dynamic time warping:} the close-to-diagonal 
warping paths show similarity between individual batteries' degradation processes.}
\label{fig:dtwresults}
\end{minipage}
\end{figure*}

The constructed fingerprint map has to be applicable for all same-model batteries, 
which can be verified with the following two statistical observations. 
First, we evaluated the similarity between the SoH degradation 
processes of the four batteries via dynamic time warping~\cite{DTW}, and the resultant warping paths 
are close to the diagonal of the degradation matrix for each battery pair (as shown in Fig.~\ref{fig:dtwresults}), exhibiting strong similarity.
Second, the SoH degradation of the four batteries used in the measurements are highly correlated, 
as shown in Table~\ref{table:correlateddegradation}.
These insights support \name's generality of training the fingerprint map with one (or more) battery 
and its application to other same-model batteries, which is reasonable as same-model batteries are expected to perform similarly --- a goal all battery manufactures aim to achieve~\cite{ZHANG2017}.
We will further evaluate the cross-battery estimation  
accuracy in Sec.~\ref{sec:evaluations}.

\begin{figure}
\begin{minipage}{1\columnwidth}
\centering
{\includegraphics[width = 0.8\columnwidth]{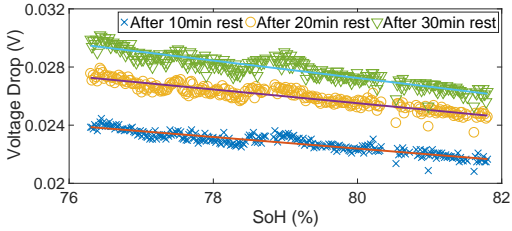}}
\caption{{\bf Linearity between voltage drop and SoH:} allows to extend the limited cycling dataset to uncovered SoH ranges, if needed.}
\label{fig:extendingdataset}
\end{minipage}
\end{figure}

\subsection{Extending Dataset}

Ideally, \nameS is to be provided by OEMs because of their accessibility to battery cycling datasets, 
e.g., covering a complete battery SoH range.
In case only a limited dataset is available, it can be extrapolated based on the linearity between voltage drop 
during resting and battery SoH.
Again, we used the cycling measurements in Fig.~\ref{fig:VoltageandSoH} to show this observation.
Fig.~\ref{fig:extendingdataset} plots the voltage drop after the battery is rested for $10$, $20$ and $30$ minutes  
during the resting period of each cycle, together with the corresponding battery SoH during that cycle. 
We can see clear linearity in all three traces of dropped voltages, with RMSE in the order of $10^{-4}$ after linear fitting.
This observation enables to identify the linear coefficients based on the available cycling dataset, 
generate relaxing voltages that correspond to uncovered SoH, and then construct the complete 
voltage fingerprint map.

\section{Collection of Relaxing Voltages on Mobile Devices}
\label{sec:voltagecollection}

We now describe how to collect local relaxing voltages on mobile devices. 

\subsection{Collection During Over-Night Charge}

\begin{figure}[ht!]
\begin{minipage}{1\columnwidth}
\begin{subfigure}{0.49\columnwidth}
\centering
{\includegraphics[width=1\columnwidth]{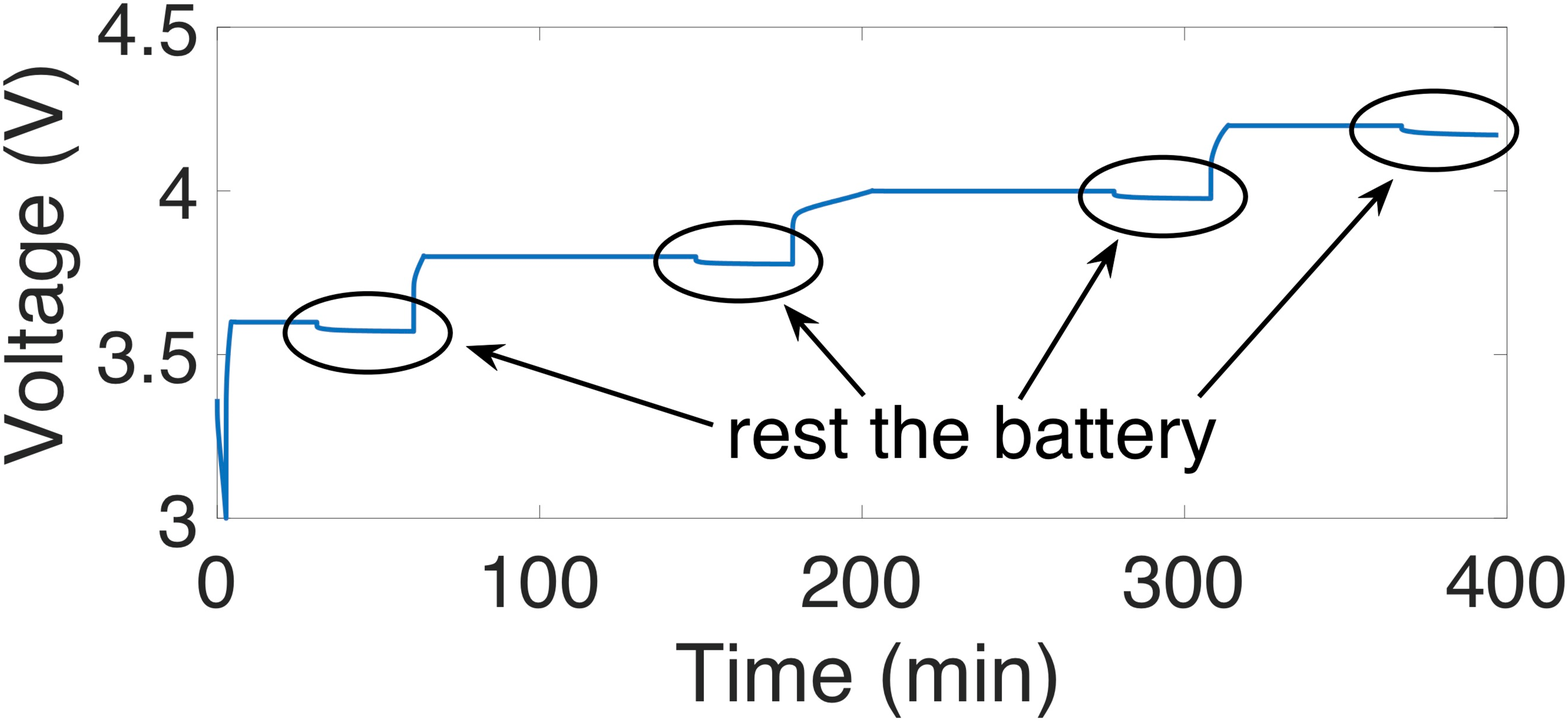}}
\caption{{\bf Voltage during measurement}}
\end{subfigure}
\hfill
\begin{subfigure}{0.49\columnwidth}
\centering
{\includegraphics[width=1\columnwidth]{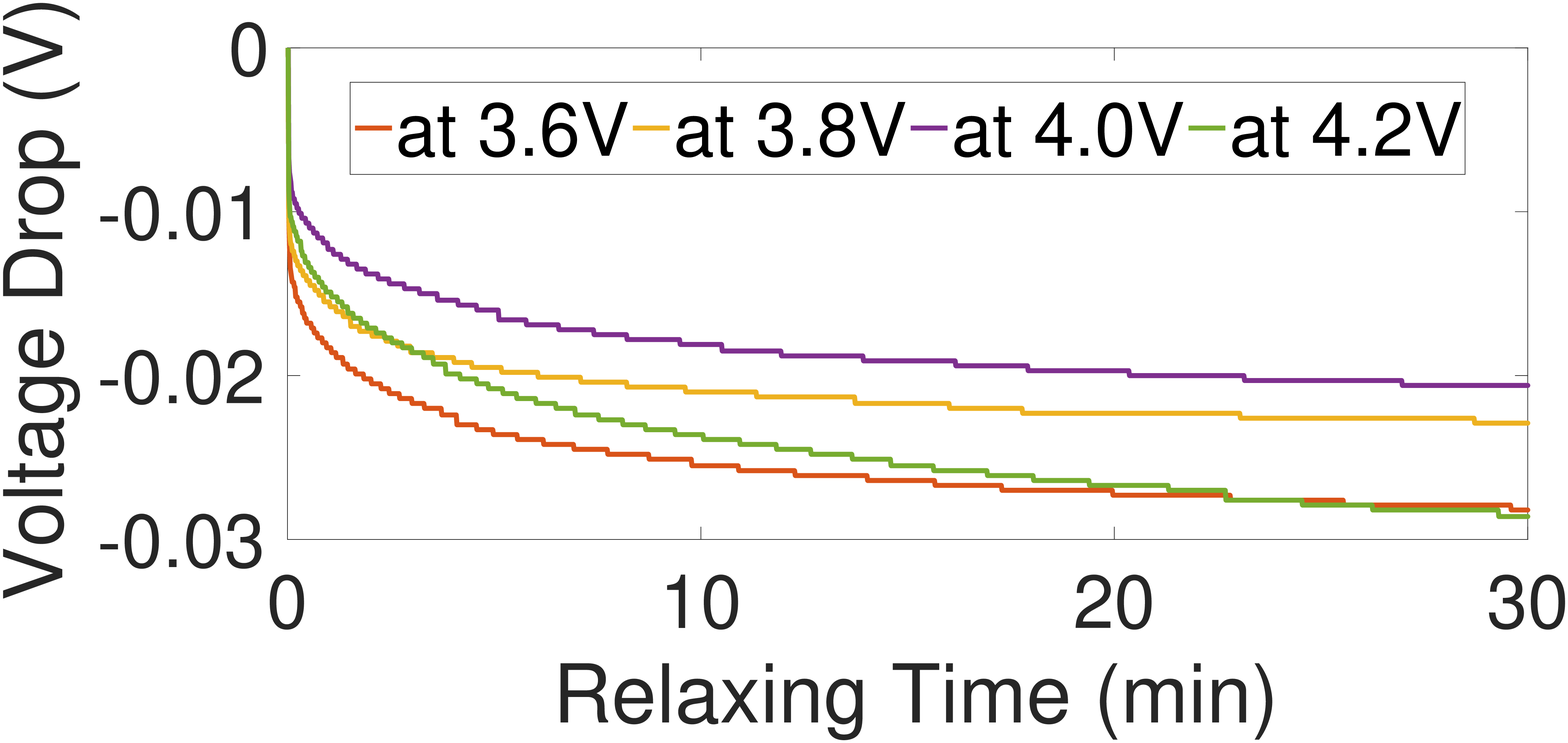}}
\caption{{\bf Relaxing voltages}}
\end{subfigure}
\caption{{\bf Starting voltage matters:} the relaxing voltage is affected by its starting voltage level, 
necessitating a unified starting voltage.}
\label{fig:relaxatdiffvoltage}
\end{minipage}
\vspace{+6pt}

\begin{minipage}{0.48\columnwidth}
\centering
{\includegraphics[width = 1\columnwidth]{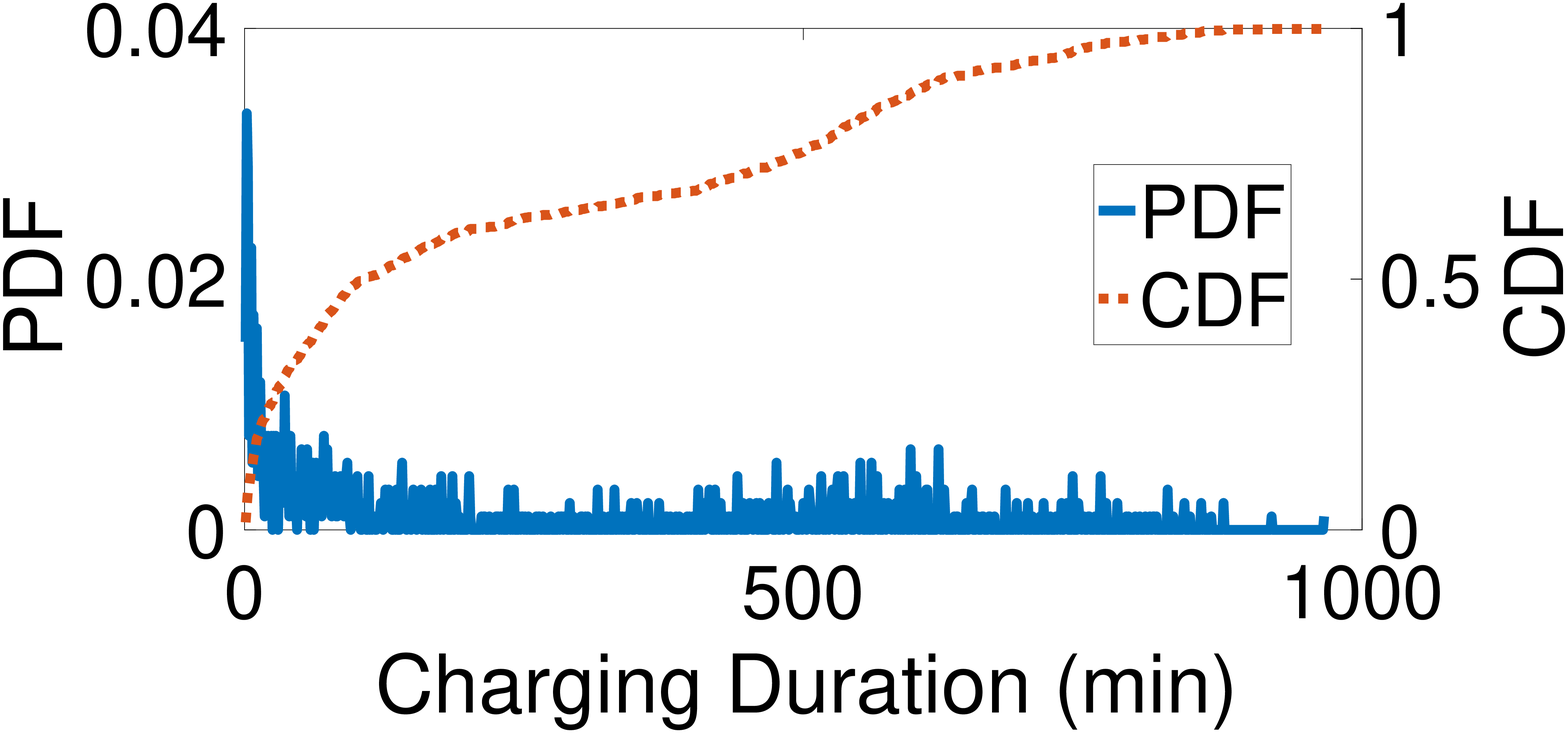}}
\caption{{\bf Users often charge devices over-night:} $34\%$ of the collected charge cases lasted 
 $>$$6$ hours.}
\label{fig:chargingtimecdf}
\end{minipage}
\hfill
\begin{minipage}{0.48\columnwidth}
\centering
{\includegraphics[width = 1\columnwidth]{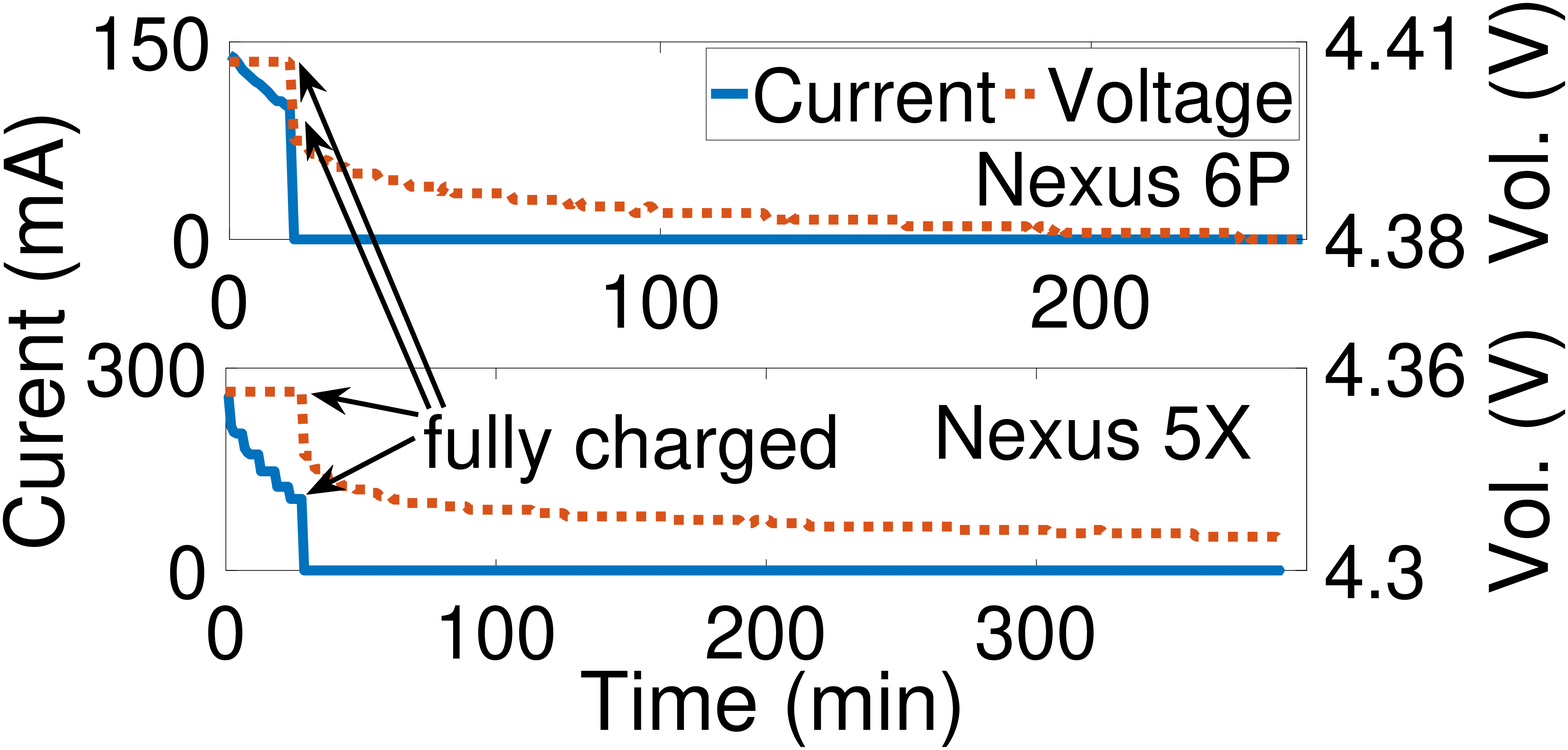}}
\caption{{\bf Over-night charge rests battery:} phone batteries are rested after reaching $100\%$ 
SoC.}
\label{fig:afterchargingtermination_Nexus6p}
\end{minipage}
\vspace{+6pt}

\centering
\begin{subfigure}{0.325\columnwidth}
\centering
{\includegraphics[width = 1\columnwidth]{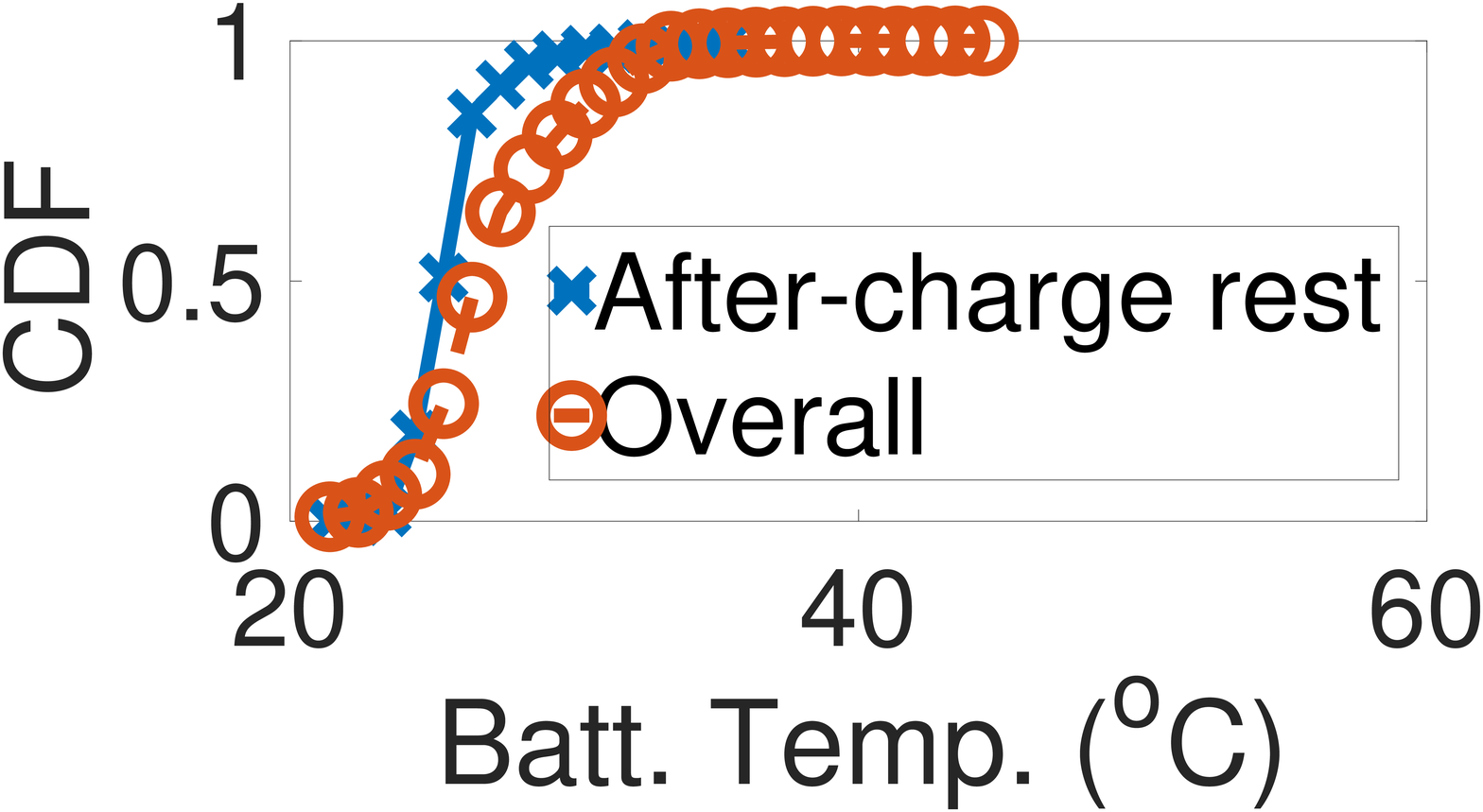}}
\caption{{\bf Galaxy S6 Edge}}
\end{subfigure}
\hfill
\begin{subfigure}{0.325\columnwidth}
\centering
{\includegraphics[width = 1\columnwidth]{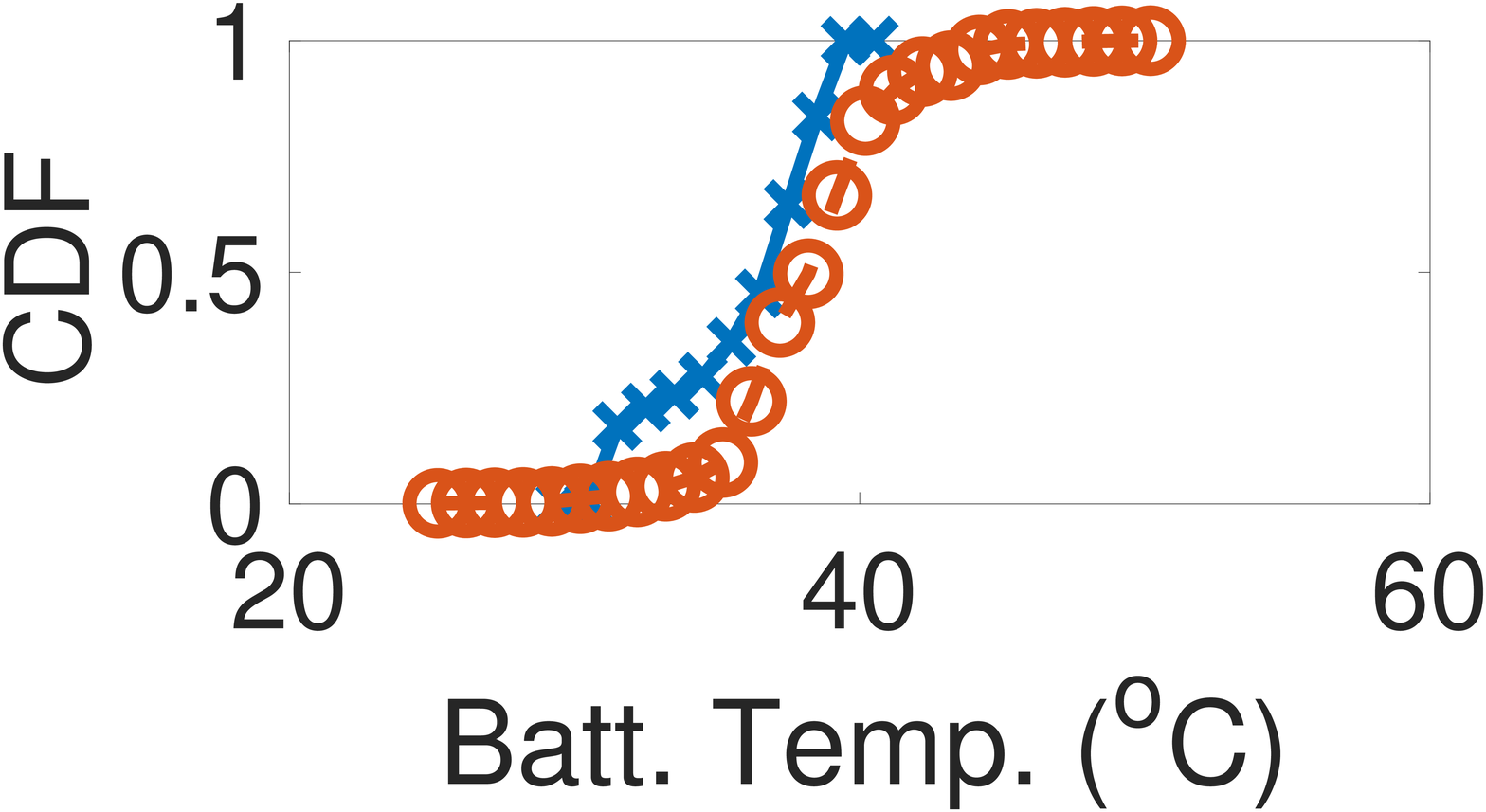}}
\caption{{\bf Nexus 5X}}
\end{subfigure}
\hfill
\begin{subfigure}{0.325\columnwidth}
\centering
{\includegraphics[width = 1\columnwidth]{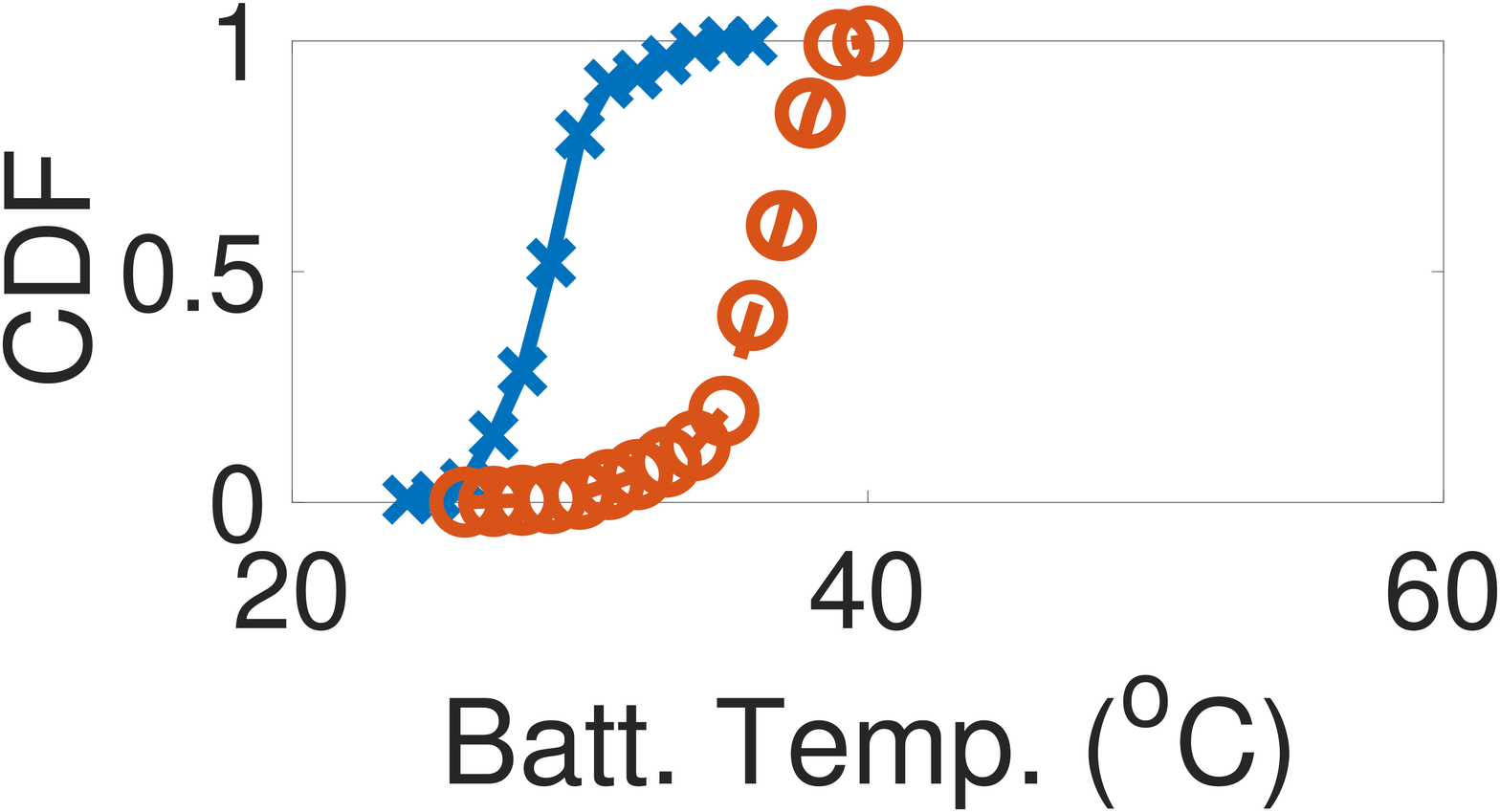}}
\caption{{\bf Nexus 6P}}
\end{subfigure}
\caption{{\bf Stable temperature during resting:} battery temperature during the after-charging resting 
period is relatively stable.}
\label{fig:temperature_5XandS6Edge}
\end{figure}

The relaxing voltages are not always collectable on mobile devices for the following reasons. 
First, the relaxing voltage requires batteries to be idle (i.e., the $30$-minute resting period 
in our cycling measurements). 
Mobile devices, however, discharge their batteries with continuous and dynamic currents even 
in idle mode, due to device monitoring and background activities~\cite{backgroundactivities,7530866,niranjan}.
Also, battery voltage is temperature-dependent~\cite{Trojan,5212025,entropycoefficient}, so a stable thermal environment 
is required to collect the relaxing voltages. This is challenging due to the well-publicized device 
overheating problem~\cite{heating}. 
Last but not the least, the relaxing voltage is affected by its starting voltage. 
Fig.~\ref{fig:relaxatdiffvoltage} compares the relaxing voltage when resting the battery at 
different voltages within [$3.6$,~$4.2$]V, showing a clear dependency between the relaxing voltage 
and its starting voltage level. Such dependency requires a unified starting voltage for the collection of
relaxing voltages.

\begin{figure*}
\centering
\begin{minipage}{2\columnwidth}
\begin{subfigure}{0.325\columnwidth}
\centering
{\includegraphics[width = 1\columnwidth]{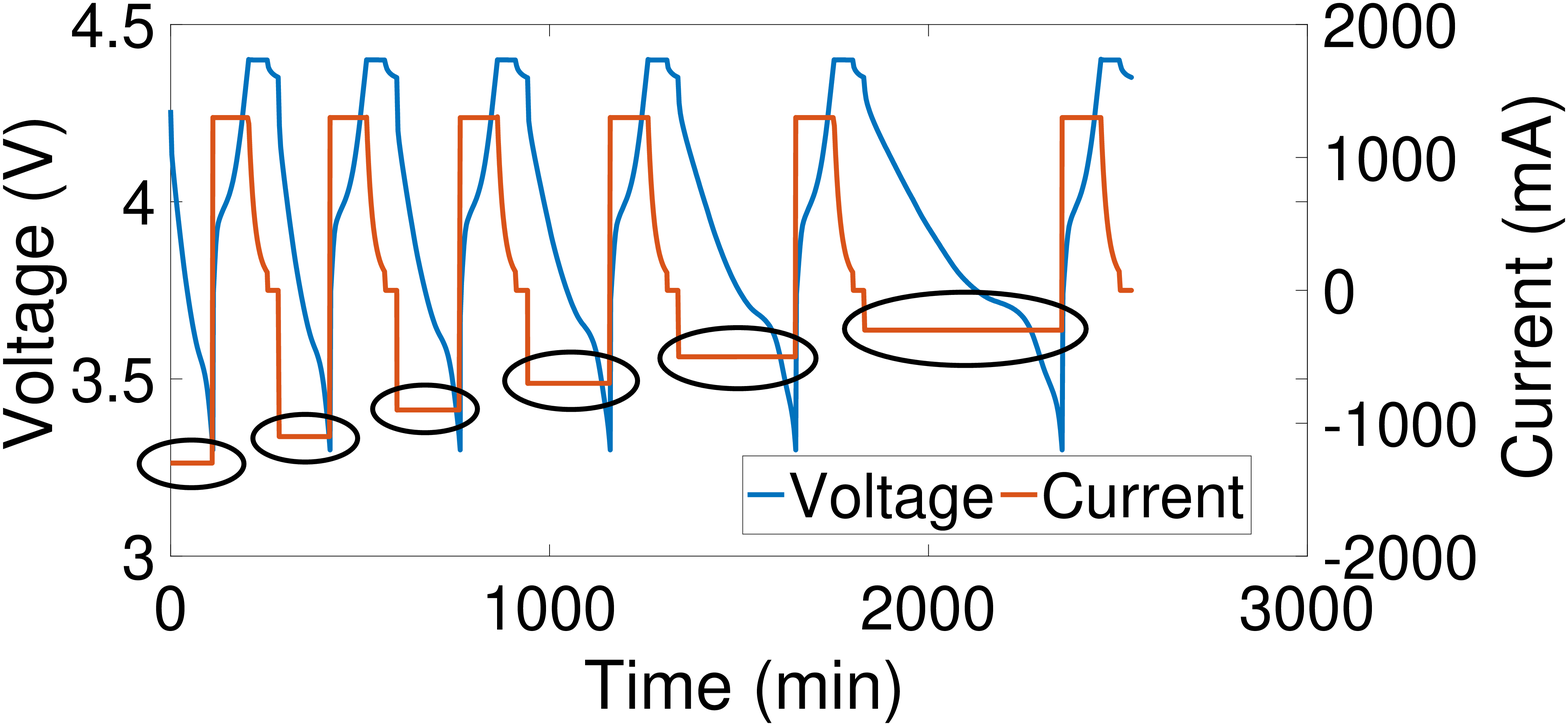}}
\caption{{\bf Discharge with different currents}}
\end{subfigure}
\hfill
\begin{subfigure}{0.325\columnwidth}
\centering
{\includegraphics[width = 1\columnwidth]{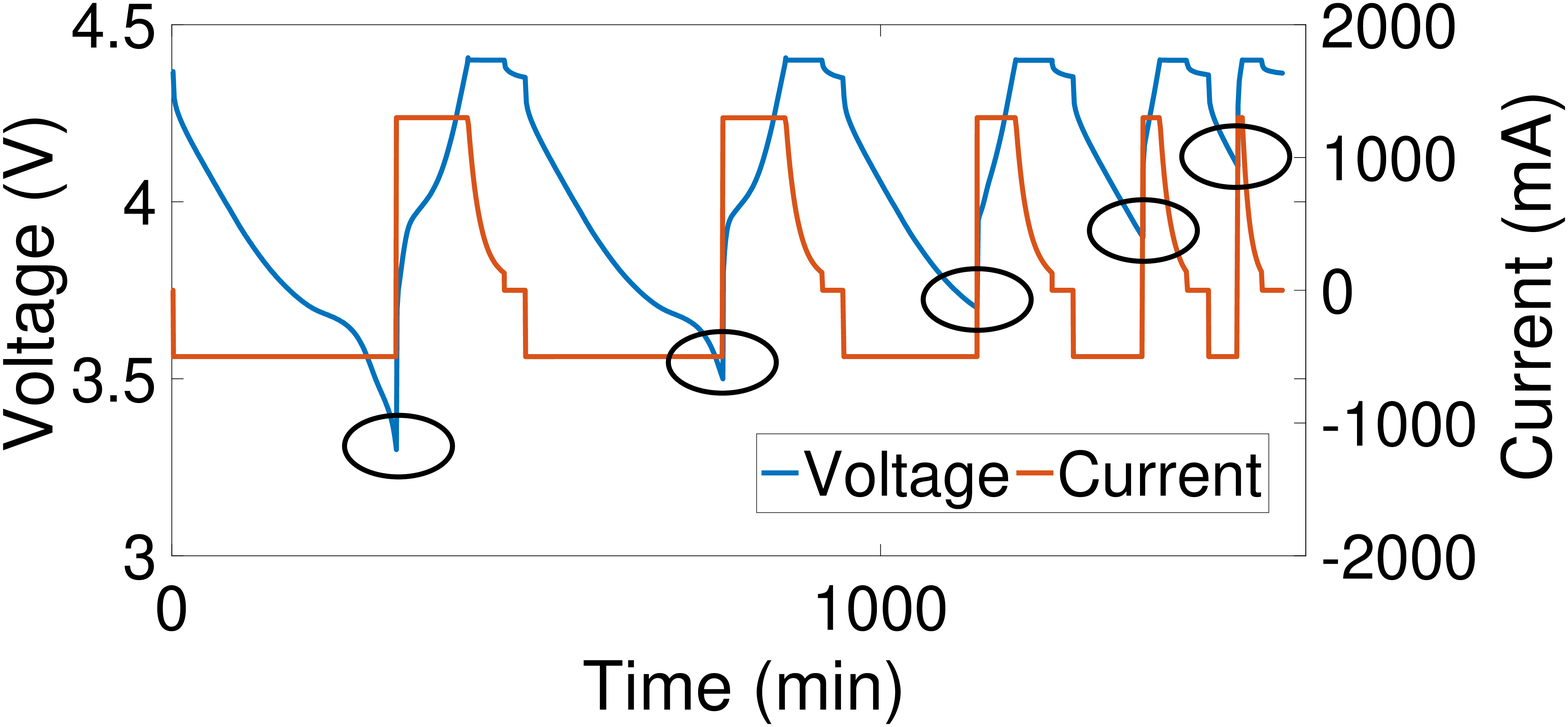}}
\caption{{\bf Discharge to different voltages}}
\end{subfigure}
\hfill
\begin{subfigure}{0.325\columnwidth}
\centering
{\includegraphics[width = 1\columnwidth]{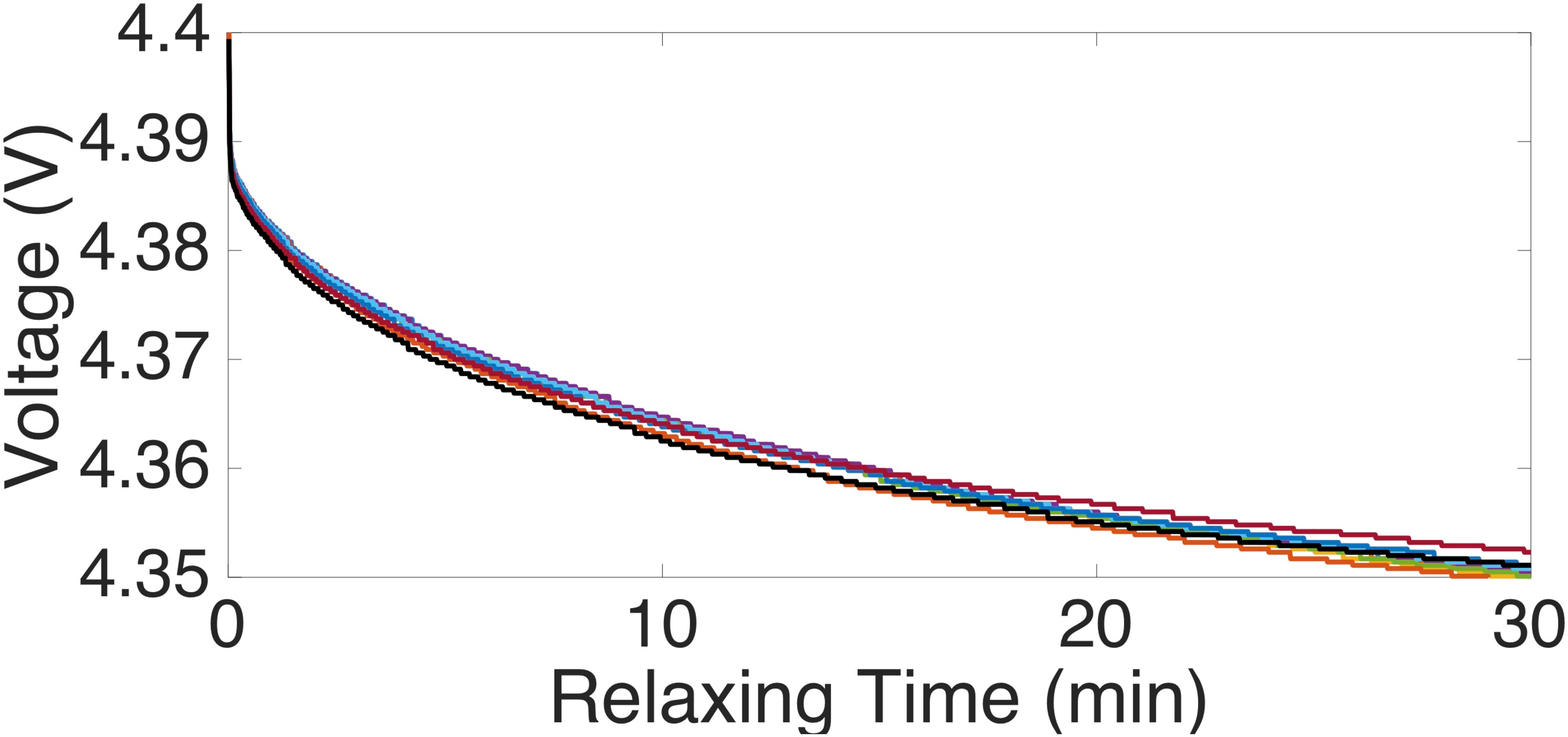}}
\caption{{\bf Collected relaxing voltages}}
\end{subfigure}
\end{minipage}
\caption{{\bf  Relaxing voltages after charging are insensitive to discharge:} relaxing voltages collected after discharging with different currents and to different cutoff voltages are close, exhibiting their insensitivity to previous discharge and thus reliability.}
\label{fig:insensitivity}
\end{figure*}

\nameS mitigates these challenges based on the fact that users often charge their devices 
over-night --- the charging duration is so long that the charger is kept connected 
even after the device is fully charged.
Fig.~\ref{fig:chargingtimecdf} plots the charging time (i.e., the time from the charger's connection to 
disconnection) distribution of $976$ charge cases collected from $7$ users over $1$--$3$ 
months,\footnote{One of the user-traces was collected from our data-collection campaign and the other 
six traces were obtained from the sample dataset of Device Analyzer from Cambridge 
University~\cite{deviceanalyzer}.} 
showing $34\%$ of them lasted over $6$ hours and are long enough to keep the charger connected 
after the device was fully charged, due to the common over-night charge~\cite{Ferreira,deviceanalyzer,Nilanjan}.
\nameS starts to collect the relaxing voltage once the battery reaches $100\%$ SoC during over-night charge, 
and stops it when the charger is disconnected. This collection of relaxing voltages mitigates all 
the above-mentioned challenges. 

First, over-night device charge rests its battery by powering the device operation with the charger.
Commodity chargers use separate power paths to charge the battery and power the device~\cite{Yevgen}, 
resting the battery if the charger is kept connected even after the battery reaches $100\%$ SoC, 
as in over-night charge. 
Fig.~\ref{fig:afterchargingtermination_Nexus6p} shows such rested batteries by keeping 
the chargers connected after fully charging a Nexus 6P and a Nexus 5X phone
--- the current reduces to, and stays at $0$mA after fully charging the battery and thus 
resting the battery; the battery voltage first instantly and then gradually drops, agreeing 
with Fig.~\ref{fig:VoltageandSoH}.
Second, over-night charge provides battery a relatively stable thermal environment. 
Most mobile devices charge their batteries with CCCV~\cite{Hoque}, during which the CV-Chg phase takes long 
at a low charging rate, thus not heating the battery much and allowing for its equilibration. 
This way, the battery operates in a stable thermal environment during the resting period after the CV-Chg
phase completes (and thus, the battery is fully charged).
To verify this, we monitor the battery temperature of a Galaxy S6 Edge, a Nexus 5X, and a Nexus 6P during 
an $8$-day real-life usage. 
Fig.~\ref{fig:temperature_5XandS6Edge} compares the temperature distribution during the resting periods
after fully charging them with that under normal usage, showing reduced thermal variations, e.g., 
the temperature range of the Nexus 5X battery is narrowed from $25$--$50$$^{\rm o}$C in normal case 
to $29$--$39$$^{\rm o}$C when resting.
Last but not the least, collecting relaxing voltages after the battery is fully charged unifies the starting voltage at the fully charged level, e.g., $4.37$V for Galaxy S6 Edge.

We must also consider if a device's usage pattern (i.e., how its battery is discharged) affects its 
after-charging relaxing voltages. To this end, we discharge, charge, and then rest a Galaxy S4 
battery for (i) $6$ cycles with different discharge currents within [$300$, $1300$]mA 
(Fig.~\ref{fig:insensitivity}(a)), and (ii) another $5$ cycles with a different cutoff voltage 
within [$3.3$, $4.1$]V (Fig.~\ref{fig:insensitivity}(b)).
The thus-collected $6+5=11$ relaxing voltage traces during each resting period are plotted in 
Fig.~\ref{fig:insensitivity}(c). These relaxing voltages are very close to each other 
(e.g., in comparison with Fig.~\ref{fig:relaxatdiffvoltage}), 
exhibiting their insensitivity to previous discharge and thus reliability --- a key advantage over~\cite{C} 
as shown in Fig.~\ref{fig:beforechargeoperation}.
Again, this is because the charge, especially CV-Chg, of the battery masks the disturbance 
caused by their previous discharge from the resting period after being fully charged.

\begin{figure}[ht!]
\begin{minipage}{1\columnwidth}
\centering
{\includegraphics[width = 0.8\columnwidth]{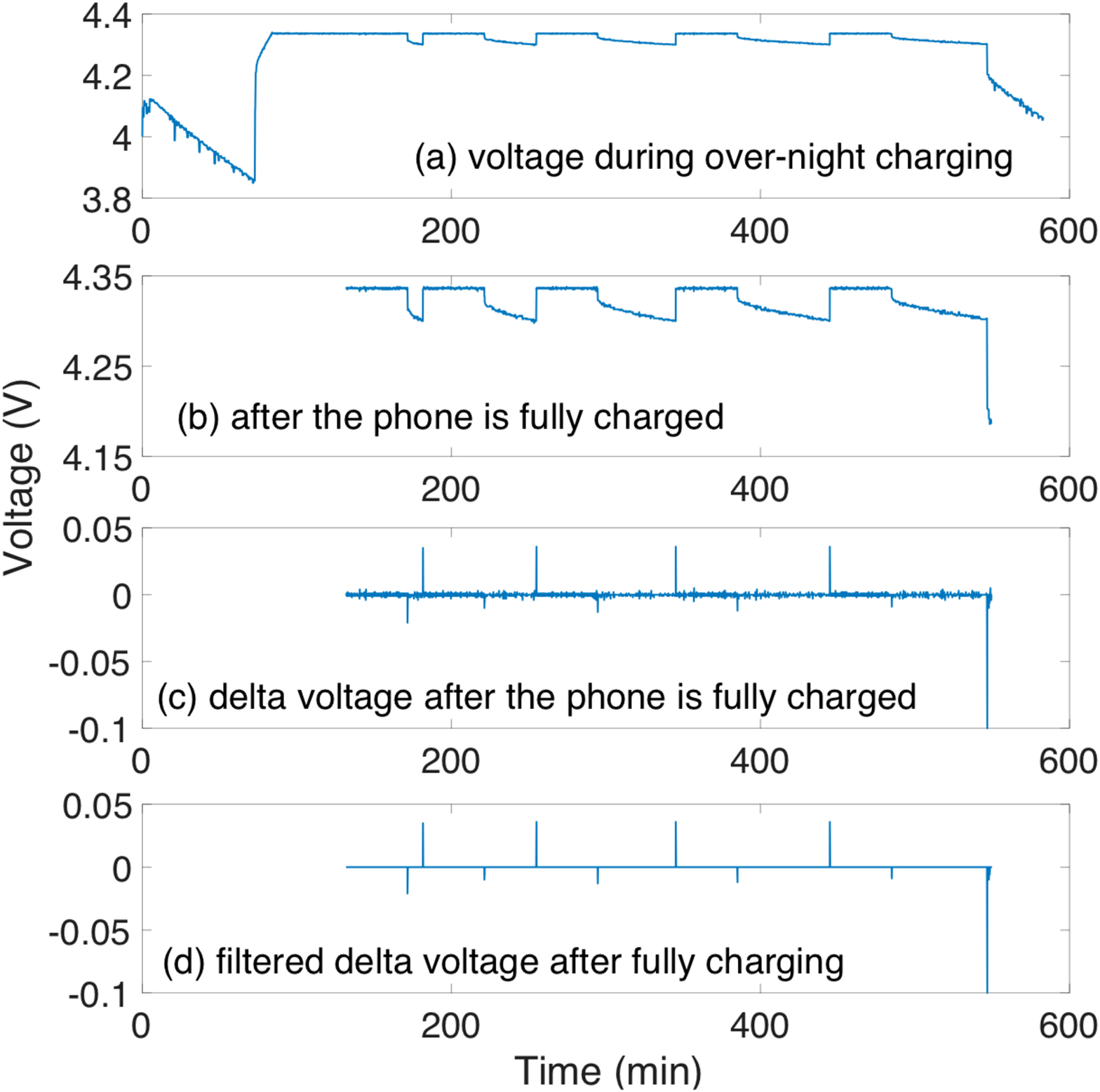}}
\caption{{\bf Mitigating trickle charge:} trickle charge pollutes the collected relaxing voltages ((a) and (b)); \nameS 
extracts sub-traces from the polluted trace by identifying the starting/stopping time instants of trickle charge ((c) and (d)).}
\label{fig:tricklechargingmitigation}
\end{minipage}
\end{figure}

\subsection{Mitigating Trickle Charge}

Certain mobile devices (e.g., Galaxy S6 Edge, Galaxy S4, etc.) use trickle charge --- charging 
a fully charged battery under no-load at a rate equal to its self-discharge rate --- to keep their battery 
at $100\%$ SoC, which invalidates the battery resting and thus pollutes the collected relaxing 
voltages.\footnote{A simple mitigation is to prevent trickle charge by disabling charge once the battery 
reaches $100\%$ SoC. For example, for Nexus 5X and Nexus 6P, this can be done by setting the {\tt battery\_charging\_enabled} flag to $0$, located at {\tt sys/class/power\_supply/battery}. This approach, however, lacks generality as the root privilege is needed.}
Specifically, these devices trigger trickle charge once the voltage of a fully-charged battery has dropped 
for a pre-defined value, e.g., $20$mV for Galaxy S6 Edge and $40$mV for Galaxy S4, and stop 
the trickle charge until the battery is fully charged again. 
Fig.~\ref{fig:tricklechargingmitigation}(a) plots the voltage of a Galaxy S4 phone during an over-night 
charge, during which trickle charge is triggered $6$ times after the phone is fully charged, 
as shown in Fig.~\ref{fig:tricklechargingmitigation}(b). 
The duration between two consecutive trickle charges increases because the battery OCV 
approaches the fully-charged level. 

Trickle charge prevents battery from resting and thus pollutes the relaxing voltages. 
\nameS extracts relaxing sub-traces from the polluted trace with a simple observation that a sudden 
increase/drop of battery voltage indicates the triggering/stopping of trickle charge. 
Specifically, \nameS calculates the 1-lag delta voltage after the device is fully charged 
(Fig.~\ref{fig:tricklechargingmitigation}(c)), and passes it through a low-pass filter 
(Fig.~\ref{fig:tricklechargingmitigation}(d)). This way, \nameS extracts the relaxing sub-traces 
by locating the peaks and valleys in the trace.

\begin{figure}[t]
\centering
\begin{subfigure}{0.49\columnwidth}
\centering
{\includegraphics[width = 1\columnwidth]{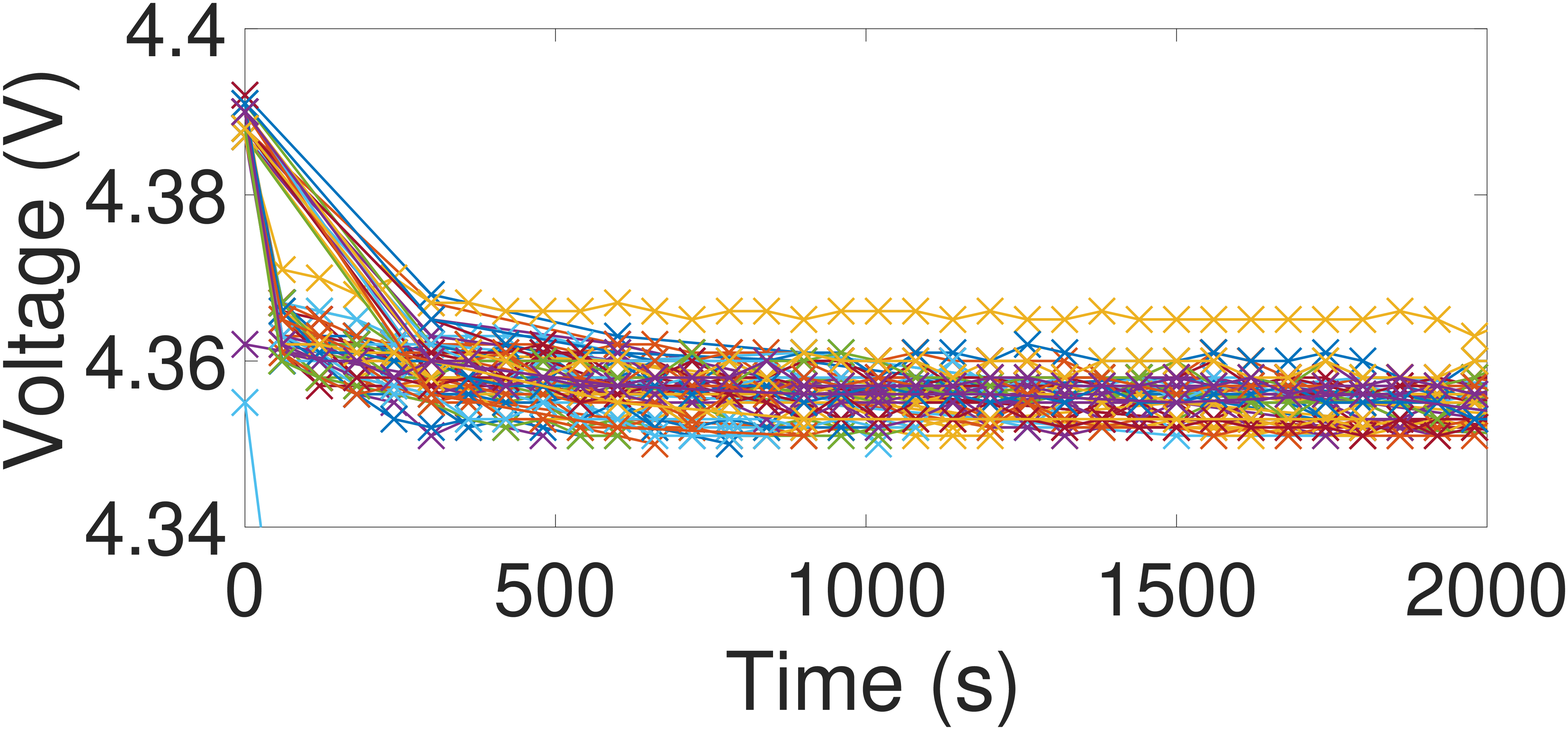}}
\caption{{\bf Raw traces}}
\end{subfigure}
\hfill
\begin{subfigure}{0.49\columnwidth}
\centering
{\includegraphics[width = 1\columnwidth]{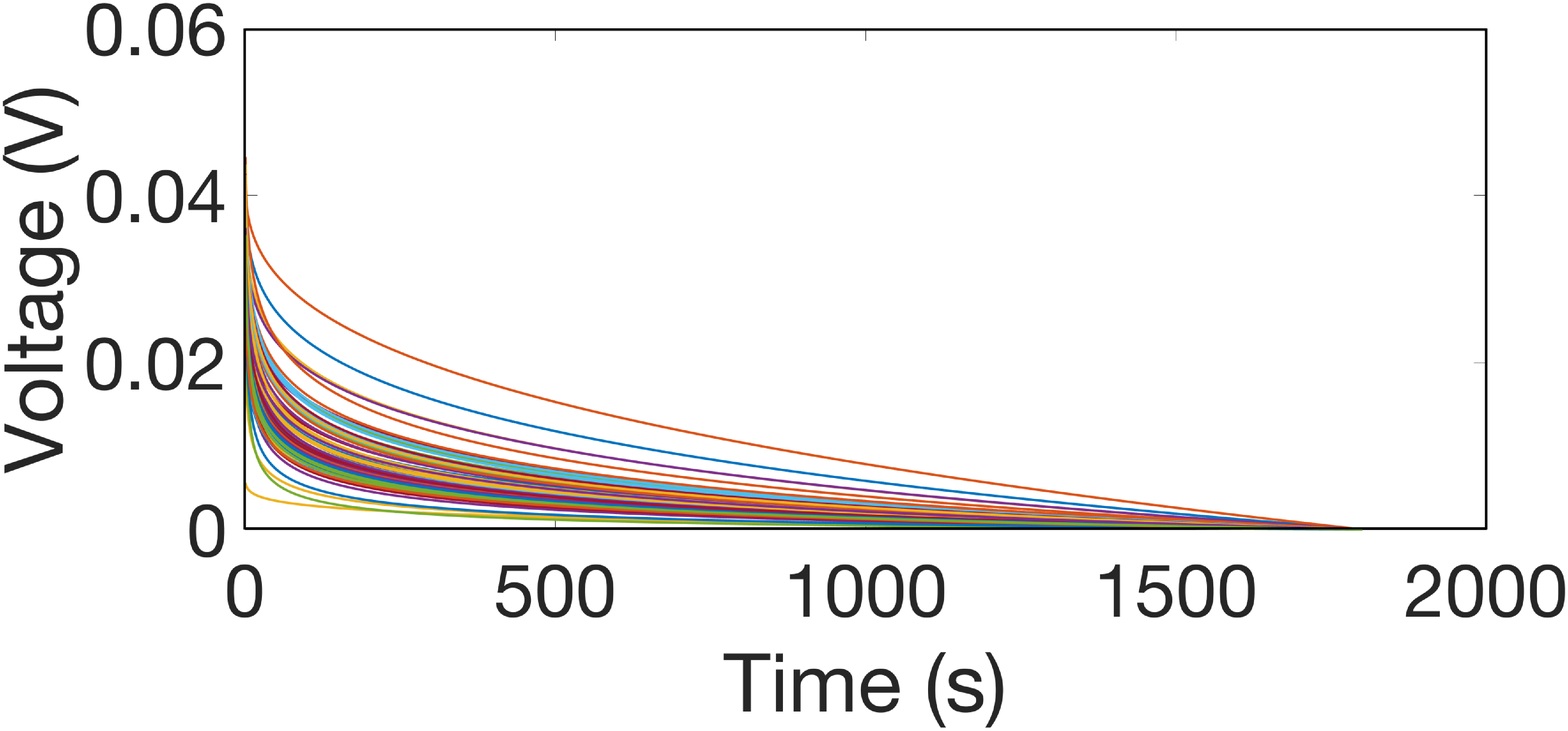}}
\caption{{\bf Processed traces}}
\end{subfigure}
\caption{{{\bf Relaxing voltages collected on a Galaxy S5 phone:} (a) raw traces after mitigating trickle charge; (b) processed traces used for fingerprint checking.}}
\label{fig:traceprocessing}
\end{figure}

Fig.~\ref{fig:traceprocessing}(a) plots $95$ of thus-extracted sub-traces with a Galaxy S5 phone, showing the power shape but with significant variance. To further improve trace quality, \nameS applies power fitting to each of these traces, concluding them to be valid if the goodness-of-fit is acceptable. 
Moreover, the sub-traces may not be long enough to form a fingerprint. 
To remedy this problem, \nameS recovers the sub-traces to, e.g., $30$-minute traces, based on the power fitting, which is then used for fingerprint checking. 
Last but not the least, \nameS uses the dropped voltages upon resting as the fingerprint to remove its dependency on the specific values of fully-charged voltage.
Fig.~\ref{fig:traceprocessing}(b) plots the processed traces based on the raw data in Fig.~\ref{fig:traceprocessing}(a).

\subsection{Post-Processing of SoH Estimations}

Multiple relaxing traces are likely to be collected and recovered during a single over-night 
charge (as in Fig.~\ref{fig:tricklechargingmitigation}), and thus multiple SoH estimations may result. 
\nameS averages such estimations as the battery SoH during that charge.
Also, there may be fluctuations among SoHs obtained from different over-night charges.
\nameS uses a first-order smoother (i.e., estimating the current SoH by linear fitting current and 
previous raw SoH estimations) to smooth such fluctuations, and reports the smoothed result as the final 
battery SoH to users. Such smoothing of fluctuations is also used in the SoC estimation of 
mobile devices~\cite{impedancetrack}.

\begin{figure}[ht!]
\centering
\begin{minipage}{1\columnwidth}
\begin{subfigure}{1\columnwidth}
\centering
{\includegraphics[width = 1\columnwidth]{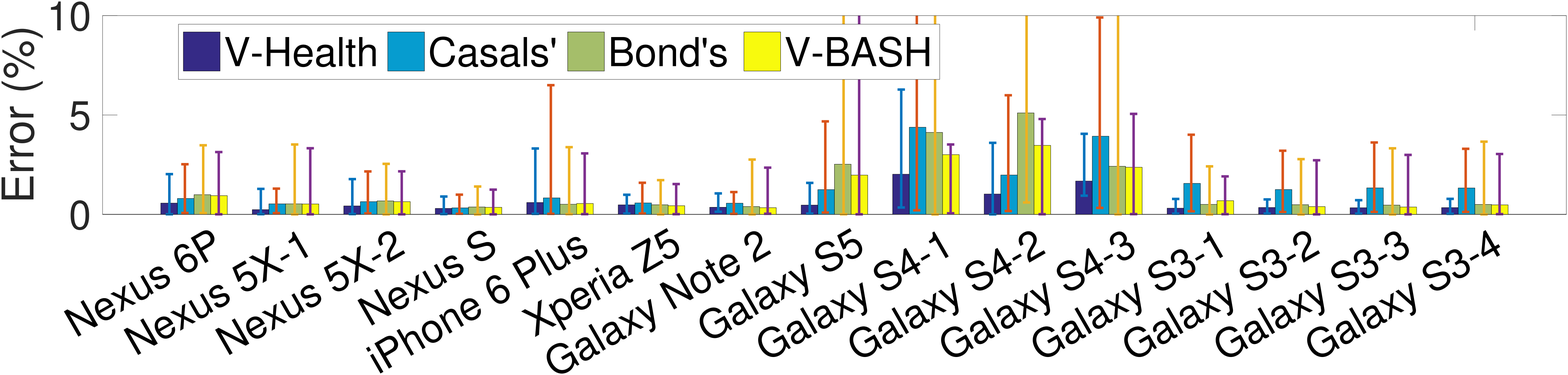}}
\caption{{\bf Same-battery validation}}
\end{subfigure}
\vspace{+6pt}

\begin{subfigure}{1\columnwidth}
\centering
{\includegraphics[width = 1\columnwidth]{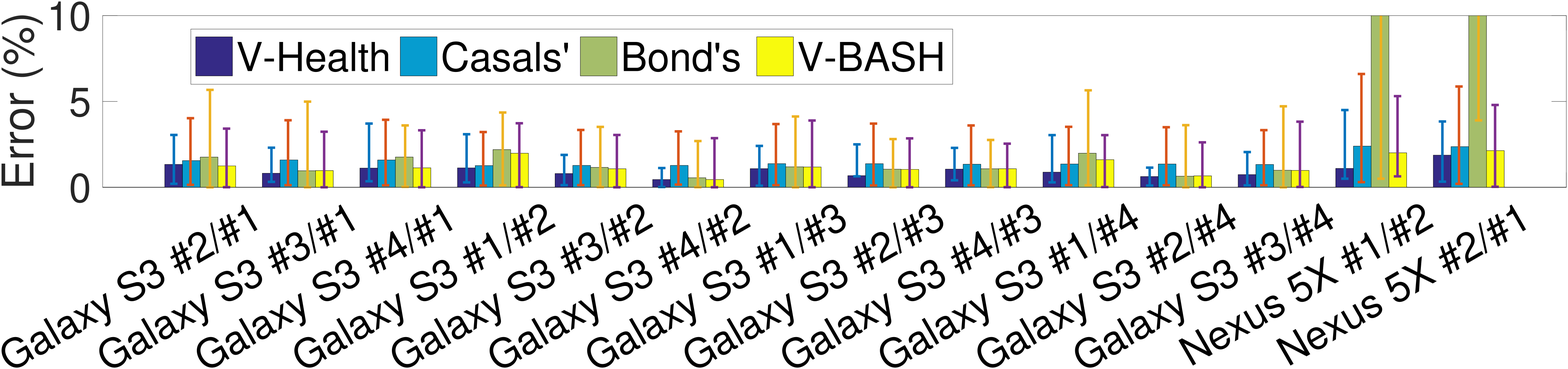}}
\caption{{\bf Cross-battery validation}}
\end{subfigure}
\vspace{+6pt}

\begin{subfigure}{1\columnwidth}
\centering
{\includegraphics[width = 1\columnwidth]{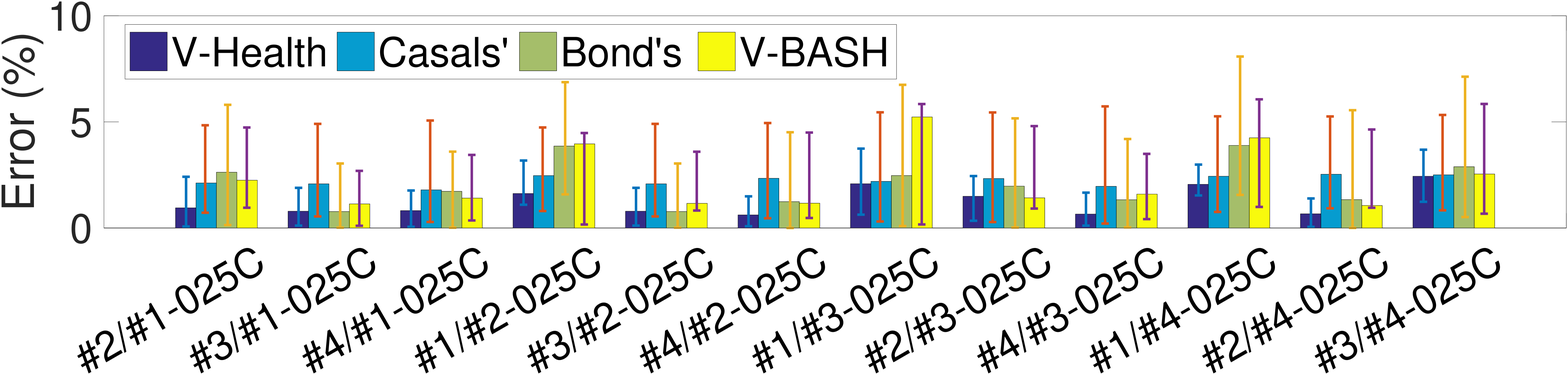}}
\caption{{\bf Cross-battery cross-profile validation}}
\end{subfigure}

\begin{subfigure}{1\columnwidth}
\centering
{\includegraphics[width = 1\columnwidth]{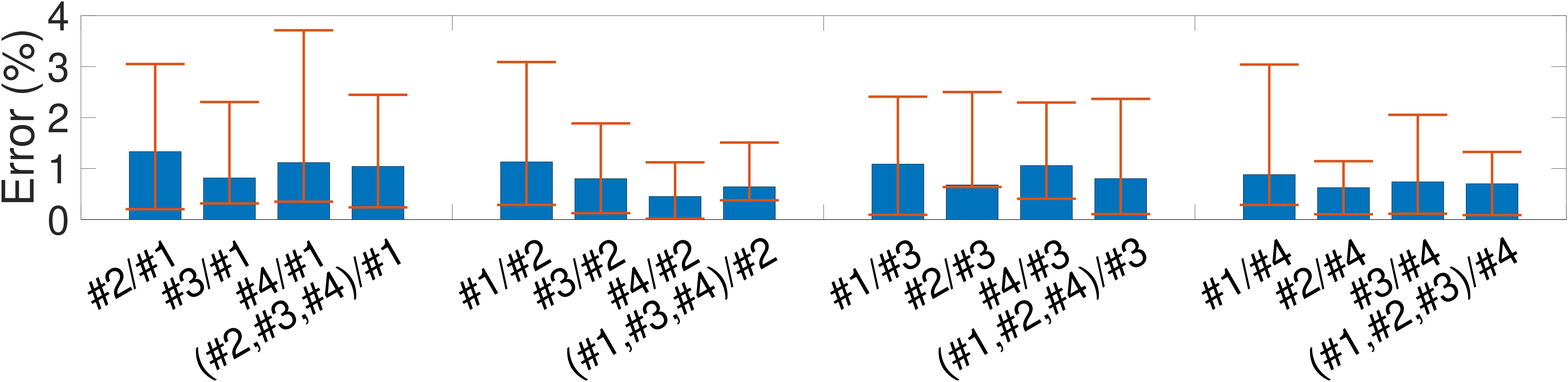}}
\caption{{\bf Training with multiple batteries}}
\end{subfigure}
\vspace{+6pt}

\begin{subfigure}{1\columnwidth}
\centering
{\includegraphics[width = 1\columnwidth]{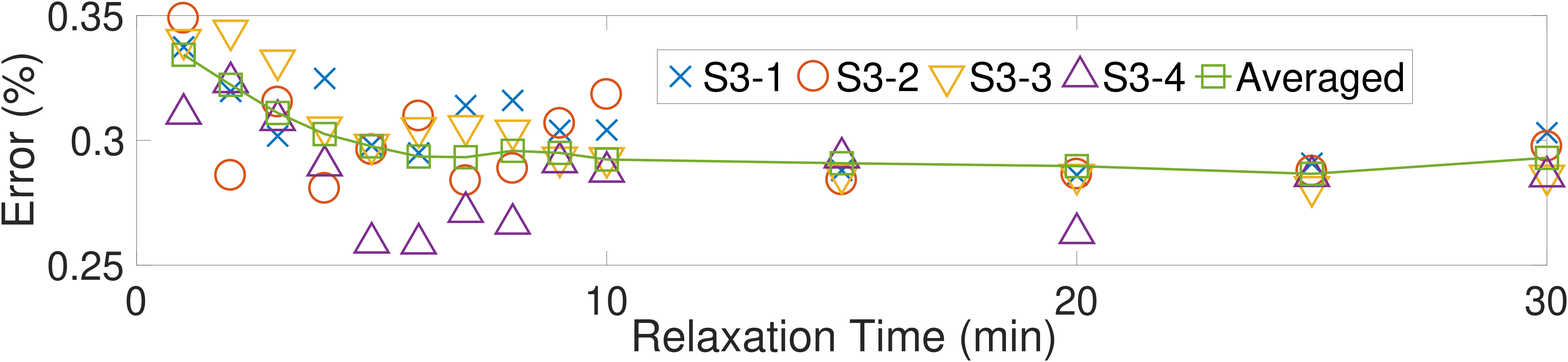}}
\caption{{\bf Impact of relaxing time}}
\end{subfigure}
\vspace{+6pt}

\begin{subfigure}{1\columnwidth}
\centering
{\includegraphics[width = 1\columnwidth]{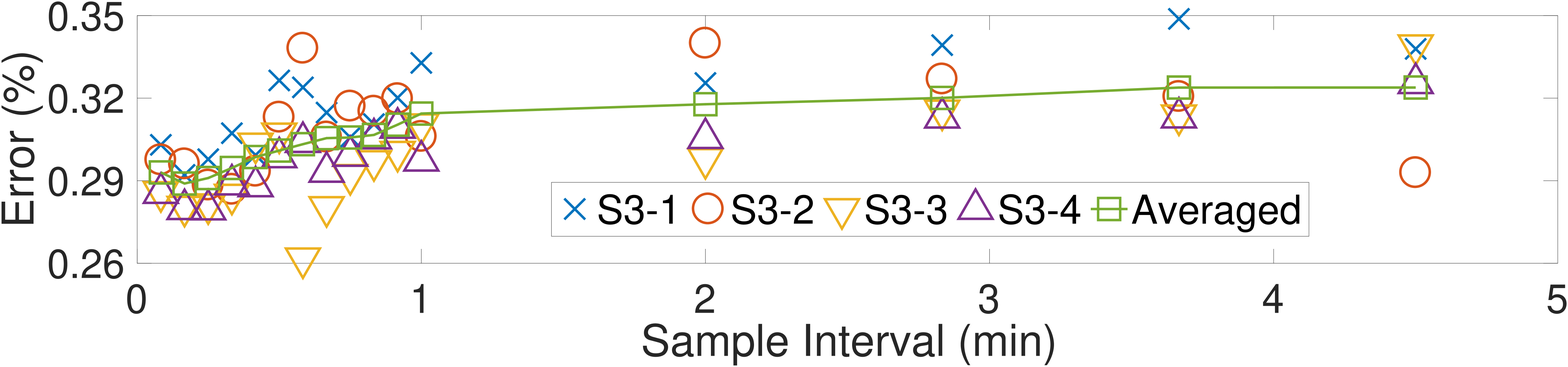}}
\caption{{\bf Impact of sampling interval}}
\end{subfigure}
\end{minipage}
\caption{{\bf Lab experiment results:} \nameS estimates battery SoH with $<$$3\%$ mean error 
and much reduced variance ((a)-(c)); training with multiple batteries increases reliability (d); 
relaxing time need not be very long but has to be logged at a high frequency ((e) and (f)).}
\label{fig:laboratoryresults}
\end{figure}

\section{Evaluation}
\label{sec:evaluations}

We evaluate \nameS using both laboratory experiments and field-tests on multiple Android phones.

\subsection{Laboratory Experiments}

We first evaluate \nameS based on the measurements summarized in Table~\ref{table:testsdetails}. 
Relaxing voltages covering a $30$-minute resting period are used as the fingerprint 
unless specified otherwise.
For the purpose of comparison, we also implement the following three baseline methods:
\begin{itemize}[noitemsep]
\item {\tt Casals'}: the final battery voltage after 5-min relaxation is linear in its SoH~\cite{H}; 
\item {\tt Bond's}: the final battery voltage after 30-min relaxation is quadratic in its SoH~\cite{M};
\item {\tt V-BASH}: the power-factor of battery voltage is linear in its SoH~\cite{iccps17}.
\end{itemize}
Note that {\tt Casals'} and {\tt Bond's} are not always feasible on phones for field-tests 
as the required voltage after a fixed-duration relaxation may not be available due to the trickle charge.

We first evaluate \nameS  based on the dataset collected with each of the batteries, 
whose results are summarized in Fig.~\ref{fig:laboratoryresults}(a), in terms of  the $5$-th and $95$-th 
percentiles of estimation errors (in absolute value) and their mean. 
\nameS estimates battery SoH with $<$$2\%$ mean error, and most of them are bounded by $0.5\%$, 
outperforming the three baselines in all the explored cases. More importantly, \nameS significantly reduces 
the variance in estimation error and thus is much more reliable when compared to the baseline methods.

We also evaluate \nameS by training the fingerprint map with a battery and validate its accuracy with the
traces collected with other same-model batteries, i.e., cross-battery validation. 
This is the real-life analogy of estimating battery SoH of local devices based on an offline-trained fingerprint map.
Fig.~\ref{fig:laboratoryresults}(b) plots the validation results with four Galaxy S3 and two Nexus 5X batteries, 
the symbol $x/y$ denotes training with battery-$x$ and validating with battery-$y$. 
The estimation error, albeit larger than the same-battery evaluation, is still bounded by $2\%$ in most cases.

Users may charge their devices with different chargers from day to day, e.g., using USB or DC chargers. 
Next we use cross-profile evaluation to verify if \nameS is tolerable in such heterogeneous charger cases, 
with the four Galaxy S3 batteries as shown in Fig.~\ref{fig:laboratoryresults}(c). 
Specifically, we train \nameS with the dataset collected when charging with $<$0.5C, 4.20V, 0.05C$>$$_{\rm cccv}$,
and validating its accuracy with the dataset collected when charging with $<$0.25C, 4.20V, 0.05C$>$$_{\rm cccv}$,
i.e., with a constant charge current of $2,200 \times 0.25 = 550$mA, approximately same as when charging 
with standard downstream USB 2.0 ports. 
Comparison of Figs.~\ref{fig:laboratoryresults}(b) and \ref{fig:laboratoryresults}(c) shows no clear evidence 
of degraded SoH estimation due to different charge profiles --- although a few cases resulting 
in $\approx$$2.5\%$ estimation error, the errors in most cases are comparable 
to Fig.~\ref{fig:laboratoryresults}(b) and some are even smaller, verifying \name's robustness 
against charger heterogeneity.

\name's reliability can be improved further by training it with multiple batteries. 
Fig.~\ref{fig:laboratoryresults}(d) plots the SoH estimation error when training \nameS with three of four 
Galaxy S3 batteries and using the fourth one for validation, and compares it with cases of single-battery training. 
The results show that training with multiple batteries reduces the variance in SoH estimation and 
thus improves \name's reliability, at the cost of slightly increased error as compared to the best case 
achieved with single-battery training. 
Note that such best cases, however, are rather random in terms of the battery used for training, 
as shown in Fig.~\ref{fig:laboratoryresults}(d).

We have also explored the impact of relaxing time duration and the voltage sampling rates on 
\name's accuracy in SoH estimation, as shown in Figs.~\ref{fig:laboratoryresults}(e) and 
\ref{fig:laboratoryresults}(f), respectively. The results show the relaxing time need not be very long, 
e.g., the estimation error converges with  $\approx$$10$-minute relaxation, but the $5$-minute 
relaxation in {\tt Casal's} is not enough. Also, \nameS prefers higher sampling rates 
for fine-grained relaxing voltages.

\begin{figure}[t]
\begin{minipage}{1\columnwidth}
\centering
{\includegraphics[width = 1\columnwidth]{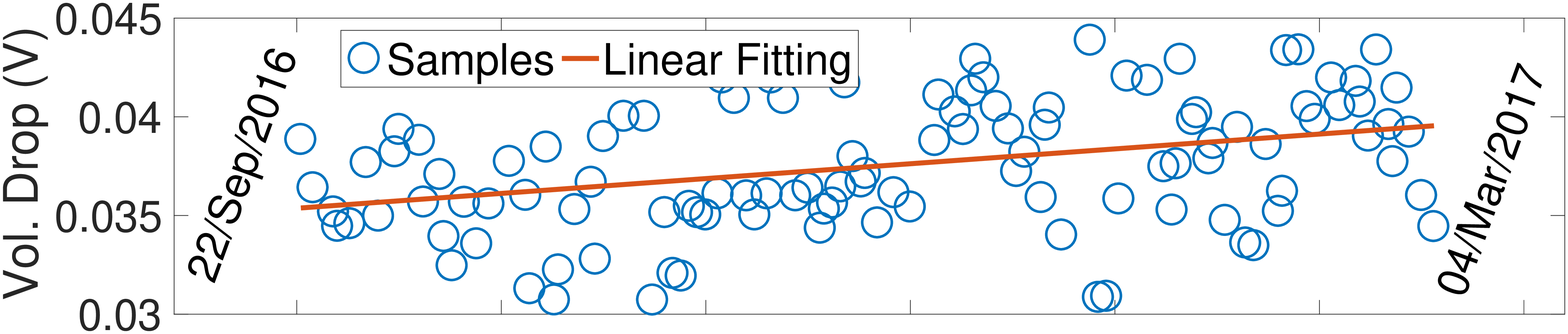}}
\caption{{{\bf Voltage drop increases over usage:} the voltage drop of a Galaxy S5 phone after 30-minute relaxation increases over usage, validating \name's basic principle in SoH estimation.}}
\label{fig:increasingofvdrop2}
\end{minipage}
\end{figure}

\subsection{Field-Tests on Android Devices}

We have also implemented \nameS on multiple Android phones, including Galaxy S5, Galaxy S4, 
Galaxy Note 2, Nexus 6P, and Nexus 5X, and evaluated them over $4$--$6$ months.
To emulate real-life usage, these devices are discharged with various combinations of Youtube, flashlight, and an app 
called {\em BatteryDrainer}~\cite{batterydrainer} that support different discharge rates, 
at an adaptive screen brightness, to a random SoC in the range of $0$--$80\%$.
The devices are then charged for $6$--$10$ hours (mostly over-night) during which the relaxing voltages are 
collected by sampling the system file {\tt /sys/class/power\_supply/battery/voltage\_now}.
We use additional batteries for each device module to train their respective fingerprint maps. 
The ground truth of the battery SoH of Galaxy S5, Galaxy S4, and Galaxy Note 2 are collected by 
removing the battery from the phones and fully charging/discharging them with the battery tester, 
with the same profile as the case of training their respective fingerprint maps. 
The SoH ground truth of Nexus 6P and Nexus 5X, whose batteries are not removable, is collected via Coulomb 
counting based on their current log during discharging, located at {\tt /sys/class/power\_supply/battery/current\_now}.
Although the thus-estimated ground truth may not be perfectly accurate due to the limitation of current sensing, 
this is the best estimation one can get as non-OEM researchers.\footnote{A non-removable battery 
can only be connected to 
the battery tester if an additional wire is soldered on to it, as we did when collecting the training traces 
with additional batteries. This, however, prevents putting the battery back to the phone.}

\begin{figure}[t]
\centering
\begin{minipage}{1\columnwidth}
\centering
\begin{subfigure}{1\columnwidth}
\centering
{\includegraphics[width = 1\columnwidth]{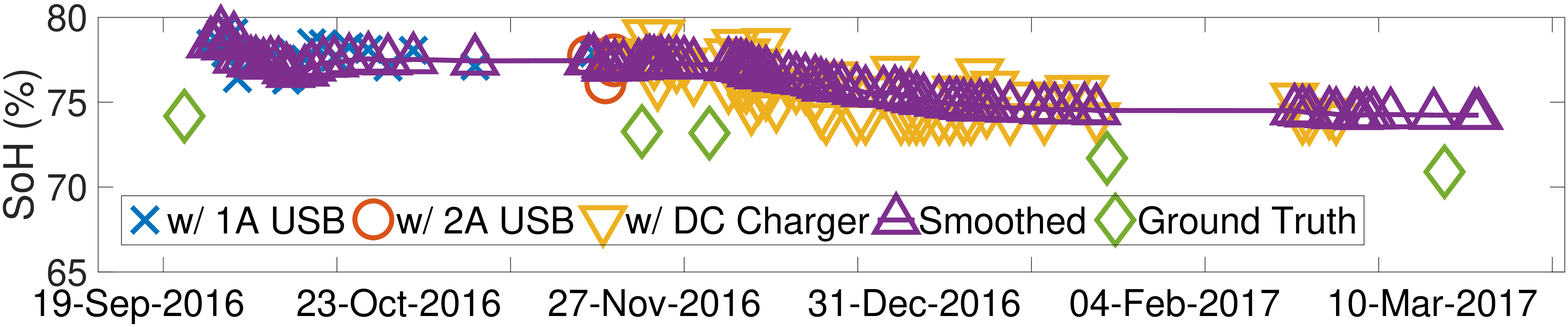}}
\caption{{\bf Galaxy S5}}
\label{fig:SoH_estimation_GalaxyS5}
\end{subfigure}
\\
\begin{subfigure}{1\columnwidth}
\centering
{\includegraphics[width = 1\columnwidth]{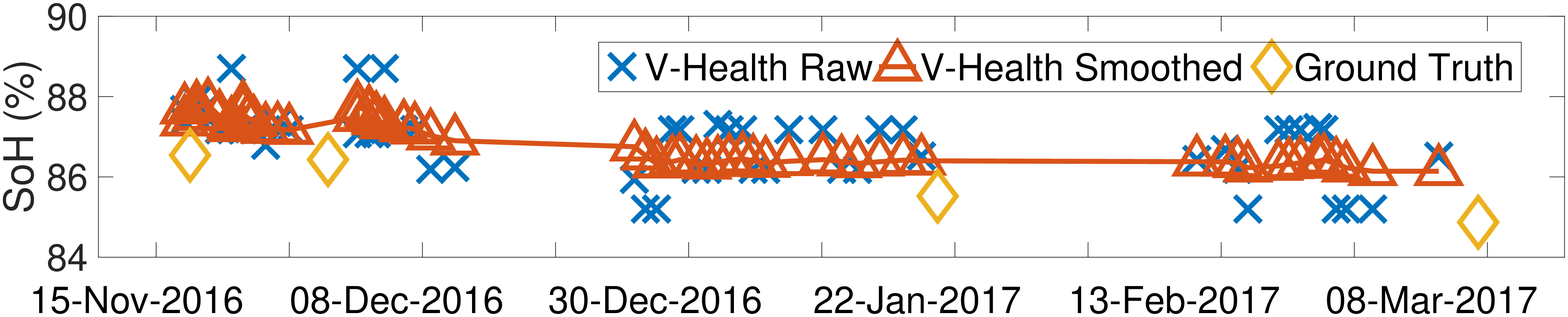}}
\caption{{\bf Galaxy S4}}
\label{fig:SoH_estimation_GalaxyS4}
\end{subfigure}
\\
\begin{subfigure}{1\columnwidth}
\centering
{\includegraphics[width = 1\columnwidth]{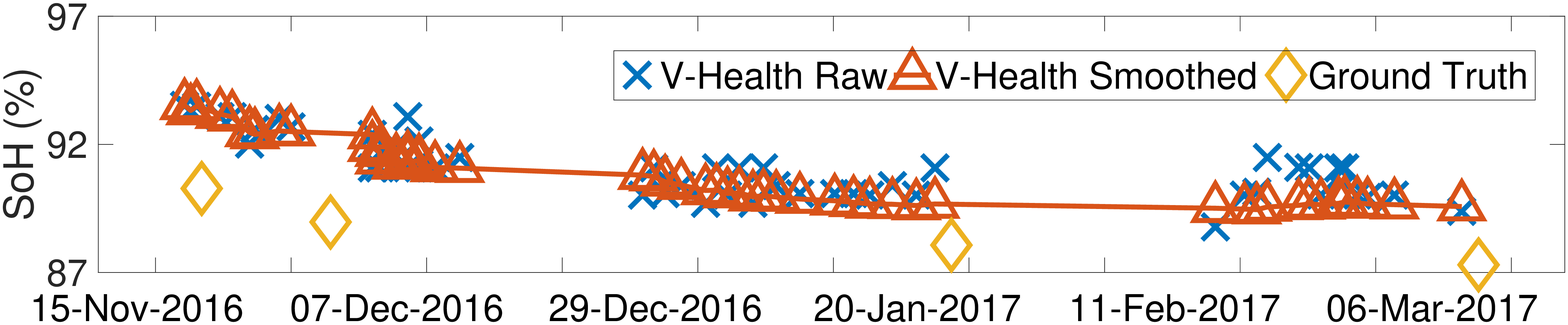}}
\caption{{\bf Galaxy Note 2}}
\label{fig:SoH_estimation_Note2}
\end{subfigure}
\\
\begin{subfigure}{1\columnwidth}
\centering
{\includegraphics[width = 1\columnwidth]{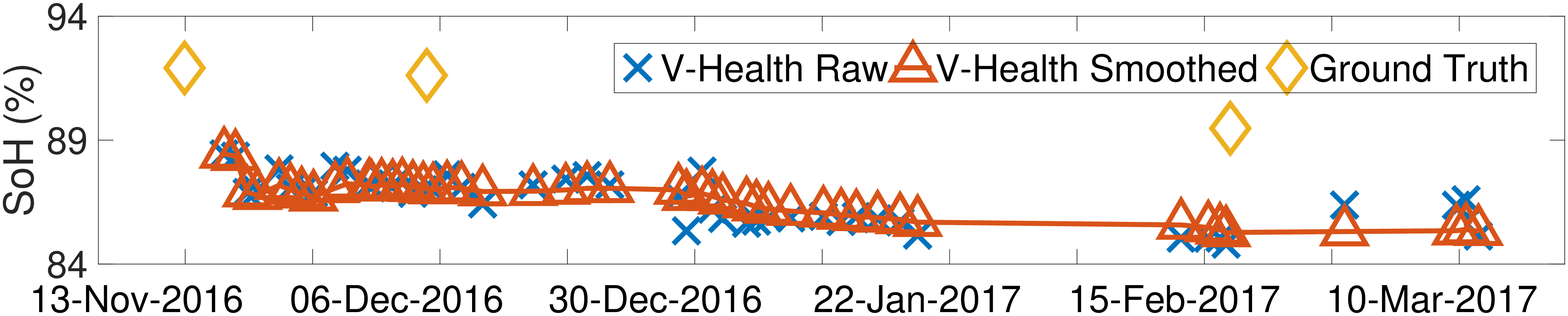}}
\caption{{\bf Nexus 6P}}
\label{fig:SoH_Estimation_Nexus6P}
\end{subfigure}
\\
\begin{subfigure}{1\columnwidth}
\centering
{\includegraphics[width = 1\columnwidth]{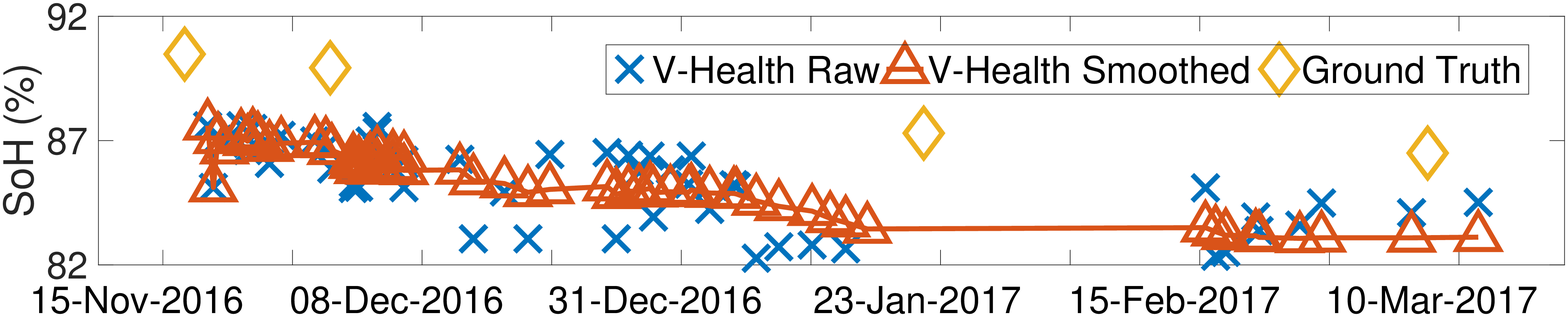}}
\caption{{\bf Nexus 5X}}
\label{fig:SoH_Estimation_Nexus5X}
\end{subfigure}
\caption{{\bf Field-test results:} \nameS estimates battery SoH with $<$$5\%$ error on multiple 
Android devices over experiment periods of $3$--$5$ months.}
\label{fig:fieldtestresults}
\end{minipage}
\end{figure}

We first examine if the voltage-SoH relationship (as in Fig.~\ref{fig:VoltageandSoH}) still holds on smartphones. 
Fig.~\ref{fig:increasingofvdrop2} plots the voltage drop of a Galaxy S5 phone after $30$-minute relaxation upon fully charged, during a period of over $5$ months. Note that the voltage after $30$-minute relaxation may not be available due to trickle charge, in which case we use power fitting to predict such voltage.
The voltage drop increases over usage, during which the battery SoH decreases, agreeing with Fig.~\ref{fig:VoltageandSoH}. 
Significant variance, however, is observed in such voltage drops, indicates methods such as {\tt Bond's} and {\tt Casals'} --- which estimate SoH based on a single voltage reading --- may be unreliable.
Also, the much pronounced variance in Fig.~\ref{fig:increasingofvdrop2} when compared with those in Fig.~\ref{fig:extendingdataset} shows a clear difference between in-laboratory measurements and field-tests on mobile devices, due to the dynamic device operation.

Next we check if \nameS can mitigate such variance and estimate SoH reliably.
Fig.~\ref{fig:fieldtestresults}(a) summarizes the estimated battery SoH with Galaxy S5 from 22/09/2016 
to 10/03/2017, together with the five ground truth SoHs measured on different dates, 
showing $<$$4\%$ errors in SoH estimation.
Also, as stated above, users may charge their devices with different chargers. To cover such cases, 
we charged the phone with different chargers during the evaluation, namely, $1$A USB (22/09/2016 -- 11/11/2016),
$2$A USB (11/11/2016 -- 17/11/2016), and its associated DC charger (18/11/2016 -- 10/03/2017). 
No clear dependency on SoH estimation accuracy and the charger selection is observed, 
demonstrating \name's robustness against heterogeneous chargers.
Finally, the first-order smoother reduces the variance and thus the fluctuations
of SoH reported to users, as compared to the per-charge estimations.
The evaluation results with Galaxy S4 and Note 2 phones are plotted in Figs.~\ref{fig:fieldtestresults}(b) and \ref{fig:fieldtestresults}(c), showing $1.5$--$4\%$ estimation error.

Figs.~\ref{fig:fieldtestresults}(d) and \ref{fig:fieldtestresults}(e) plot the evaluation results with Nexus 6P and Nexus 5X, showing $4$--$5\%$ error 
in SoH estimation. 
This relatively large error could be due partially, besides the inaccurate PMIC-provided current information, 
to battery's rate-capacity effect --- batteries deliver more capacity when discharged with 
less currents~\cite{Rakhmatov,SOSP15}. The two phones have an average discharge 
current of $\approx$$300$mA when collecting their SoH ground truth, much less than the $0.5$C 
discharge rate (i.e., $1,725$mA for Nexus 6P and $1,350$mA for Nexus 5X) used in training the fingerprint maps, thus leading to the 
over-estimation of the batteries' full charge capacity and their SoH. 
Note that the first-order smoother needs at least $3$ samples, causing the initial fluctuation in the smoothed SoH in Fig.~\ref{fig:fieldtestresults}(e).

\begin{table}
\centering
    \caption{{\tt Casals'} and {\tt Bond's} are unreliable on phones.}
  \label{table:vdropresults}
  \scriptsize
  \begin{tabular}{|c|*5c|}
    \hline
    \textbf{\textbf{}} & \textbf{Galaxy S5} &\textbf{Galaxy S4} & \textbf{Note 2} & \textbf{Nexus 5X} & \textbf{Nexus 6P}\\
    \hline
    \hline
            \multirow{1}{*}{\textbf{{Casals'}}}  &52.5\% & $>$$400\%$  & $47.3\%$  & $<$$-1,000\%$ & $>$$900\%$\\
	\hline
	    \multirow{1}{*}{\textbf{{Bond's}}}  &59.3\% & $>$$1,000\%$  & $136.2\%$  & $>$$1,000\%$ & $>$$1,000\%$\\
	\hline
  \end{tabular}
\end{table}

We have also tried to estimate these phones' battery SoH with the two baseline methods {\tt Casals'} and {\tt Bond's} based on the same sets of 
collected relaxing voltages, as summarized in Table~\ref{table:vdropresults}. 
Again, note that the required voltage after $5$- or $30$-minute relaxation may not be available due to trickle charge, 
in which case we use power fitting to predict such voltage and then use it
to estimate SoH.
The SoHs estimated by the two baseline methods have much larger error than \name, and even exceed $100\%$ 
or below $0\%$ in many cases, showing their unreliability on phones.

 \begin{figure*}[t]
\begin{minipage}{2.09\columnwidth}
\begin{subfigure}{0.32\columnwidth}
\centering
{\includegraphics[width=1\columnwidth]{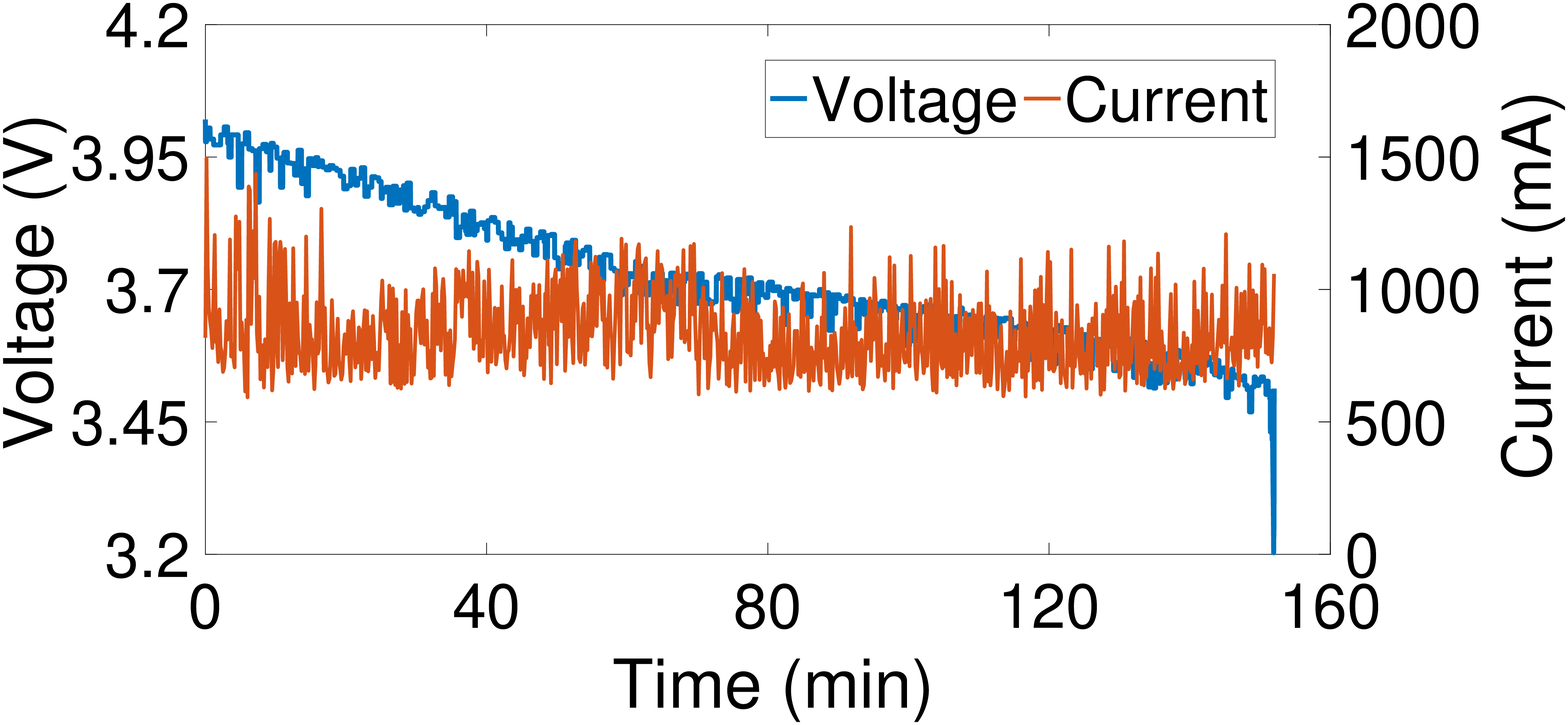}}
\caption{{\bf Discharge an Xperia Z}}
\end{subfigure}
\hfill
\begin{subfigure}{0.32\columnwidth}
\centering
{\includegraphics[width=1\columnwidth]{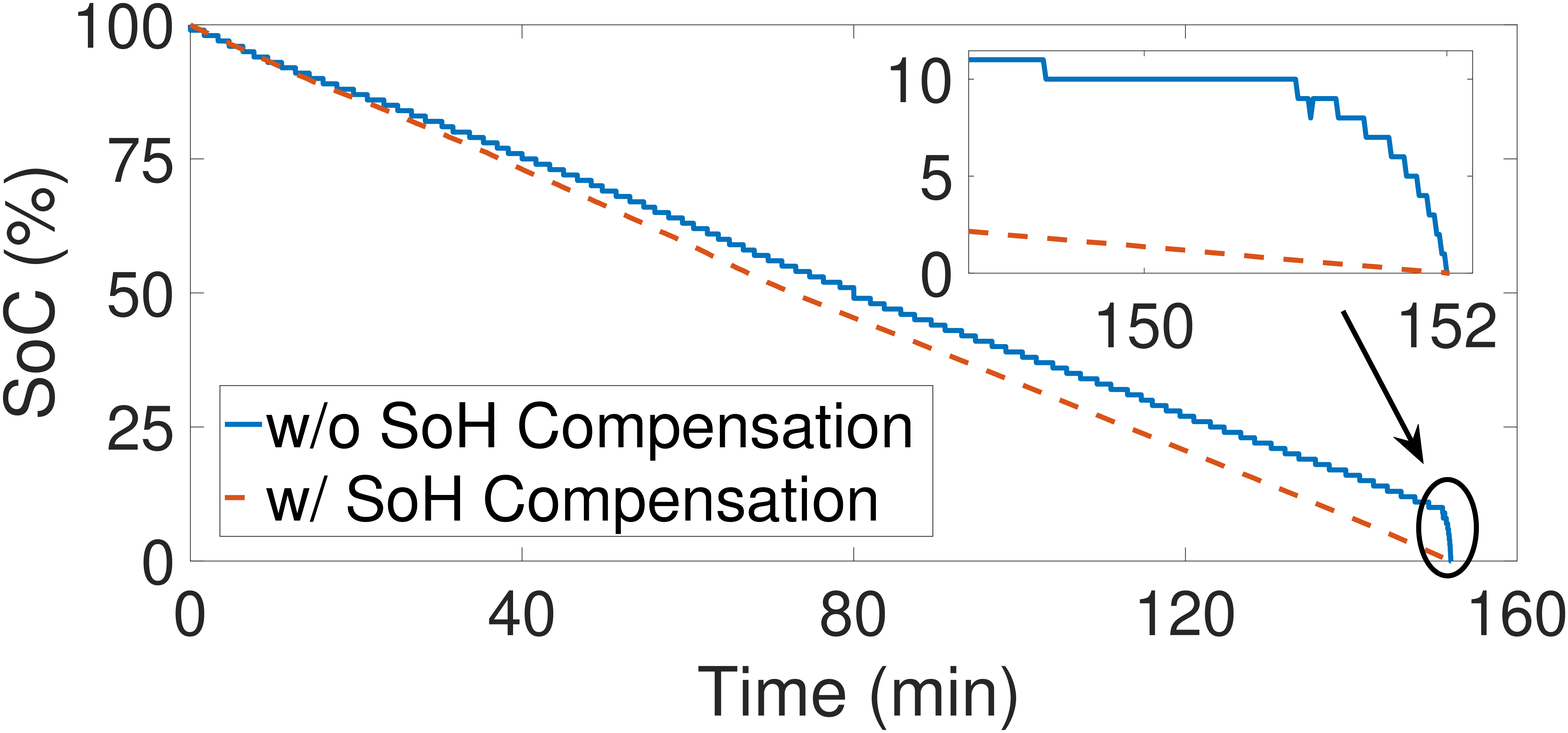}}
\caption{{\bf SoC estimation}}
\end{subfigure}
\hfill
\begin{subfigure}{0.32\columnwidth}
\centering
{\includegraphics[width=1\columnwidth]{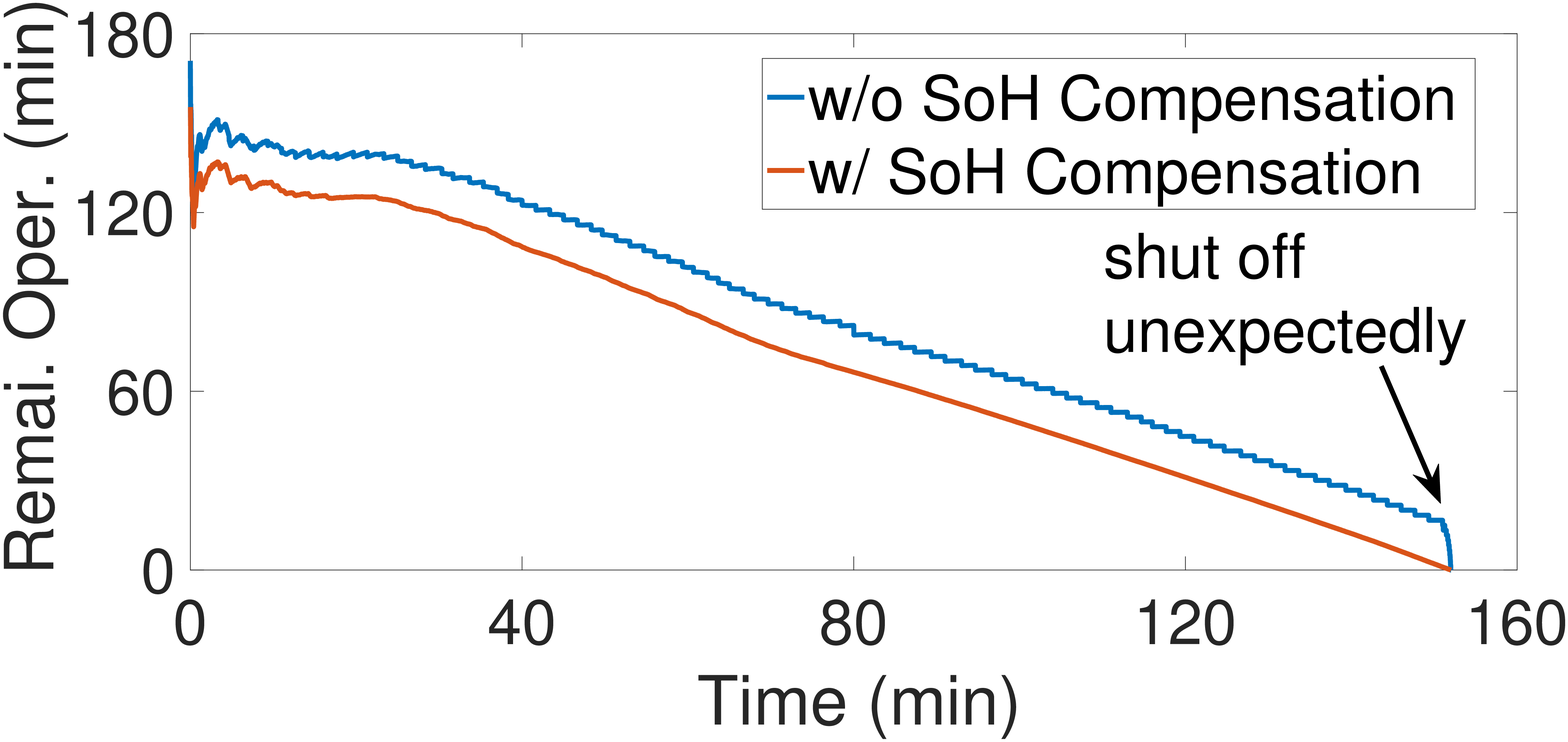}}
\caption{{\bf Remaining operation time}}
\end{subfigure}
\caption{{\bf SoH-compensated SoC estimation:} the knowledge on battery SoH allows more accurate estimation on its SoC and thus remaining operation time, alleviating unexpected device shut-off.}
\label{fig:sohcompensatedsoc}
\end{minipage}
\end{figure*}

\begin{figure*}[t]
\begin{minipage}{0.66\columnwidth}
\centering
{\includegraphics[width=1\columnwidth]{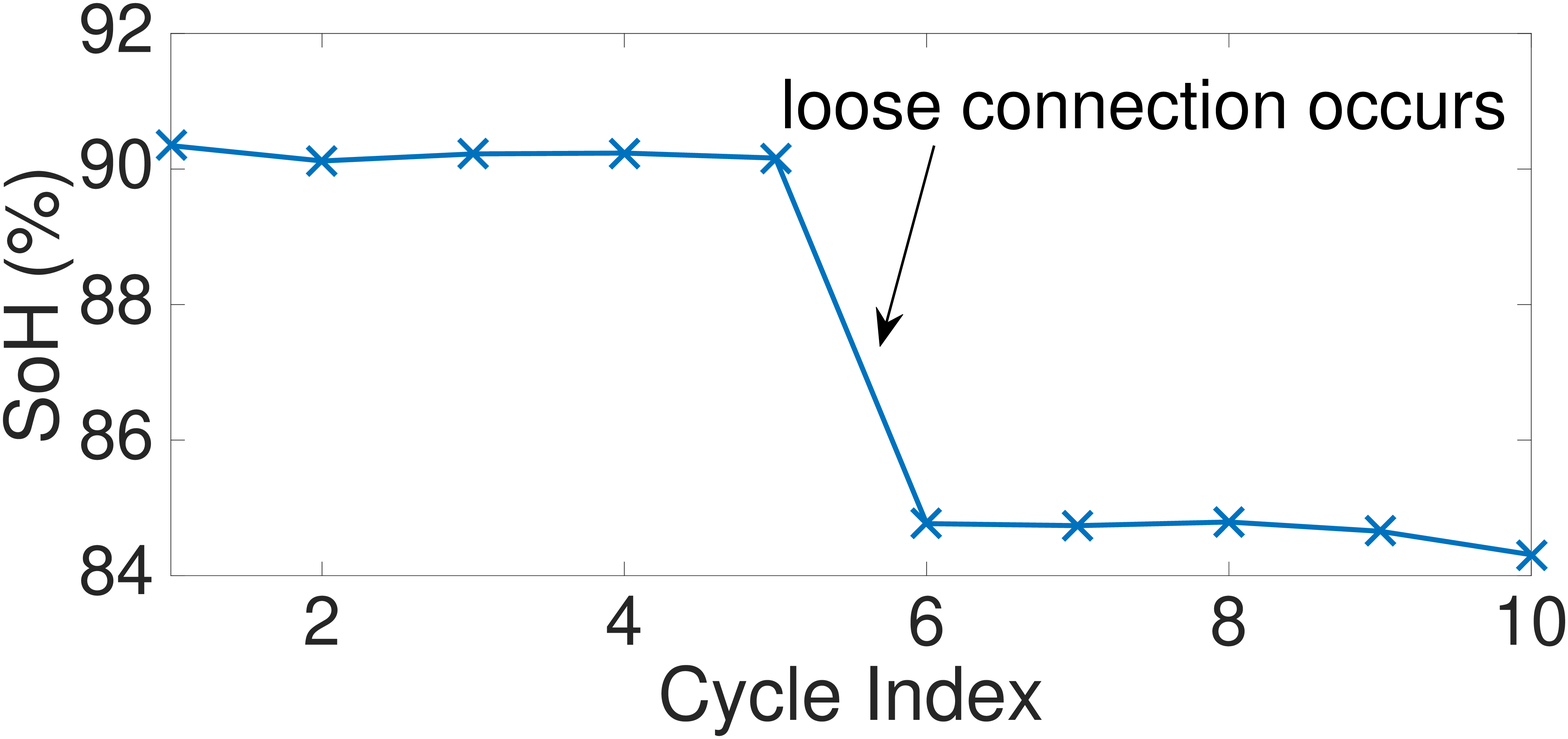}}
\caption{{\bf Abnormal detection:} \nameS detects the loose connection as unusual SoH drop.}
\label{fig:looseconnection_s5}
\end{minipage}
\hfill
\begin{minipage}{0.66\columnwidth}
\centering
{\includegraphics[width=1\columnwidth]{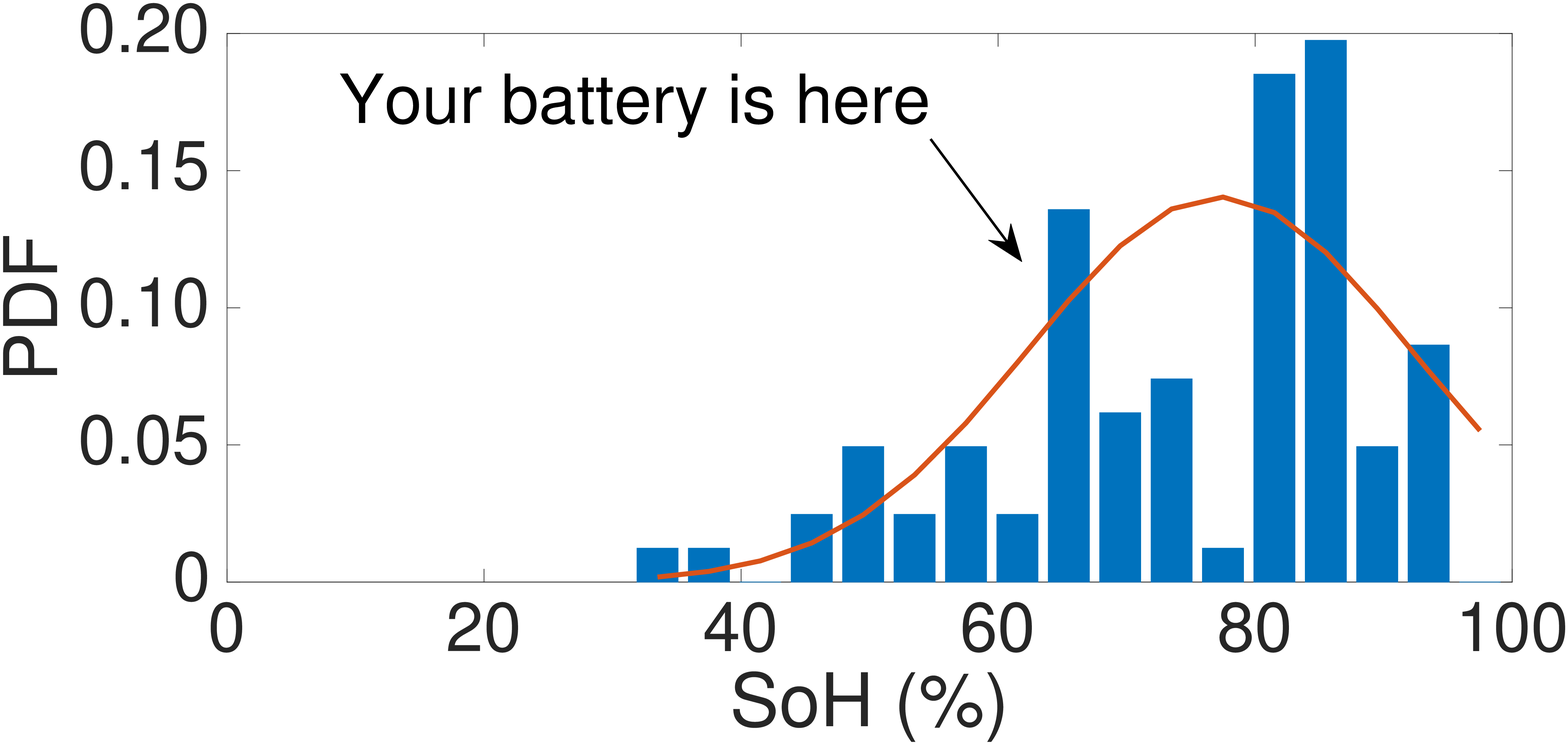}}
\caption{{\bf Cross-user battery comparison:} \nameS allows to compare batteries with others.}
\label{fig:SoHdis}
\end{minipage}
\hfill
\begin{minipage}{0.66\columnwidth}
\centering
{\includegraphics[width = 1\columnwidth]{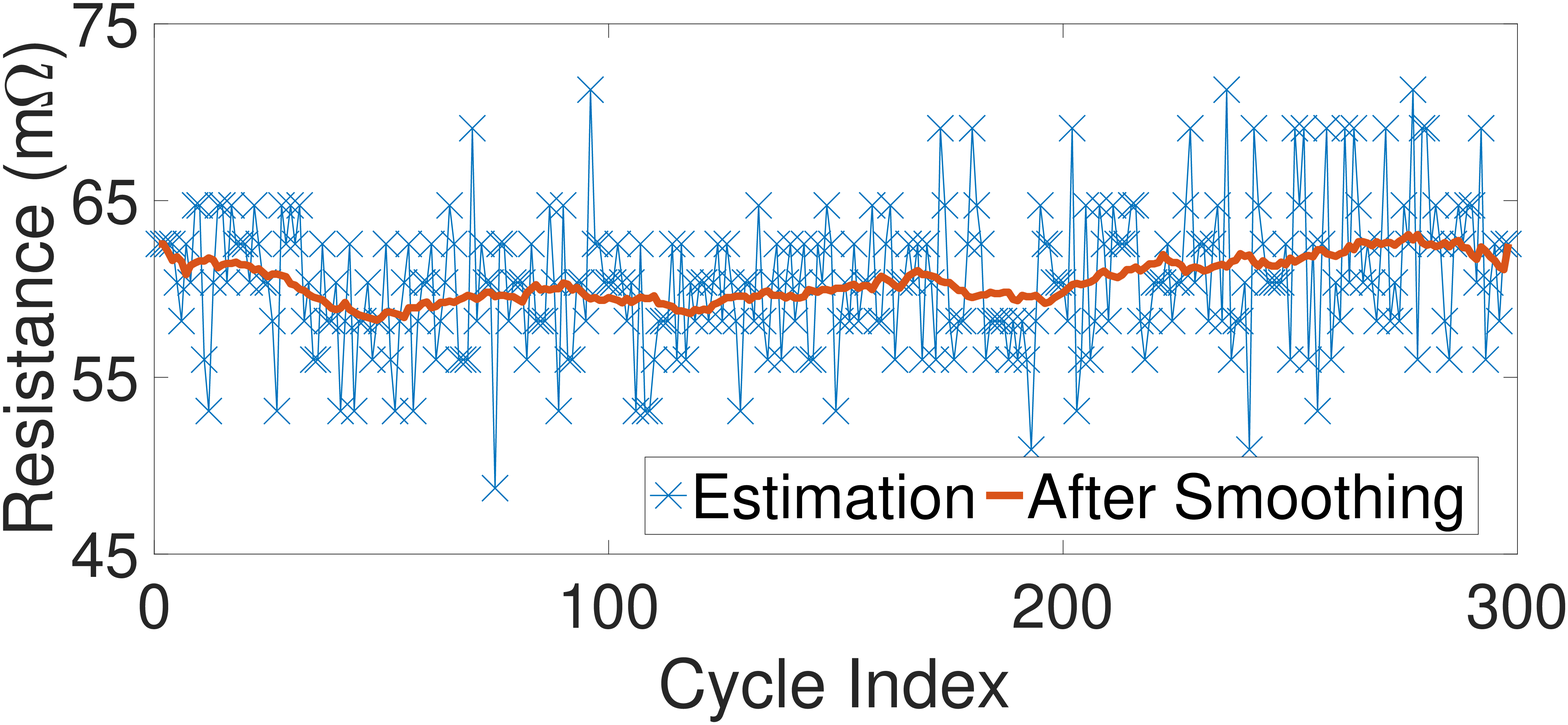}}
\caption{{\bf Battery resistance monitoring:} \nameS facilitates to monitor battery resistance.}
\label{fig:resistance}
\end{minipage}
\end{figure*}

\section{Use-Cases}
\label{sec:usecases}

\nameS also enables four novel use-cases that improve user experience from different perspectives.

\subsection{SoH-Compensated SoC Estimation}

Besides answering the question {\em ``how long will my phone battery last?"} with the interpretation of 
battery lifetime, \nameS also addresses this question in the remaining device operation time, by facilitating 
the SoH-compensated SoC estimation and thus the accurate estimation on phones' remaining power supply.
Fig.~\ref{fig:sohcompensatedsoc}(a) plots the voltage and current when running a fully-charged 
Xperia Z phone with the {\em BatteryDrainer} until shutting off, delivering $2,117$mAh capacity 
in total and thus indicating an SoH of $2,117/2,330 = 90.9\%$. 
Fig.~\ref{fig:sohcompensatedsoc}(b) plots the battery SoC shown to the user during the same discharge 
process --- the phone shuts off with $\approx$$10\%$ remaining SoC. 
Also plotted in Fig.~\ref{fig:sohcompensatedsoc}(b) is the battery SoC compensated with the captured 
SoH degradation, e.g., by \name, based on Eq.~(\ref{equ:socandsoh}), which provides users more 
accurate SoC estimation and thus alleviating shutting the phone off unexpectedly.
Fig.~\ref{fig:sohcompensatedsoc}(c) plots the thus-estimated remaining operation time based on 
the same approach used in TI's {\em Impedance Track}~\cite{impedancetrack} --- the phone shuts off 
when {\em thinking} it can operate $20$ minutes longer due to battery degradation, which can be 
reliably mitigated with the SoH-compensated SoC estimation, enabled by \name.

\subsection{Abnormal Battery Behavior Detection}

The battery SoH monitoring, enabled by \name, also allows to detect battery's abnormal behavior. 
We show this with the example of detecting the loose connection between 
battery and the device, an issue found on devices such as 
Lumia 920~\cite{lumia920looseconnection}, iPhone 5~\cite{iphone5looseconnection}, 
and Note 4~\cite{note4looseconnection}.
Such loose connection increases the connecting resistance and thus device heating, pronouncing the risks of thermal runaway and even battery explosion if not detected in time~\cite{batteryconnection}.
The increased connecting resistance reduces battery's usable capacity; in \name, this is observed as an unusual SoH drop and thus detectable.
We charge/rest/discharge a Galaxy S5 battery for $10$ cycles to validate this: the battery 
is firmly connected to the tester in the first $5$ cycles; in the last $5$ cycles, 
a $100$m$\Omega$ resistor is inserted between the battery and the tester to emulate their loose connection. 
Fig.~\ref{fig:looseconnection_s5} plots the battery SoH reported by \nameS during these $10$ cycles. 
A clear SoH drop is observed when switching from the firm- to loose-connection settings, 
validating its detectability of \name.
Such an unusual battery SoH drop is also explored and verified by Sood {\em et al.}~\cite{K}, 
but with the assistance of an additional ultrasonic pulser and a nanofocus radiographic system.

\subsection{Cross-User Battery Comparison}

Another use-case enabled by \nameS is the cross-user comparison among batteries of same-model devices, as illustrated in Fig.~\ref{fig:SoHdis} based on $82$ Li-ion batteries used in our laboratory. 
Such comparison not only allows users to locate their batteries' strength among others, but also facilitates 
characterization of battery-friendly/harmful usage patterns, when coupled with energy diagnosis services that monitor devices' daily usage, e.g., Carat~\cite{Carat}.

\subsection{Battery Resistance Monitoring}

Batteries' internal resistance increases as they age, reducing their usable capacity and pronouncing device heating. 
Battery resistance is traditionally estimated based on the voltage change when the current switches between two stable levels, i.e., $r = dV/dI$~\cite{vedge,resistancemeasurements}.  The requirements on stable current is to eliminate the influence of dynamic current on voltage response, which, however, does not hold on mobile devices in most cases. 
\name, by collecting the relaxing voltage --- the current before the relaxation  changes gradually and is small (i.e., during CV-Chg) and no current is applied to the battery after entering relaxation, allows for estimation of
battery resistance as a by-product. 
Fig.~\ref{fig:resistance} plots the estimated resistance of a Galaxy S3 battery based on $dV/dI$ after $1$s relaxation~\cite{resistancemeasurements}, according to the relaxing voltages collected in Fig.~\ref{fig:VoltageandSoH}. The battery resistance increases from $58$m$\Omega$ to $63$m$\Omega$ during the measurements, agreeing with the $68$m$\Omega$ ground truth measured with a BVIR battery resistance tester~\cite{BVIR} afterwards these measurements. 
This resistance information helps users/OEMs diagnose their device batteries from another angle.

\section{Conclusions}
\label{sec:conclusions}

In this paper, we have designed, implemented, and evaluated \name, a low-cost user-level 
battery SoH estimation service for mobile devices based solely on their voltage, and thus is deployable 
on all commodity mobile devices. 
\nameS is inspired by our empirical finding that the relaxing battery voltage fingerprints its SoH, 
and is steered by \numTests battery measurements, consisting of \numCycles charging/resting/discharging cycles in total and lasting over \cumMonth months cumulatively.
We have also presented four novel use-cases enabled by \name, improving mobile users' experience in SoC estimation, abnormal behavior detection, cross-user comparison, and resistance monitoring.
We have evaluated \nameS using both laboratory experiments and field-tests with multiple Android 
devices over $4$-$6$ months, showing $<$$5\%$ error in SoH estimation. 
A main takeaway from \nameS is the necessity to integrate physical battery properties with device 
usage behaviors in the battery management of user-centric systems such as smartphones.

\end{spacing}

\small
\bibliographystyle{IEEEtran}\balance
\bibliography{reference}\balance
\normalsize

\end{sloppypar}

\end{document}